\documentclass[11pt,letterpaper]{article}
 
\usepackage{amsmath,amssymb}
\usepackage{array}
\usepackage{cite}

\usepackage{float}
\usepackage{amsmath,amsfonts,graphicx,bm}
\usepackage[usenames]{color}
\usepackage{multirow}

\usepackage{amsmath,amsfonts,graphicx,bbm,multicol}

\usepackage{subfigure}
\usepackage{titlesec}
\newcommand*{\justifyheading}{\centering}
\titleformat{\chapter}[display]
  {\normalfont\huge\bfseries\justifyheading}{\chaptertitlename\ \thechapter}
  {20pt}{\Huge}
\titleformat{\section}
  {\normalfont\Large\bfseries\justifyheading}{\thesection}{1em}{}
\titleformat{\subsection}
  {\normalfont\Large\bfseries\justifyheading}{\thesubsection}{1em}{}

\newcolumntype{x}[1]{
{\centering\hspace{0pt}}p{#1}}

\def\etmiss{E\!\!\!\!\slash_{T}}
\def\ptmiss{p\!\!\!\slash_{T}}

\begin{document}

\baselineskip=18pt

\def\thesection{\Roman{section}} 
\def\thesubsection{\indent\Alph{subsection}}

\newcommand{\eps}{\epsilon}
\newcommand{\pslash}{\!\not\! p}
\newcommand{\I}{\rm 1\kern-.24em l} 
\newcommand{\Tr}{\mathop{\rm Tr}}

\def\wpL{W_L^\prime}
\def\wpR{W_R^\prime}

\def\zpri{Z^\prime}
\def\wpri{W^\prime}

\def\wpLt{\tilde{W}_{L}'}
\def\wpRt{\tilde{W}_{R}'}
\def\mwpL{M_{W_{L}'}}
\def\mwpR{M_{W_{R}'}}

\def\mwpri{M_{W'}}

\def\beq{\begin{equation}}
\def\eeq{\end{equation}}
\def\bea{\begin{eqnarray}}
\def\eea{\end{eqnarray}}
\def\bmat{\begin{pmatrix}}
\def\emat{\end{pmatrix}}
\def\to{\rightarrow}
\newcommand{\chk}[1]{{\bf Check: {\bf #1}}}
\newcommand{\todo}[1]{{\bf To do: {\bf #1}}}
\newcommand{\discuss}[1]{{\bf Discuss: {\it #1}}}
\def\gev{\rm GeV}
\def\tev{\rm TeV}
\def\fbi{\rm fb^{-1}}
\def\lsim{\mathrel{\raise.3ex\hbox{$<$\kern-.75em\lower1ex\hbox{$\sim$}}}}
\def\gsim{\mathrel{\raise.3ex\hbox{$>$\kern-.75em\lower1ex\hbox{$\sim$}}}}
\newcommand{ \slashchar }[1]{\setbox0=\hbox{$#1$}   
   \dimen0=\wd0                                     
   \setbox1=\hbox{/} \dimen1=\wd1                   
   \ifdim\dimen0>\dimen1                            
      \rlap{\hbox to \dimen0{\hfil/\hfil}}          
      #1                                            
   \else                                            
      \rlap{\hbox to \dimen1{\hfil$#1$\hfil}}       
      /                                             
   \fi}                                             %
\def\ptmiss{\slashchar{p}_{T}}
\def\etmiss{\slashchar{E}_{T}}
\providecommand{\tabularnewline}{\\}


\thispagestyle{empty}
\vspace{20pt}
\font\cmss=cmss10 \font\cmsss=cmss10 at 7pt

\begin{flushright}
PITT-PACC-1211
\end{flushright}

\hfill
\vspace{20pt}

\begin{center}
{\Large \textbf
{Lepton Number Violation and $W'$ Chiral Couplings at the LHC
}}

\vspace{5pt}

{\large  
Tao Han$\, ^{a,f}$\footnote{than@pitt.edu}, 
Ian Lewis$\, ^{b}$\footnote{ilewis@bnl.gov},
Richard Ruiz$\, ^{a,c}$\footnote{rer50@pitt.edu},
Zong-guo Si$\, ^{d,e}$\footnote{zgsi@sdu.edu.cn}
}

\vspace{10pt}
$^{a}$\textit{Pittsburgh Particle physics, Astronomy, and Cosmology Center\\
Department of Physics $\&$ Astronomy, University of Pittsburgh, 
Pittsburgh, PA 15260, USA}
\\
$^{b}$\textit{Department of Physics, Brookhaven National
Laboratory, Upton, NY 11973, USA}
\\
$^{c}$\textit{Department of Physics, University of Wisconsin,
Madison, WI 53706, USA}
\\
$^{d}$\textit{Department of Physics, Shandong University,
Jinan, Shandong 250100, P.R.~China}
\\
$^{e}$\textit{Center for High-Energy Physics, Peking University,
Beijing 100871, P.R.~China}
\\
$^{f}$\textit{Center for High Energy Physics, Dept. of Physics, Tsinghua University, 
Beijing, P.R.~China}
\\
\vspace{10pt}
\end{center}

\vspace{15pt}

\begin{center}
\textbf{Abstract}
\end{center}

We study the observability for a heavy Majorana neutrino $N$ along with a new charged gauge boson $W'$ at the LHC. 
We emphasize the complementarity of these two particles in their production and decay to unambiguously determine their properties. 
We show that the Majorana nature of $N$ can be verified by the lepton-number violating like-sign dilepton process, and by polar and azimuthal angular distributions. 
The chirality of the $W'$ coupling to leptons and to quarks can be determined by a polar angle distribution in the reconstructed frame and an azimuthal angle distribution. 

\vfill\eject
\noindent


\section{Introduction}

Neutrino experiments, over the past decade, have shown undeniably that neutrinos are massive and have large mixing angles~\cite{BargerReview}.
In the  Standard Model (SM) of particle physics, neutrino masses can be accommodated by a non-renormalizable dimension-5 operator containing left-handed (L.H.) neutrinos, $\nu_{L}$ \cite{dim6}.  Such an operator can be generated at low energy by including heavy right-handed (R.H.) neutrinos, $\nu_{R}$. 
However, the R.H.~neutrinos are gauge singlets and so Majorana mass terms should also be present without violating any gauge symmetry.
The consequences of massive Majorana neutrinos are well-known~\cite{seesaw,TypeIII,Inverse}, 
and have been incorporated into many models, such as left-right symmetric theories~\cite{LRModels}; 
supersymmetric (SUSY) $SO(10)$ grand unified theories (GUTs)~\cite{SO10SUSYGUT} and other GUTs~\cite{MGUT}; 
R-parity violating SUSY~\cite{RparitySUSY}; and extra dimensions~\cite{ExtraDim}.  
A recent review of TeV scale neutrino mass models can be found in Ref.~\cite{Chen:2011de}.

Many of the aforementioned models contain an extended gauge group or Keluza-Klein (KK) excitations of SM gauge bosons.
We refer to additional vector bosons charged under the $U(1)_{EM}$ gauge group collectively as ``$\wpri$''.
If the masses of the $\wpri$ and the lightest heavy neutrino mass eigenstate, $N$, 
are both on the order of a few TeV, then they can be produced in tandem at the Large Hadron Collider (LHC).
As first observed by Ref.~\cite{Keung:1983uu}, 
a $\wpri$ with mass greater than a Majorana neutrino's mass allows the possibility of observing the spectacular lepton number ($L$) violating process 
\begin{equation}
 pp\rightarrow \wpri\rightarrow \ell^\pm N\rightarrow \ell^\pm\ell^\pm jj.
\label{eq:WprN}
\end{equation}

If a $\wpri$ is discovered at the LHC~\cite{schmaltz:2011lrsm}, it is obviously imperative to measure its chiral coupling to fermions.
In a previous work~\cite{Gopalakrishna:2010}, three of the present authors proposed measuring the $W'$ chiral couplings to quarks by studying the process
\begin{equation}
pp\rightarrow\wpri\rightarrow t\bar{b}\rightarrow \ell^+\nu_\ell b\bar{b}.
\end{equation}
It was found that the couplings could be establish as being purely left- or purely right-handed by analyzing the 
polar angle of the charged lepton in the top's rest-frame with respect to the top's direction of motion in the partonic center of momentum (c.m.) frame.

We now extend this prior analysis into the leptonic sector via the $L$-violating cascade decay of Eq.~(\ref{eq:WprN}).
More specifically, by reconstructing the polar angle of the lepton originating from the neutrino decay in the 
neutrino rest-frame and with respect to the direction of motion of the neutrino in the partonic c.m.~frame, 
it can be uniquely determined if the $\wpri$ coupling to leptons is purely left-handed, purely right-handed, 
or a mixture of the two.
We show that the distribution of the angle made between $N$'s production plane and its sequential decay 
plane is sensitive to the $\wpri$ chiral coupling with the initial-state quarks but $independent$ of the $\wpri$ coupling to leptons.  
These results are demonstrated through a combination of analytical calculations and event simulations, assuming nominal LHC parameters.

Majorana neutrinos can decay into either leptons or antileptons, and so $\wpri$ and $N$ may also contribute to the $L$-conserving collider signature
\begin{equation}
 pp\rightarrow \wpri\rightarrow \ell^\pm N\rightarrow \ell^+\ell^- jj.
\label{eq:WprNpm}
\end{equation}
For completeness, we have analyzed the polar angular distributions of the unlike-sign process and comment on the important differences between the $L$-conserving and $L$-violating cases.

This paper is structured as follows:  
First, in section~\ref{ThFrame.SEC}, we present our notation for the $\wpri$ couplings to SM particles and neutrino mass eigenstates, and list current constraints on both $\wpri$'s and $N$'s.
In section~\ref{WpLHC.SEC}, we discuss the production and decay of $\wpri$'s and $N$'s at the LHC.
The like-sign lepton signature, $pp\rightarrow\ell^\pm\ell^\pm jj$, its reconstruction, and suppressed background are fully analyzed in section~\ref{Like.SEC}.
In \ref{WpriAC.SEC}, we propose methods to measure independently the chiral couplings of the $\wpri$ to leptons and to the initial-state quarks.
Finally, in section~\ref{Opp.SEC}, we provide a few comments on the contribution of $\wpri$ and $N$ to the $L$-conserving process $pp\rightarrow \wpri\rightarrow \ell^+\ell^- jj$ regarding the difference between the Majorana and Dirac neutrinos. 
We conclude and summarize our results in section~\ref{Conc.SEC}.
Two appendices are additionally included. 
The first addresses neutrino mass mixing in the context of $W'$ couplings, and the second presents a derivation of the matrix element and angular distributions for our like- and unlike-sign dilepton signals.

\section{Theoretical Framework and Current Constraints}
\label{ThFrame.SEC}
There are many Beyond the Standard Model (BSM) theories containing additional  
vector bosons that couple to SM fermions,
for example: left-right symmetric theories~\cite{Pati:1974yy} with a new $SU(2)_R$ symmetry and an associated $W'_R$; 
Little Higgs models with enlarged gauge symmetries~\cite{ArkaniHamed:2002qy}; 
extra dimensional theories with KK excitations~\cite{Appelquist:2000nn,Csaki:2003zu,Agashe:2003zs}.  
Heavy Majorana neutrinos in BSM theories~\cite{LRModels,SO10SUSYGUT,MGUT,RparitySUSY,ExtraDim}, 
and in particular those with TeV-scale masses~\cite{Dorsner:2006fx,Bajc:2007zf,deGouvea:2006gz,de Gouvea:2007uz}, are just as common.

In this analysis, we assume the existence of a new heavy electrically charged vector boson, $W^{'\pm}$ with mass $M_{W'}$, and a right-handed neutrino, $N_{R}$.
We denote the corresponding heavy neutrino mass eigenstate as $N$ with mass $m_{N}$.
We stipulate that $M_{W'}$ is of the order of a few TeV and  $M_{W'}>m_{N}$ 
 so as the $W'\rightarrow N\ell$ decay is kinematically accessible by the LHC, but do not otherwise tailor to a specific theory.
Regarding the parameterization of mixing between neutrino mass eigenstates with SM flavor eigenstates, we adopt the notation of Ref.~\cite{Atre:2009rg}, 
and extend it to include coupling to a model-independent $W'$ in Appendix~\ref{appendNeu.APP}.
This parameterization is accomplished with a minimum amount of parameters.

\subsection{$\wpri$ Chiral Coupling to Fermions}
The model-independent Lagrangian that governs the interaction between SM quarks and a new, massive, electrically charged vector boson, $W'$, is given by
\begin{equation}
 \mathcal{L}=-\frac{1}{\sqrt{2}}\sum_{i,j=1}^{3}W_{\mu}^{'+}\overline{u_{i}}V_{ij}^{CKM'}\gamma^{\mu}\left[g_{R}^{q}P_{R}+g_{L}^{q}P_{L}\right]d_{j}+{\it h.c.},
\end{equation}
where $u_i\ (d_j)$ denotes the  
Dirac spinor of an up-(down-)type quark with flavor $i\ (j)$; 
$V^{CKM'}$ parameterizes the mixing between flavors $i$ and $j$ for the new charged current interactions 
just as the Cabibbo-Kobayashi-Maskawa (CKM) matrix does in the SM;
$g_{R,L}^{q}$ is the $W'$'s universal coupling strength to right-(left-) handed quarks; 
and $P_{R,L}=\frac{1}{2}\left(1\pm\gamma_5\right)$ denotes the $R,L$-handed chiral projection operator.
\label{wpriNcoup.SEC}

We parameterize the new boson's coupling to charged leptons with flavor $\ell$ and neutral leptons with mass $m_{m}$ (for the three light states) or $m_{N}$ (for the heavy state) in the following way:
\begin{eqnarray}
\mathcal{L}=
&-&\sum_{\ell=e}^{\tau}\frac{g_{R}^{\ell}}{\sqrt{2}}W_{\mu}^{'+}\left[\sum_{m=1}^{3}\overline{\nu_{m}^{c}}X_{\ell m} + \overline{N}Y_{\ell N}\right]\gamma^{\mu}P_{R}\ell^{-}\nonumber\\
&-&\sum_{\ell=e}^{\tau}\frac{g_{L}^{\ell}}{\sqrt{2}}W_{\mu}^{'+}\left[\sum_{m=1}^{3}\overline{\nu_{m}}U_{\ell m}^{*} + \overline{N^{c}}V_{\ell N}^{*}\right]\gamma^{\mu}P_{L}\ell^{-}+{\it h.c.}
\label{wpLagrangian.EQ}
\end{eqnarray}
$g_{R}^{\ell}~(g_{L}^{\ell})$ is the $W'$'s coupling strength to R.H. (L.H.) leptons; 
$X_{\ell m}~(U_{\ell m})$ parameterizes the mixing between light neutrino mass eigenstates and R.H. (L.H.) interactions; and
$Y_{\ell N}~(V_{\ell N})$ parameterizes the mixing between the heavy neutrino mass eigenstate and R.H. (L.H.) interactions.
Lastly, $\psi^{c}=\mathcal{C}\overline{\psi}^{T}$ denotes the charge conjugate of the field $\psi$, 
with $\mathcal{C}$ being the charge conjugate operator, and the chiral states satisfy $P_{L}(\psi^{c})=(P_{R}\psi)^{c}.$
In Appendix~\ref{appendNeu.APP}, our choice of parameterization is discussed in detail. 
From a viewpoint of the model construction as discussed in Refs.~\cite{BargerReview,Keung:1983uu,Atre:2009rg}, one may expect that 
$UU^{\dagger},~YY^{\dagger}\sim\mathcal{O}(1)$ and 
$VV^{\dagger},~XX^{\dagger}\sim\mathcal{O}(m_{m}/m_{N}).$ 
Since we prefer a model-independent approach, we will not follow rigorously the above argument and will take the parameters as 
\begin{equation}
UU^{\dagger},~YY^{\dagger}\sim\mathcal{O}(1),\quad\text{and}\quad
VV^{\dagger},~XX^{\dagger}\sim\mathcal{O}(10^{-3}),
\label{mixingFromUnit.EQ}
\end{equation}
which is guided by the current constraints as presented later in this section.

In Eq.~(\ref{wpLagrangian.EQ}), the $\wpri$ is allowed to have both independent right-handed $(g_R^{q,\ell})$ and left-handed $(g_L^{q,\ell})$ couplings.
Subsequently, the pure gauge states $W'_R$ and $W'_L$ are special cases of $W'$ when
\begin{equation}
g_{R}^{q,\ell} \ne 0\quad\text{and}\quad g_{L}^{q,\ell}=0,
\label{pureGuageRH.EQ}
\end{equation}
and
\begin{equation}
g_{R}^{q,\ell}=0\quad\text{and}\quad g_{L}^{q,\ell} \ne 0,
\label{pureGuageLH.EQ}
\end{equation}
respectively. Additionally, the SM $W$ coupling to leptons can be recovered from Eq.~(\ref{wpLagrangian.EQ}) by setting
\begin{equation}
g_{R}^{\ell}=0,\quad\text{and}\quad g_{L}^{\ell}=g.
\end{equation}
Here, $g$ is the usual SM SU$(2)_L$ coupling constant.

\subsection{Current Constraints on $W'$}
\label{Constraints.SEC}
We list only the most stringent, most relevant constraints to our analysis here and refer the reader to Ref.~\cite{PDG:2012,Constraints} for a more complete review.

\begin{itemize}
\item $\bf{Bounds~from~CMS}$: 
The CMS Experiment has searched for $W_{R}$ and heavy $N$, where $M_{W_{R}}>m_{N}$, with the $\ell^{\pm}\ell^{\pm}jj$ collider signature~\cite{Chatrchyan:2011WR}, 
assuming $g_{R}=g$. With 5.0 fb$^{-1}$ of 7 TeV and 3.7 fb$^{-1}$ of 8 TeV $pp$ collisions, the present mass bounds for $W'_{R}$ and $N$ are
\begin{equation}
 M_{W_{R}} > 2.9\text{ TeV }(m_{N}\approx0.8\text{ TeV})\quad\text{and}\quad m_{N}>1.9\text{ TeV }(M_{W_{R}}\approx2.4\text{ TeV}.)
\end{equation}
The search for the sequential SM $W'$, $W'_{SSM}$, decaying into a charged SM lepton plus~$\slash\!\!\!\!E_{T}$, with $g'=g$, has also been performed. 
With 3.7 fb$^{-1}$ of 8 TeV $pp$ collisions \cite{Chatrchyan:2012Wssm}, the present mass bound is
\begin{equation}
 M_{W_{SSM}} > 2.85\text{ TeV.}
\end{equation}

\item $\bf{Bounds~from~ATLAS}$: The ATLAS Experiment has also searched for $W_{R}$ and heavy $N$, under the same stipulations as the CMS Experiment~\cite{Aad:2012WR}. 
With 2.1 fb$^{-1}$ of 7 TeV $pp$ collisions, the present mass bounds for $W'_{R}$ and $N$ are
\begin{equation}
 M_{W_{R}} > 2.5\text{ TeV }(m_{N}\approx0.8\text{ TeV})\quad\text{and}\quad m_{N}>1.6\text{ TeV }(M_{W_{R}}\approx1.8\text{ TeV}.)
\end{equation}
\item $\bf{Global~Fit~Analysis}$: The effects of a generic $Z'$ boson on EW precision observables place bounds~\cite{Cacciapaglia:2006pk} of 
\begin{equation}
 M_{Z'}/g_{Z'}\gtrsim 2.7-6.7~\text{TeV}.
\end{equation}
For $Z'$ and $W'$ bosons originating from the same broken symmetry, we expect similar constraints on $M_{\wpri}/g_{\wpri}$ since
\begin{equation}
 M_{W'}\sim M_{Z'}\times\mathcal{O}(1).
\end{equation}
\item $\bf{Bounds~on~W_{L}-W_{R}~Mixing}$: Non-leptonic Kaon decays~\cite{Donoghue:1982} 
and universality in Weak decays~\cite{Wolf:1984} constrain $W_{L}-W_{R}$ mixing. 
The present bound for the L-R mixing angle $\zeta$~\cite{LRModels} is
\begin{equation}
 \vert \zeta \vert \leq 1\sim4\times 10^{-3}.
\end{equation}
\end{itemize}

\subsection{Current Constraints on $N$}
More complete lists of constraints on low and high mass neutrinos, respectively, are available~\cite{Atre:2009rg,PDG:2012}.
\begin{itemize}
\item $\bf{Bounds~from~0\nu\beta\beta}$: For $m_{N}\gg 1$ GeV, a lack of evidence for neutrinoless double beta decay bounds the mixing between heavy neutrino states and the electron-flavor state at~\cite{Belanger:1995nh}
\bea
\displaystyle\sum_{m'}\frac{|V_{em'}|^2}{m_{m'}}< 5\times10^{-5}\,{\rm TeV}^{-1},
\label{zerovtwobBound.EQ}
\eea
where the sum is over all heavy Majorana neutrinos.
\item $\bf{Bounds~from~EW~Precision~Data}$: A TeV scale singlet neutrino mixing with the SM flavor states is constrained~\cite{delAguila:2008pw} by
\begin{equation}
 |V_{eN}|^2,|V_{\mu N}|^2<0.003\quad\text{and}\quad|V_{\tau N}|^2<0.006
 \label{emutauMixingBounds.EQ}
\end{equation}
\end{itemize}
%
\section{$W'$ and $N$ Production and Decay at the LHC}
\label{WpLHC.SEC}

For the remainder of this analysis, we consider for our various benchmark calculations only the 
pure gauge states $W'_{R}$ and $W'_{L}$, respectively given by Eq.~(\ref{pureGuageRH.EQ}) and Eq.~(\ref{pureGuageLH.EQ}), and with SM coupling strength
\begin{equation}
 g^{q,\ell}_{R,L}=g.
\end{equation}
More general results can be obtained by simple scaling. Unless explicitly stated otherwise, we take
\begin{equation}
\label{benchParam.EQ}
 M_{W'}=3\text{ TeV},\quad 
 m_{N}=500\text{ GeV},\quad
\vert V^{CKM'}_{ud}\vert^{2}=1,
 \end{equation}
 and use the CTEQ6L1 parton distribution functions (pdfs)~\cite{Pumplin:2002vw} for all hadronic-level cross section calculations.
 Explicitly, we consider only the $ud\rightarrow W'$ production mode.

 Regarding our choice of neutrino mixing parameters, for mixing between L.H. gauge states and light mass eigenstates, 
 we use the Pontecorvo–Maki–Nakagawa–Sakata (PMNS) matrix with mixing angles taken from Ref.~\cite{PDG:2012}, which includes recent measurements of 
 $\theta_{13}$, and take $\delta_{CP},\alpha_{1},\alpha_{2}=0$.
 The bounds from $0\nu\beta\beta$ decay are quite severe and discourage collider searches for $L-$violation in the electronic channel.
However, neutrino mixing between the mu- or tau-flavor state and lightest heavy mass eigenstate can still be considerably larger in L.H. interactions.
Therefore, we use 
\begin{equation}
\vert V_{eN}\vert^{2}=2.5\times10^{-5},
\quad
\vert V_{\mu N}\vert^{2}=1\times10^{-3},
\quad\text{and}\quad
\vert V_{\tau N}\vert^{2}=1\times10^{-3}.
\label{mixingHeavyLH.EQ}
\end{equation}
These numerical values are in line with Eqs.~(\ref{zerovtwobBound.EQ}), (\ref{emutauMixingBounds.EQ}), and (\ref{benchParam.EQ}); 
and  furthermore, mimic the observed $\mu-\tau$ symmetry seen in mixing between flavor states and light mass eigenstates. 
Where necessary, for mixing between R.H gauge states and light mass eigenstates, we apply the unitarity condition
\begin{equation}
\sum_{m=1}^{3}\vert X_{\ell m}\vert^{2} = 1 - \sum_{m=1}^{3}\vert U_{\ell m}\vert^{2},
\quad\text{for}\quad
\ell=e,\mu,\tau.
\end{equation}
For mixing between R.H. gauge states and the lightest, heavy mass eigenstate, we apply Eq.~(\ref{mixingFromUnit.EQ}) and take
\begin{equation}
\vert Y_{\ell N}\vert^{2}=1,
\quad\text{for}\quad
\ell=e,\mu,\tau.
\label{mixingHeavyRH.EQ}
\end{equation}
\subsection{$\wpri$ Production and Decay}

\begin{figure}[tb]
\centering
\subfigure[]{
	\includegraphics[width=0.45\textwidth]{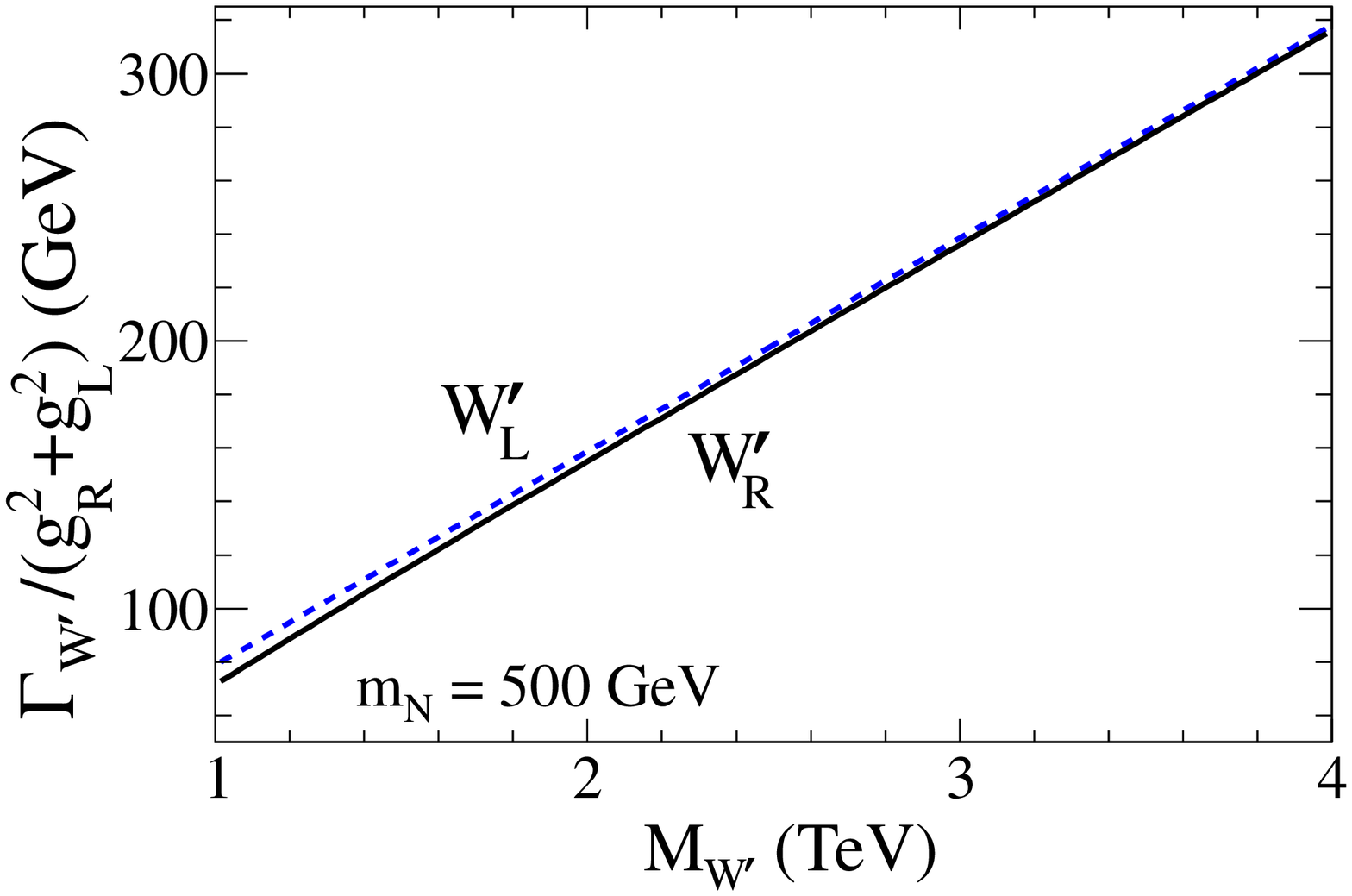}
	\label{wpwidth.FIG}
}
\subfigure[]{
	\includegraphics[width=0.45\textwidth]{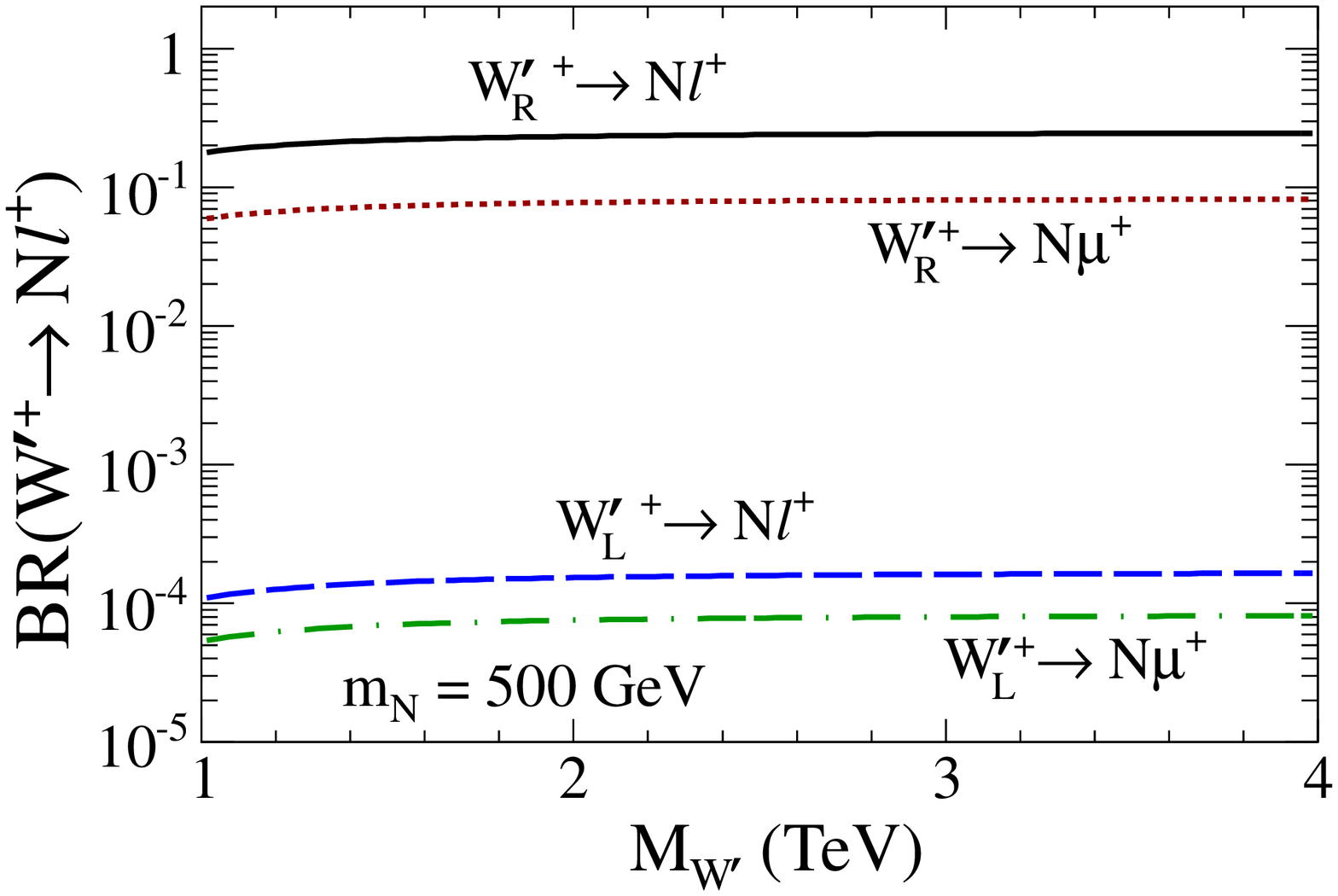}
	\label{BR.FIG}
}\vspace{.15in}\\
\subfigure[]{
	\label{mwprod_8TeV.FIG}
	\includegraphics[width=0.45\textwidth]{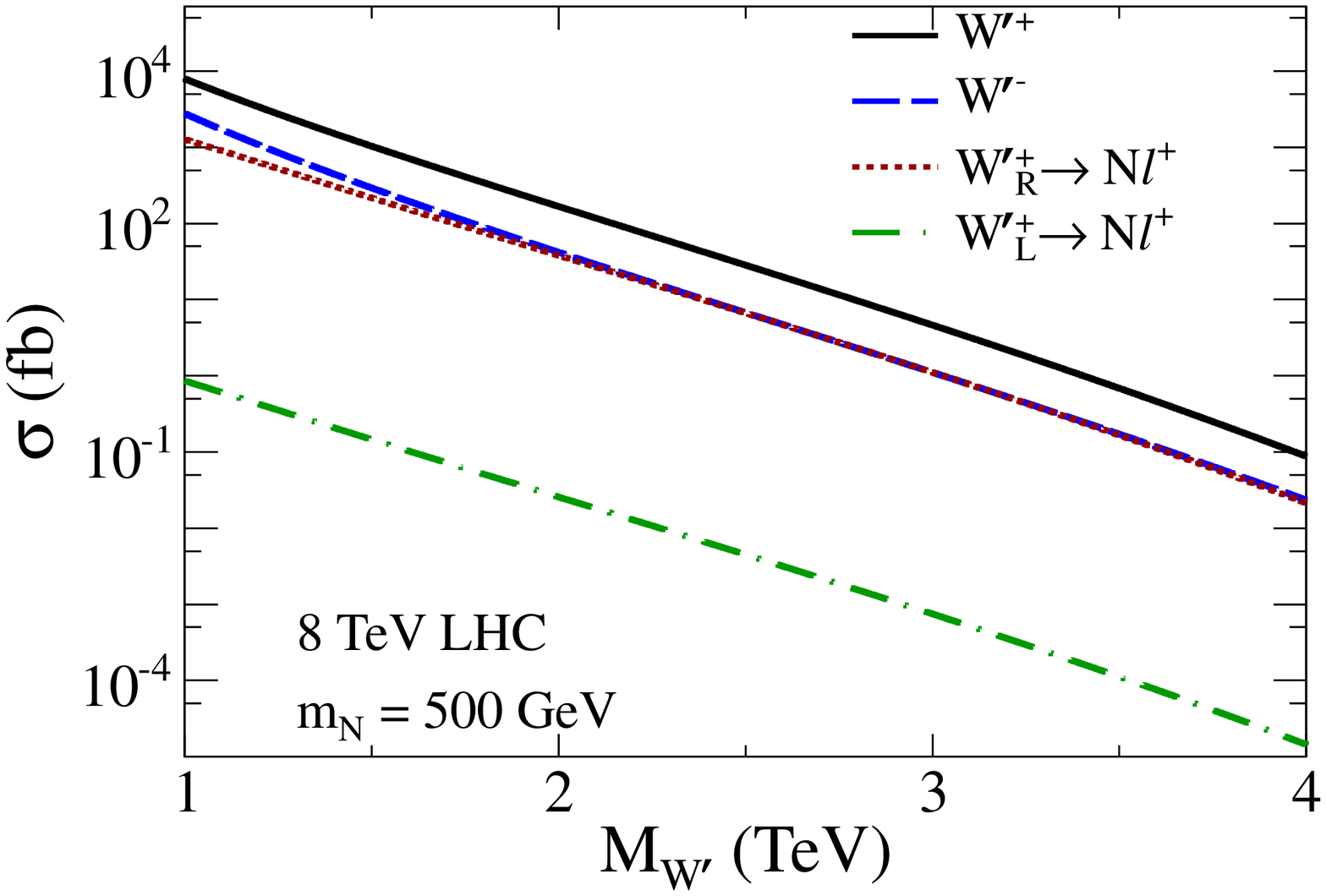}
}
\subfigure[]{
	\label{mwprod_14TeV.FIG}
	\label{mwprod.FIG}
	\includegraphics[width=0.45\textwidth]{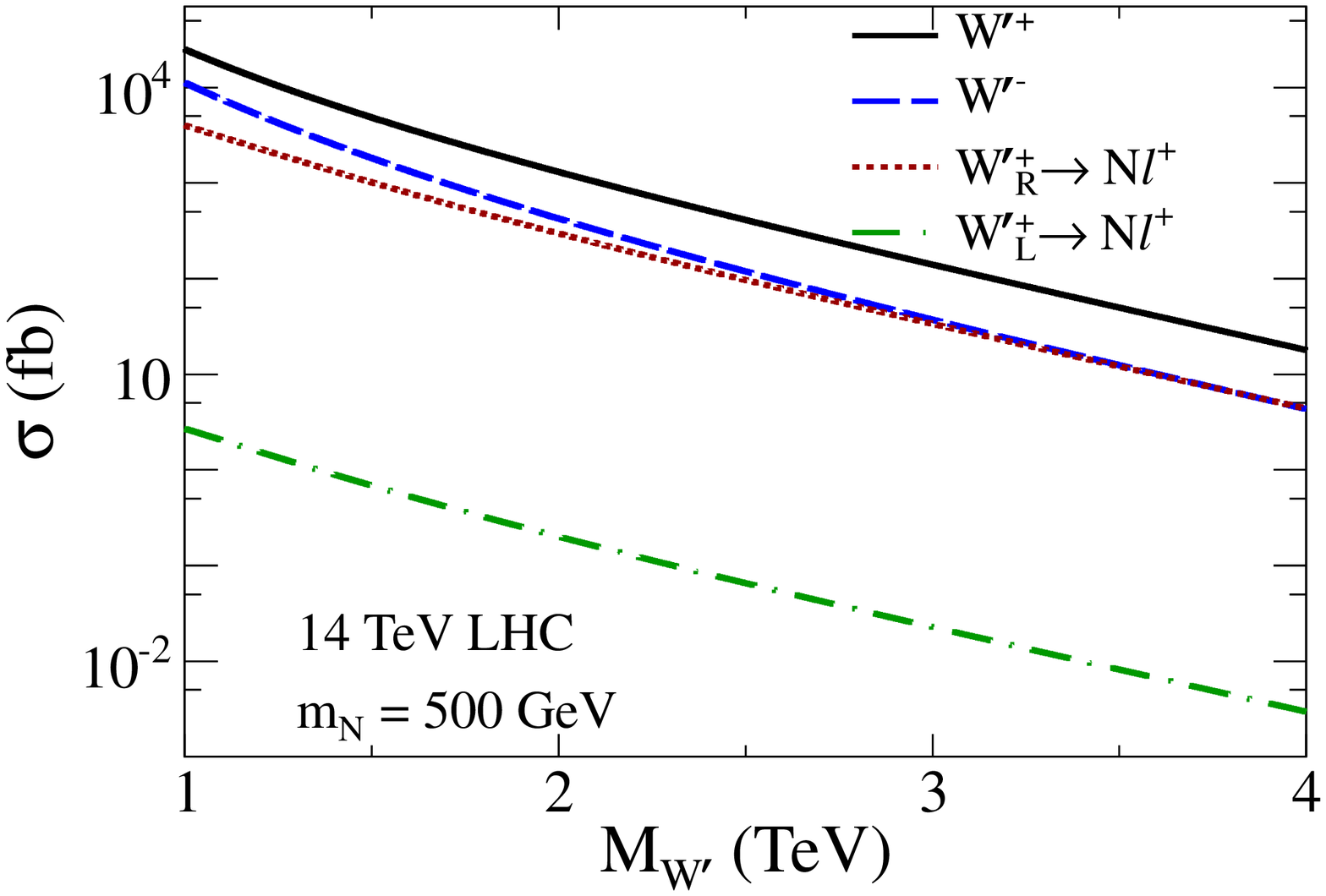}
}	
\caption{(a) The total decay width for $W'_{R}$ (solid) and $W'_{L}$ (dash); 
(b) the branching ratio of $\wpri_{R,L}\rightarrow N\ell^{+}$,
with subsequent $\wpri_{R}\rightarrow N\mu^{+}$ (dot) and $\wpri_{L}\rightarrow N\mu^{+}$ (dash-dot) ratios;
and the production cross sections at the (c) 8 and (d) 14 TeV LHC of $W'_{R}$ (solid), $W'_{L}$ (dash), 
$W'_{R}\rightarrow N\ell^{+}$ (dot), and $W'_{L}\rightarrow N\ell^{+}$ (dash-dot).}
\label{proddec.FIG}
\end{figure}

Under our parameterization, the partial widths for $\wpri$ decaying into a pairs of quarks are
\bea
\nonumber
\Gamma(W'\rightarrow \bar{q}q') &=& 3|{V^{\rm CKM}_{qq'}}'|^{2}(g_L^{q~2} + g_R^{q~2})\frac{\mwpri}{48\pi},
\qquad \\
\Gamma(W'\rightarrow tb) &=& 3|{V^{\rm CKM}_{tb}}'|^{2}(g_L^{q~2} + g_R^{q~2})\frac{\mwpri}{48\pi}\bigg{(}1-x_{t}^{2}\bigg{)}^2\bigg{(}1+\frac{1}{2}x_{t}^{2}\bigg{)}, \qquad 
\label{parttop.EQ}
\eea
where $x_{i}=m_{i}/M_{W'}$, and the factors of three represent color multiplicity. Likewise, the partial widths of the $W'$ decaying to leptons are
\bea
\Gamma(W'\rightarrow \ell \nu_m)&=&\left(g^{\ell~2}_R |X_{\ell m}|^2 + g^{\ell~2}_L|U_{\ell m}|^{2}\right)\frac{\mwpri}{48\pi},\qquad \\
\Gamma(W'\rightarrow \ell N)        &=&\left(g^{\ell~2}_R |Y_{\ell N}|^2 + g^{\ell~2}_L|V_{\ell N}|^{2}\right)\frac{\mwpri}{48\pi}\bigg{(}1-x_{N}^{2}\bigg{)}^2\bigg{(}1+\frac{1}{2}x_{N}^{2}\bigg{)}. \qquad 
\eea
Summing over the partial widths, the full widths are found to be
\begin{eqnarray}
\Gamma_{W'_{R}} &=& \frac{M_{W'}}{32\pi} \left[ 4 + (1-x_{t}^{2})^2(2+x_{t}^{2}) + (1-x_{N}^{2})^2(2+x_{N}^{2})\frac{1}{3}\sum_{\ell=e}^{\tau}\vert Y_{\ell N}\vert^{2} 
+ \frac{2}{3}\sum_{m=1,\ell=e}^{3,\tau}\vert X_{\ell m}\vert^{2}\right] \\
\Gamma_{W'_{L}} &=& \frac{M_{W'}}{32\pi} \left[ 4 + (1-x_{t}^{2})^2(2+x_{t}^{2}) + (1-x_{N}^{2})^2(2+x_{N}^{2})\frac{1}{3}\sum_{\ell=e}^{\tau}\vert V_{\ell N}\vert^{2}  
+\frac{2}{3}\sum_{m=1,\ell=e}^{3,\tau}\vert U_{\ell m}\vert^{2}\right].
\end{eqnarray}

As a function of $M_{W'}$, Fig.~\ref{proddec.FIG} shows (a) the total $\wpri$ decay width; 
(b) the branding ratio (BR) of $\wpri\rightarrow N\ell$, for $\ell=e,\mu,\tau$, defined as the ratio of the partial width to the total $\wpri$ width, $\Gamma_\wpri$:
\bea
{\rm BR}(\wpri \rightarrow \ell N)=\frac{\Gamma(\wpri \rightarrow \ell N)}{\Gamma_{\wpri}};
\eea
and the production cross sections for the pure gauge eigenstates $W'_{R,L}$, 
along with $pp\rightarrow {\wpri}^+_{R,L}\rightarrow N\ell^+$ in (c) 8 TeV and (d) 14 TeV $pp$ collisions.

The production cross section of the $W'$ and its subsequent decay to $N$ is calculated in the usual fashion~\cite{Denner:1992}.
The treatment of our full $2\rightarrow4$ process, on the otherhand, is addressed in Appendix~\ref{appendME.APP}.
Since the $u$-quark is more prevalent in the proton than the $d$-quark, 
and since the dominate subprocess of $W'^+$ ($W'^-$) production at the LHC is $u\bar{d}\rightarrow {\wpri}^+$ ($d\bar{u} \rightarrow {\wpri}^-$), 
the production cross section of ${\wpri}^+$ is greater than the ${\wpri}^-$ cross section.
In a similar vein, the mixing between L.H. interaction states and heavy neutrino mass eigenstates is suppressed by $\vert V_{\ell N}\vert^{2}\sim\mathcal{O}(10^{-3})$,
whereas the mixing between R.H. interaction states and heavy neutrino mass eigenstates is proportional to $\vert Y_{\ell N}\vert^{2}\sim\mathcal{O}(1)$.
Consequently, the $W'_{L}\rightarrow N\ell$ branching ratio, and hence the $pp\rightarrow W'_{L}\rightarrow N\ell$ cross section, 
is roughly three orders of magnitude smaller than the $W'_{R}$ rates.

\begin{figure}[tb]
\centering
\subfigure[]{
      \includegraphics[clip,width=0.45\textwidth]{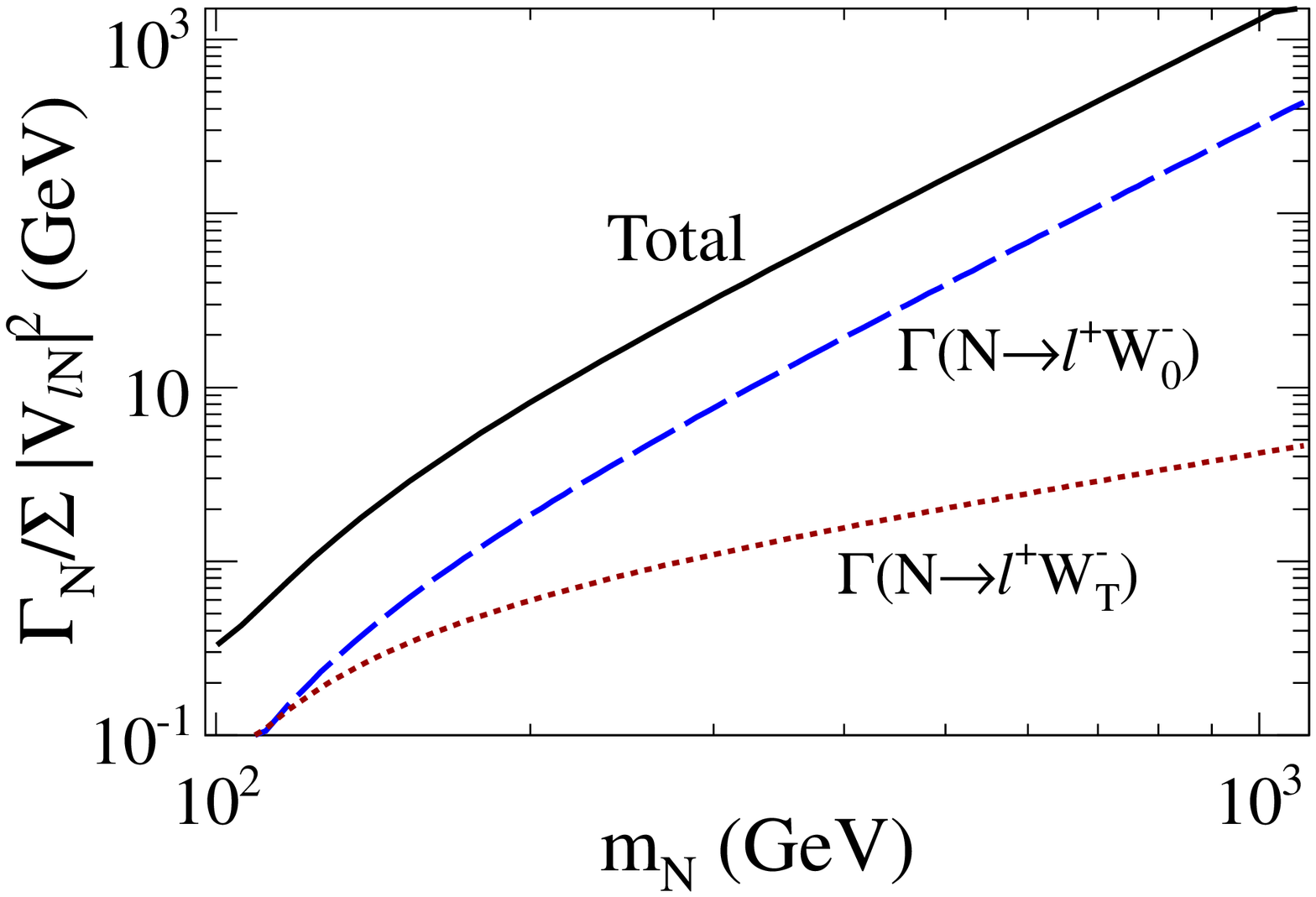}
\label{Nwidth.FIG}
}
\subfigure[]{
      \includegraphics[clip,width=0.45\textwidth]{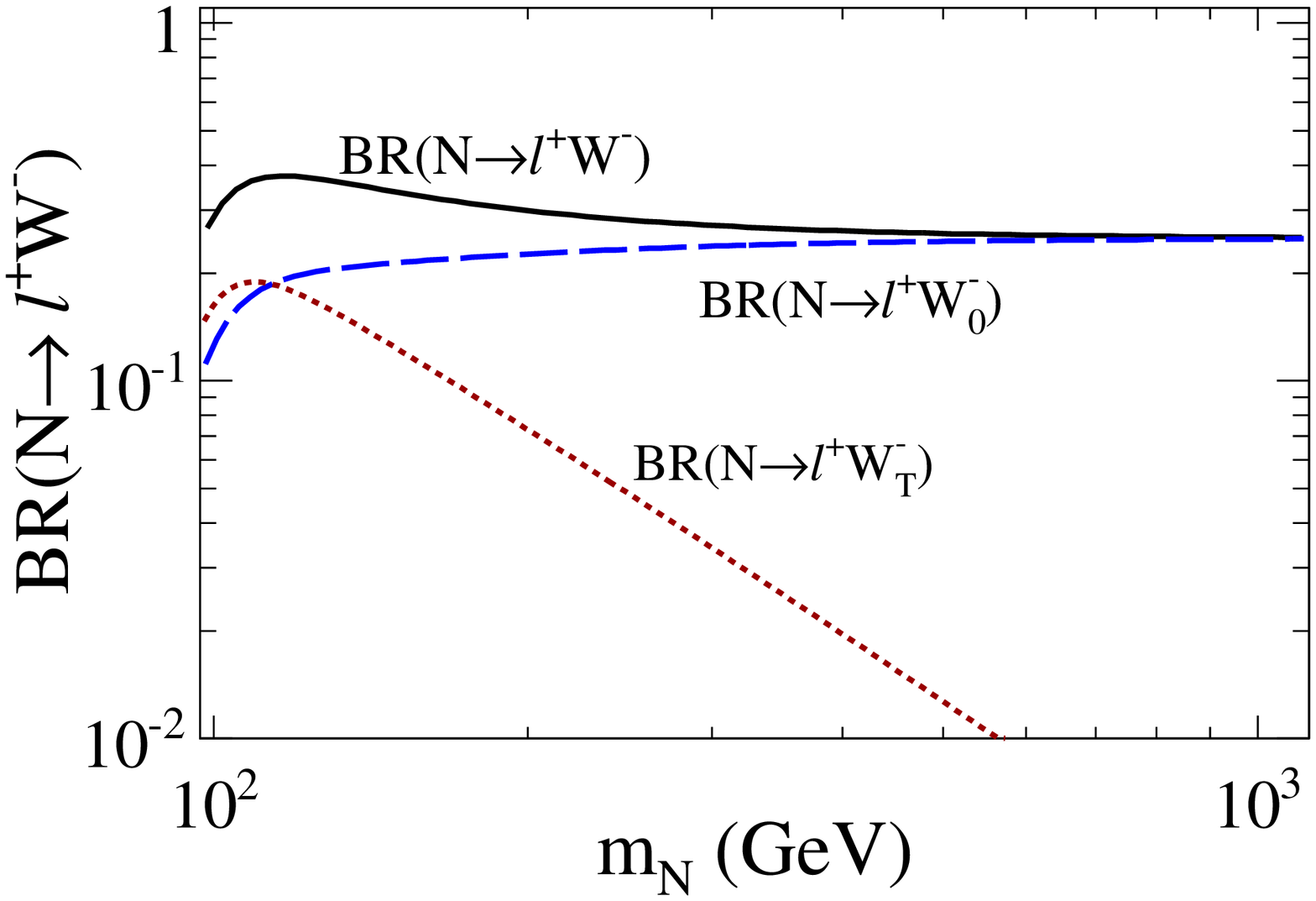}
\label{NBR.FIG}
}
\caption{As a function of heavy neutrino mass, (a) the total $N$ width and the $N\rightarrow\ell^{+}W^{-}_{\lambda}$ partial widths,
and (b) the combined $N\rightarrow \ell^{+} W^{-}$ and individual $N\rightarrow \ell^{+} W^{-}_{\lambda}$ branching ratios 
for longitudinal $(\lambda=0)$ and transverse $(\lambda=T)$ $W$ polarizations.} 
\label{NW.FIG}
\end{figure}
\subsection{Heavy Neutrino Decay}
\label{NeutDec.SEC}

A heavy neutrino with mass of a few hundred GeV or more can decay through on-shell SM gauge and Higgs bosons.  
The partial widths of the lightest heavy neutrino are
\bea
\Gamma(N\rightarrow \ell^{\pm} W^\mp_0)&\equiv\Gamma_0=&\frac{g^2}{64\pi M^2_W}|V_{\ell N}|^2m^3_N(1-y^2_W)^2\nonumber\\
\Gamma(N\rightarrow \ell^{\pm} W^\mp_T)&\equiv\Gamma_T=&\frac{g^2}{32\pi}|V_{\ell N}|^2m_N\left(1-y^2_W\right)^2\nonumber\\
\Gamma(N\rightarrow \nu_\ell Z)&\equiv\Gamma_Z=&\frac{g^2}{64\pi M_{W}^2}|V_{\ell N}|^2m_N^3(1-y^2_Z)^2\left(1+2y^2_{Z}\right)\nonumber\\
\Gamma(N\rightarrow \nu_\ell H)&\equiv\Gamma_H=& \frac{g^2}{64\pi M^2_W}|V_{\ell N}|^2  m^3_N(1-y^2_H)^2\label{NWidth.EQ}
\eea
where $W_{0,T}$ are longitudinally and transversely polarized $W$'s, respectively, and $y_i=M_i/m_N$.
The decays of the heavy neutrino through a $\wpri$ are not kinematically accessible. 
The total width is
\begin{eqnarray}
\Gamma_{\rm N}=\sum^\tau_{\ell = e}\left(2(\Gamma_0+\Gamma_T)+\Gamma_Z+\Gamma_H\right)
\end{eqnarray}
where the factor of two in front of $\Gamma_{0,T}$ is from the sum over positively and negatively charged leptons.

Figure \ref{Nwidth.FIG} shows the total decay width (solid) and the partial decay widths to positively charged lepton (dashed) normalized to the sum over the mixing matrices.  
For this plot the mass of the SM Higgs boson is set to $125$~GeV.  
The normalized width grows dramatically with mass due to decays into longitudinally polarized $W$'s and $Z$'s and the Higgs boson.  
Although the width appears to be large at high neutrino mass, for mixing angles on the order of a percent or less the width is still narrow.

\begin{figure}[tb]
\centering
\subfigure[]{
      \includegraphics[width=0.45\textwidth,angle=0]{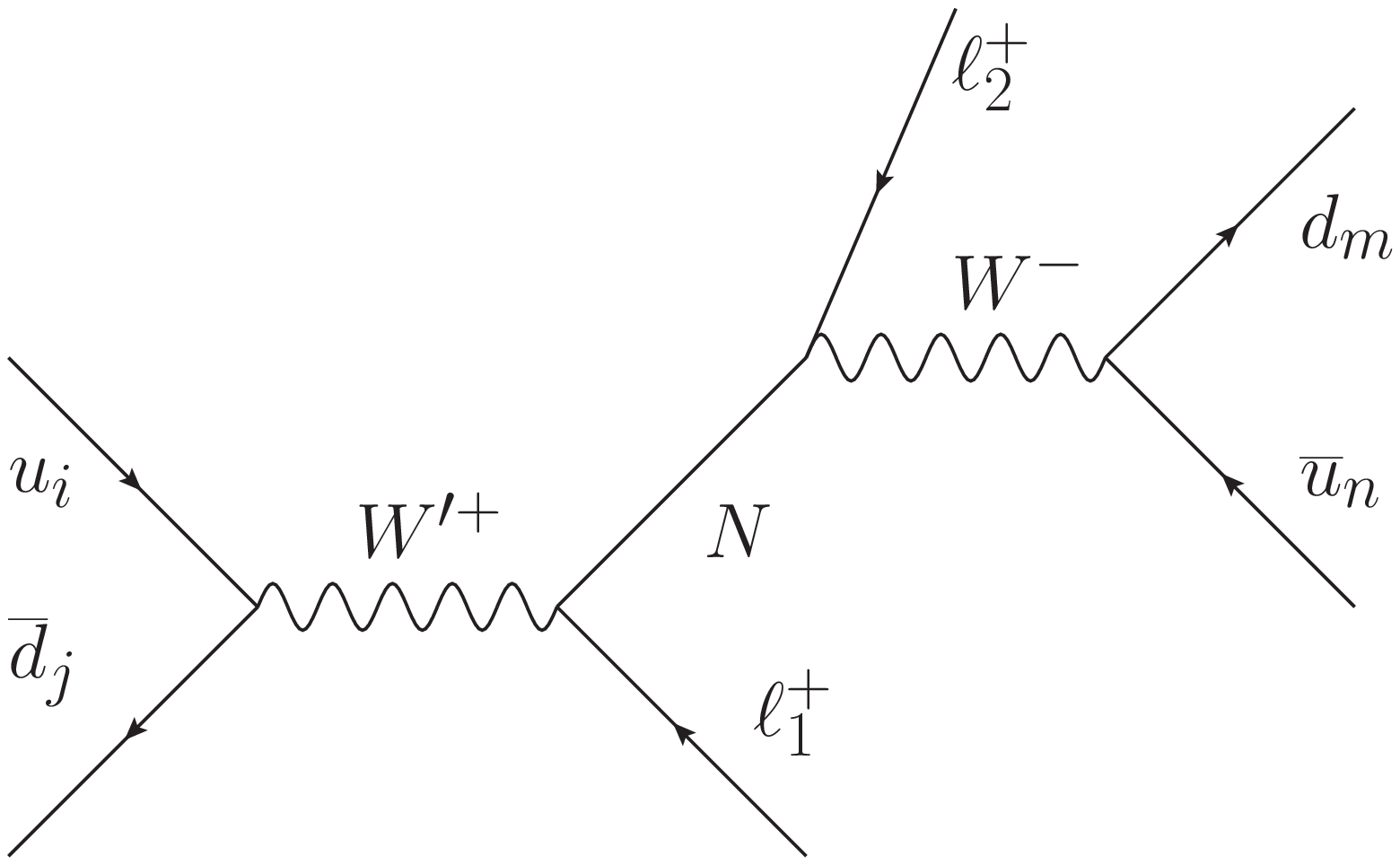}
\label{qq2Wp2llqq1.FIG}
}
\subfigure[]{
    \includegraphics[width=0.45\textwidth,angle=0]{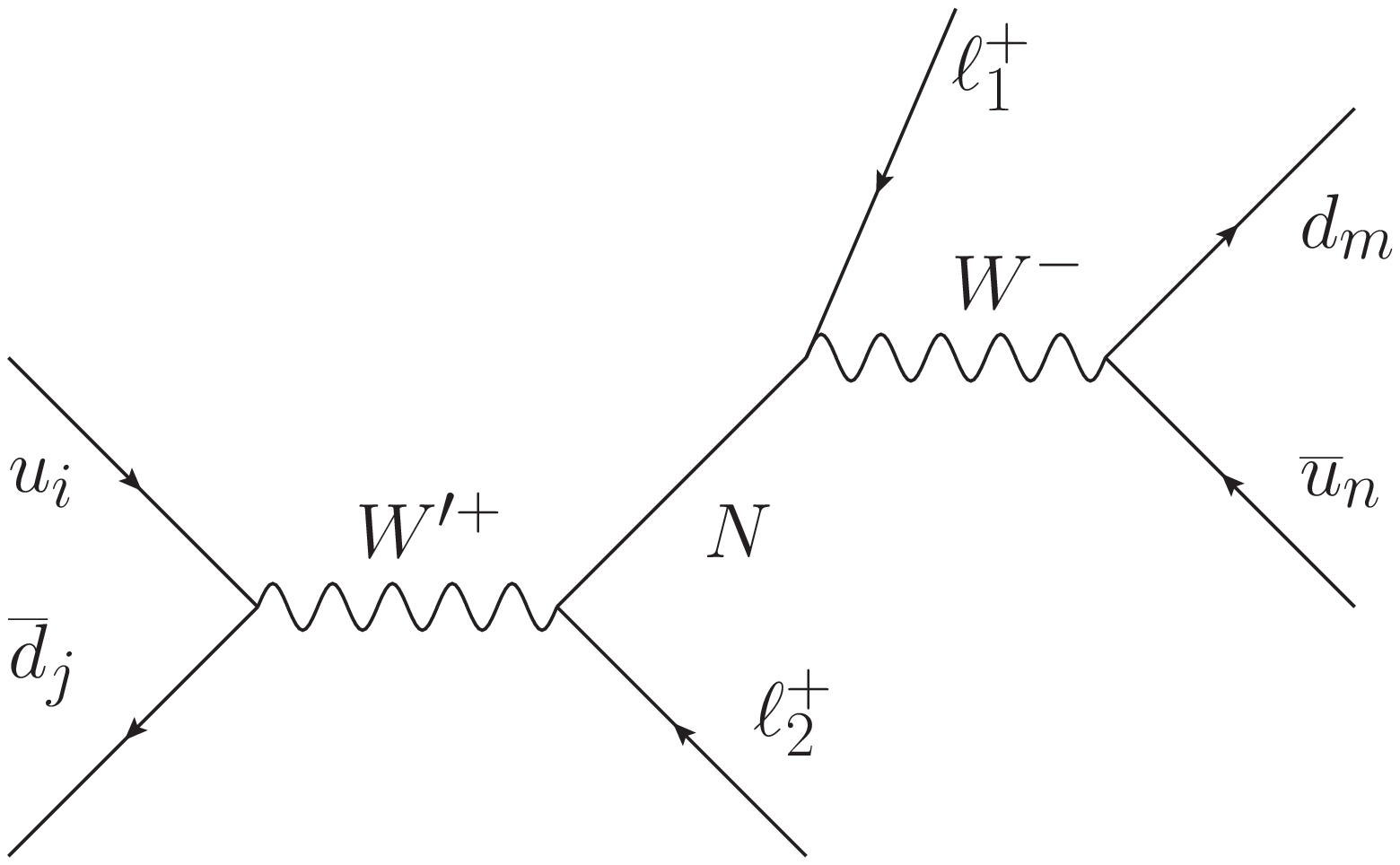}
}
\caption{
The partonic level process for a heavy ${\wpri}^+$ production and decay to like sign leptons in hadronic collisions. 
}
\label{qq2Wp2llqq.FIG}
\end{figure}

Also of interest is the branching ratio (BR) of heavy neutrinos into charged leptons:
\bea
\rm{BR}\left(N\rightarrow \ell^\pm W^\mp\right)=\frac{\sum^\tau_{\ell=e} \left(\Gamma_0+\Gamma_T\right)}{\Gamma_{\rm Tot}}
\label{NBR.EQ}
\eea
Figure \ref{NBR.FIG} shows the total BR of the heavy neutrino into positively charged leptons (solid) and individually the BR into longitudinally (dashed) and transversely (dotted) polarize $W$'s as a function of neutrino mass.  The BR's into negatively charged leptons are the same.  As the mass of the neutrino increases the $Z$ and Higgs decay channels open, hence the branching ratio into charged leptons decreases.  Since $\Gamma_0$ grows more quickly with neutrino mass than $\Gamma_T$, for $m_N\gg M_W$ the total BR converges to the BR into longitudinally polarized $W$'s.
  Also, at high neutrino masses
\bea
\Gamma_0\approx\Gamma_H\approx \Gamma_Z
\eea
Hence the total width approaches $4\Gamma_0$ and, from Eq.~(\ref{NBR.EQ}), the branching ratio into a positively charged leptons is approximately $0.25$.
This is a manifestation of the Goldstone Equivalence Theorem when taking $m_N$ and $V_{\ell N}$ as independent parameters.

\section{Like-Sign Dilepton Signature}
\label{Like.SEC}

\begin{figure}[tb]
\centering
\subfigure[]{
      \includegraphics[clip,width=0.45\textwidth]{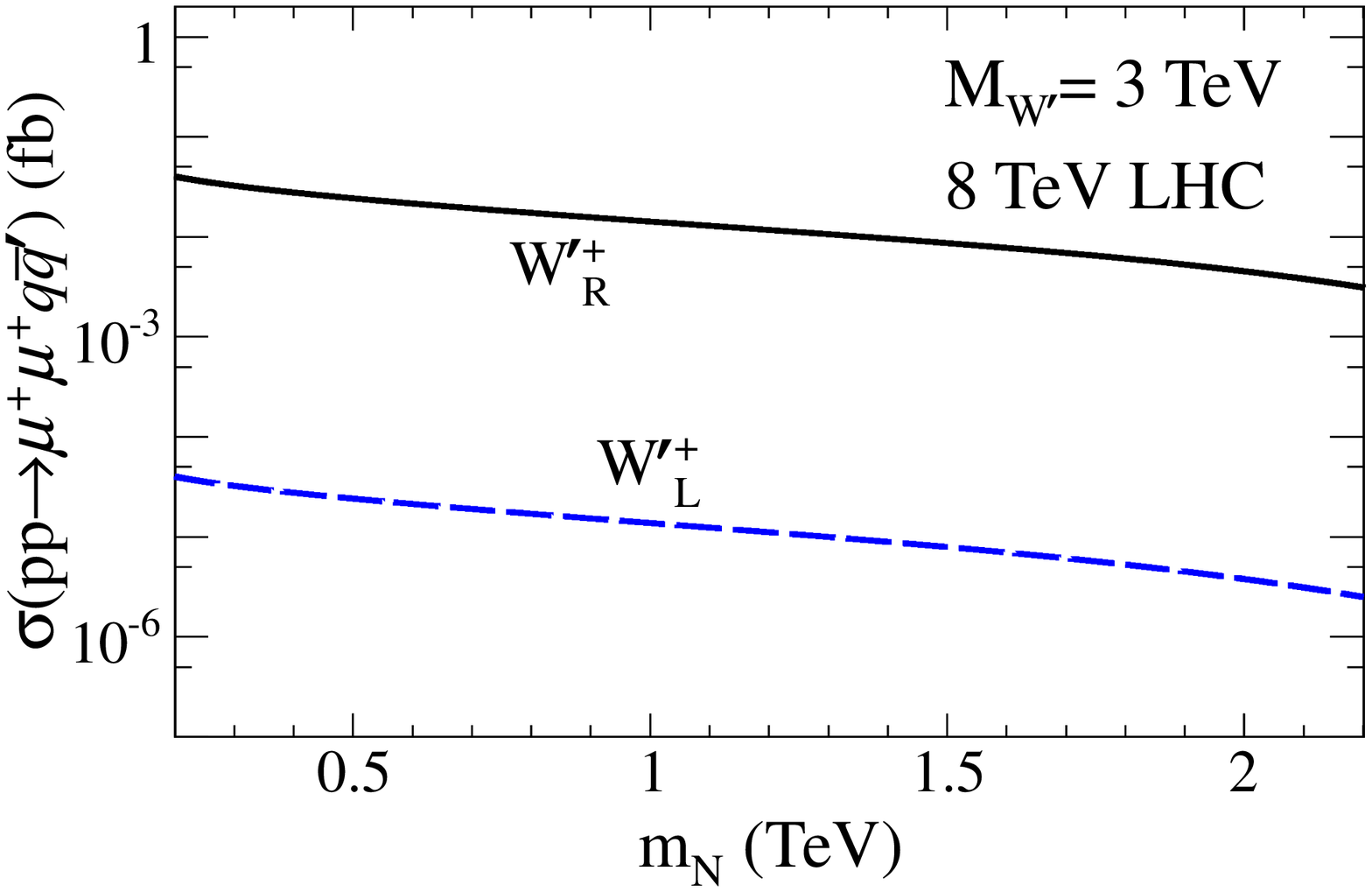}
}
\subfigure[]{
      \includegraphics[clip,width=0.45\textwidth]{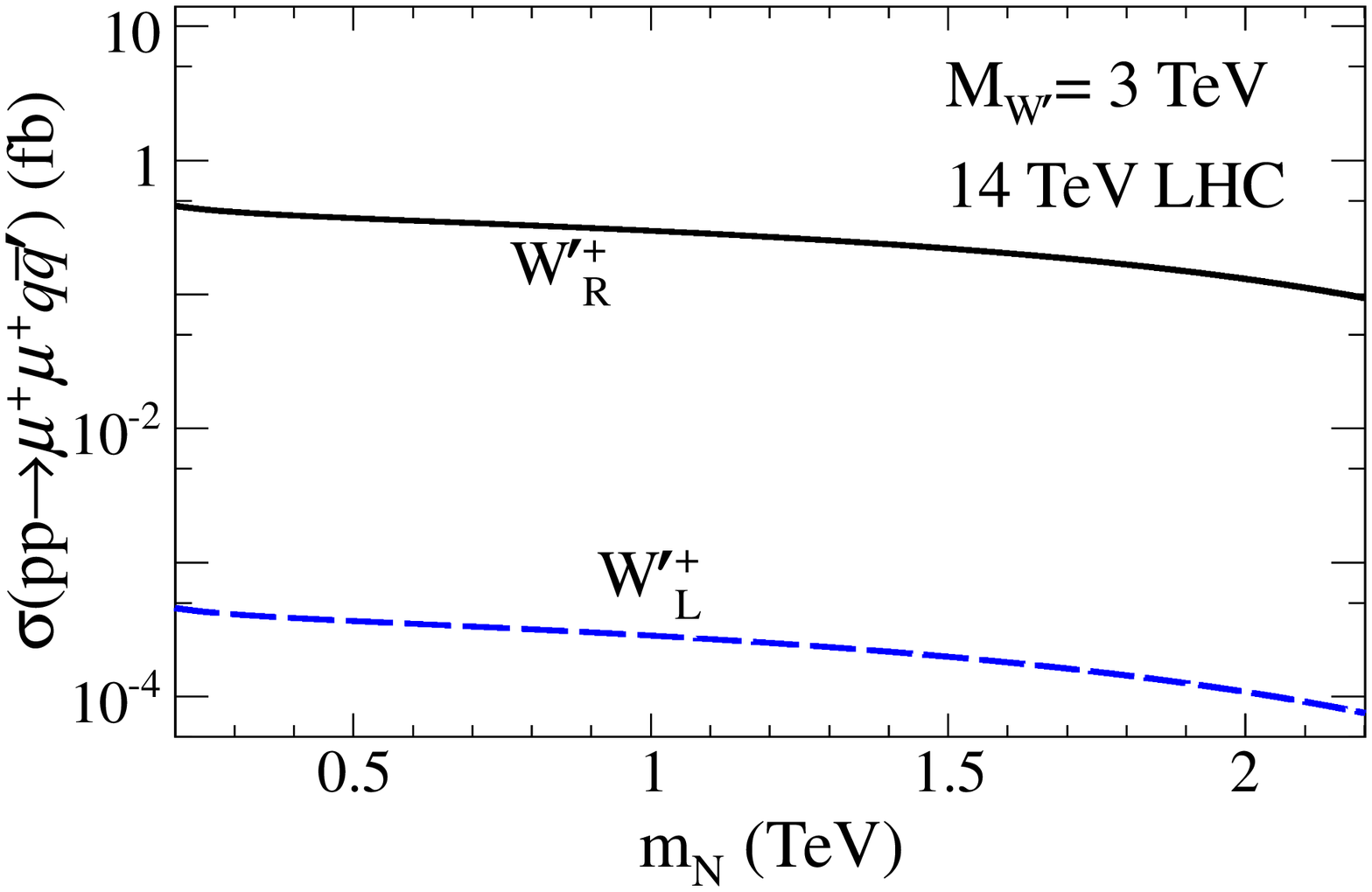}    
}
\caption{Total cross section of $pp\rightarrow {W'}^+\rightarrow \mu^+\mu^+W^{-}$ times $W^{-}\rightarrow q\bar{q}'$ branching ratio versus heavy neutrino mass at (a) 8 and  (b) 14 TeV. 
Solid (dashed) line corresponds to $W'_R$ ($W'_L$) gauge state.}
      \label{NeutXsect.fig}
\end{figure}

A distinctive feature of Majorana neutrinos is that they facilitate $L$-violating processes,
and to study this behavior at the LHC we consider the $L$-violating cascade
\begin{eqnarray}
u(p_A)~\bar{d}(p_B)\rightarrow W_{R,L}^{'+} (q)\rightarrow \ell^+_1(p_1)~N(p_{N}) \rightarrow \ell^+_1(p_1)~\ell^+_2(p_2) ~q(p_3)~ \bar{q}'(p_4).
\label{sig.EQ}
\end{eqnarray}
The two diagrams that contribute to this process are shown in Fig.~\ref{qq2Wp2llqq.FIG}. Figure \ref{NeutXsect.fig} 
shows the total production cross section for the like-sign dimuon process as a function of $m_{N}$. 
In it, the solid line denotes the pure $W'_R$ gauge state while the dashed line represents the pure $W'_L$ state.  
Since the $W'_R\rightarrow N\mu$ branching ratio is larger than $W'_L\rightarrow N\mu$ ratio, the cross section for $W'_R$ is systematically larger than for $W'_L$. 
Additionally, as the neutrino mass approaches the $W'$ mass the cross section drops precipitously due to phase space suppression.

In principle, the conjugate process, $\bar{u}d\rightarrow W^{'-}$, should also be possible at the LHC.
However, it will possess a much smaller production rate because the $\bar{u}d$ initial-state has a smaller parton luminosity than $u\bar{d}$.  
Despite this, all reconstruction methods and observables discussed below are applicable to both processes. 

\subsection{Event Selection}
For simplicity, we restrict our study to like-sign muons.
There is no change in the analysis if extended to electrons; however, $\etmiss$ requirements must be reassessed for inclusion of unstable $\tau$'s~\cite{aguilar:2012lrsm}.
Consequently, our signal consists strictly of two positively charged leptons and two jets, a fact that allows for considerable background suppression.
In simulating this like-sign leptons plus dijet signal, to make our analysis more realistic, 
we smear the lepton and jet energies to emulate real detector resolution effects.
These effects are assumed to be Gaussian and parameterized by
\bea
\frac{\sigma(E)}{E}=\frac{a}{\sqrt{E}}\oplus b,
\label{eq:smear}
\eea
where $\sigma(E)/E$ is the energy resolution, $a$ is a sampling term, 
$b$ is a constant term, $\oplus$ represents addition in quadrature, and all energies are measured in GeV.  
For leptons we take $a=5\%$ and $b=0.55\%$, and for jets we take $a=100\%$ and $b=5\%$ \cite{Ball:2007zza}.  

After smearing, we define our candidate event as two positively charged leptons and two jets passing the following 
basic kinematic and fiducial cuts on the transverse momentum, $p_T$, and pseudorapidity, $\eta$:
\begin{eqnarray}
p_T^j\geq 30~{\rm GeV},~ p_T^\ell\geq~{\rm 20~GeV},~ \eta_j\leq 3.0,~  \eta_\ell\leq 2.5.
\label{cuts1.EQ}
\end{eqnarray}
Table~\ref{WpNxsect.TAB} lists the cross sections for Eq.~(\ref{sig.EQ}) assuming the pure $\wpri_{R,L}$ gauge states at the 8 and 14 TeV LHC without smearing or acceptance cuts (row 1), 
and with smearing plus acceptance cuts from Eq.~(\ref{cuts1.EQ}) (row 2).
Here and henceforth, we assume a 100$\%$ efficiency for lepton and jet identification.

 \begin{table}
\begin{center}
\begin{tabular}{|c|c|c|c|c|} \hline \hline
                         \multirow{2}{*}{ $\displaystyle\sigma$(fb)}&\multicolumn{2}{c|} {8 TeV}& \multicolumn{2}{c|} {14 TeV} \\\cline{2-5}
                                  &$\wpri_L$             &$\wpri_R$& $\wpri_L$            & $\wpri_R$   \\ \hline \hline
~~Reco.~without Cuts or Smearing  & $4.6\times10^{-5}$   & $0.046$ & $9.3\times10^{-4}$   & $0.95$ \\ \hline
~~~+~Smearing~+~Fiducial~+~Kinematics (Eq.~(\ref{cuts1.EQ})) & $4.0\times10^{-5}$& 0.035 & $8.2\times10^{-4}$ & 0.71 \\ \hline
~+~Isolation~(Eq.~(\ref{cuts2.EQ}))&$2.1\times10^{-5}$   & 0.027   & $3.2\times10^{-4}$   & 0.50 \\ \hline
~~~~~~+$\not\!\!E_{T}$~+~$m_{jj} $~Requirements~(Eq.~(\ref{cuts3.EQ})) & $1.7\times10^{-5}$ & 0.023 & $2.6\times10^{-4}$ & 0.42 \\ \hline 
~~~~+~Mass~Req.~(Eq.~(\ref{cuts4.EQ})) & $7.2\times10^{-6}$ & 0.012 & $2.0\times10^{-4}$ & 0.35  \\ \hline\hline
~$\sigma$(All~Cuts)/$\sigma$(Smearing~+~Fid.~+~Kin.)&  18\% & 35\% & 25\%  & 49\% \\   \hline\hline
\end{tabular} 
\caption{Cross section for $pp\rightarrow {W'}_{L,R}^+\rightarrow \mu^+\mu^+ q\overline{q}'$ after consecutive cuts for 8 and 14 TeV LHC. 
}
\label{WpNxsect.TAB}
\end{center}
\end{table}

The goal of this analysis is to unambiguously determine the properties of $W'$ and $N$.
To do so, our candidate leptons and jets must be well-defined and well-separated, 
that latter of which is measured by
\bea
\Delta R_{ij} = \sqrt{(\Delta \phi_{ij})^2+(\Delta\eta_{ij})^2},
\eea
where $\Delta\phi_{ij}$ and $\Delta\eta_{ij}$ are the difference in the azimuthal angles and rapidities, respectively, of particles $i$ and $j$. 
Subsequently, we apply isolation cuts on our candidate objects: 
\begin{eqnarray}
\Delta R^{\rm min}_{\ell j} \ge 0.4,~~\Delta R_{jj}\ge 0.3
\label{cuts2.EQ}
\end{eqnarray}
for all lepton and jet combinations, where $\Delta R^{\rm min}_{\ell j}$ is defined as
\bea
\Delta R^{\rm min}_{\ell j} = \min_{i=\wpri, N} \Delta R^{\rm min}_{\ell_i j}.
\label{delRMinDef.EQ}
\eea
In Eq.~(\ref{delRMinDef.EQ}), the subscript $i=W',N$ on $\ell_{i}$ denotes the identified parent particle of $\ell_{i}$.
 The effects of the isolation cuts applied at both the 8 and 14 TeV LHC are shown in the third row of Table~\ref{WpNxsect.TAB}.  
To understand the origin of these precise numbers and parent-particle identification, we digress  
to succinctly connect properties of our chiral Lagrangian to the final-state kinematical distributions.

\subsection{Characteristics of Kinematical Distributions}
\label{Recon.SEC}

Our signal suffers from a very evident ambiguity: either lepton can originate from the neutrino decay.
The origin of each lepton must thus be determined in order to fully reconstruct an event.
As noted in section~\ref{WpLHC.SEC}, the width of $N$ is narrow.
Consequently, there is a very small probability for the phase space of each diagram in Fig.~\ref{qq2Wp2llqq.FIG} to overlap,
meaning that the interference of the two diagrams is negligible.
In fact, in the $\wpri_R$ case, the interference is exactly zero because the charged lepton from the $N$ decay is left-handed while the charged lepton from the $\wpri_R$ is right-handed.  
Furthermore, since the two diagrams add incoherently, it is reasonable to expect that only one diagram contributes at a time. 
Intuitively, this means that only one of the two following momentum combinations will closely reconstruct the heavy neutrino mass:
\begin{eqnarray}
m^2_{1jj}=(p_1+p_3+p_4)^2~~~{\rm or}~~~m^2_{2jj}=(p_2+p_3+p_4)^2,
\label{mNPermu.EQ}
\end{eqnarray}
where $p_{3}$ and $p_{4}$ are the momenta of our final-state jets.

\begin{figure}[tb]
\centering
\subfigure[]{
        \includegraphics[width=0.48\textwidth]{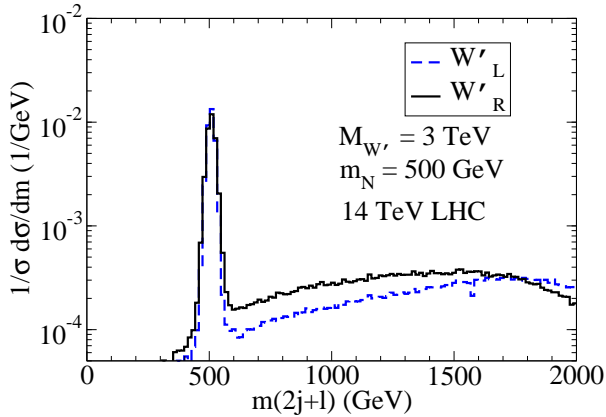}
}
\caption{Invariant mass distribution of $m_{\ell_{1}jj}$ and $m_{\ell_{2}jj}$, where $\ell_{i}$ for $i=1,2$ could originate from either the $W'$ or $N$. 
The cuts in Eqs.~(\ref{cuts1.EQ}) and (\ref{cuts2.EQ}) as well as the energy smearing have been applied. }
\label{m2jl.FIG}
\end{figure}

After calculating both permutations of $m_{N}$ (Fig.~\ref{m2jl.FIG}), the appearance of the $N$ mass peak is stark. 
Using the central value of the mass peak, $m_{N}^{Reco.}$, we identify the charged lepton from the $N$ decay as the 
charged lepton from our candidate event that most closely recovers $m_{N}^{Reco.}$, i.e.,
\begin{equation}
\Delta m_{min} = \min_{i=1,2} \vert m_{ijj} - m_{N}^{Reco.} \vert,
\end{equation}
where $m_{ijj}$ for $i=1,2$ is defined by Eq.~(\ref{mNPermu.EQ}).

\begin{figure}[tb]
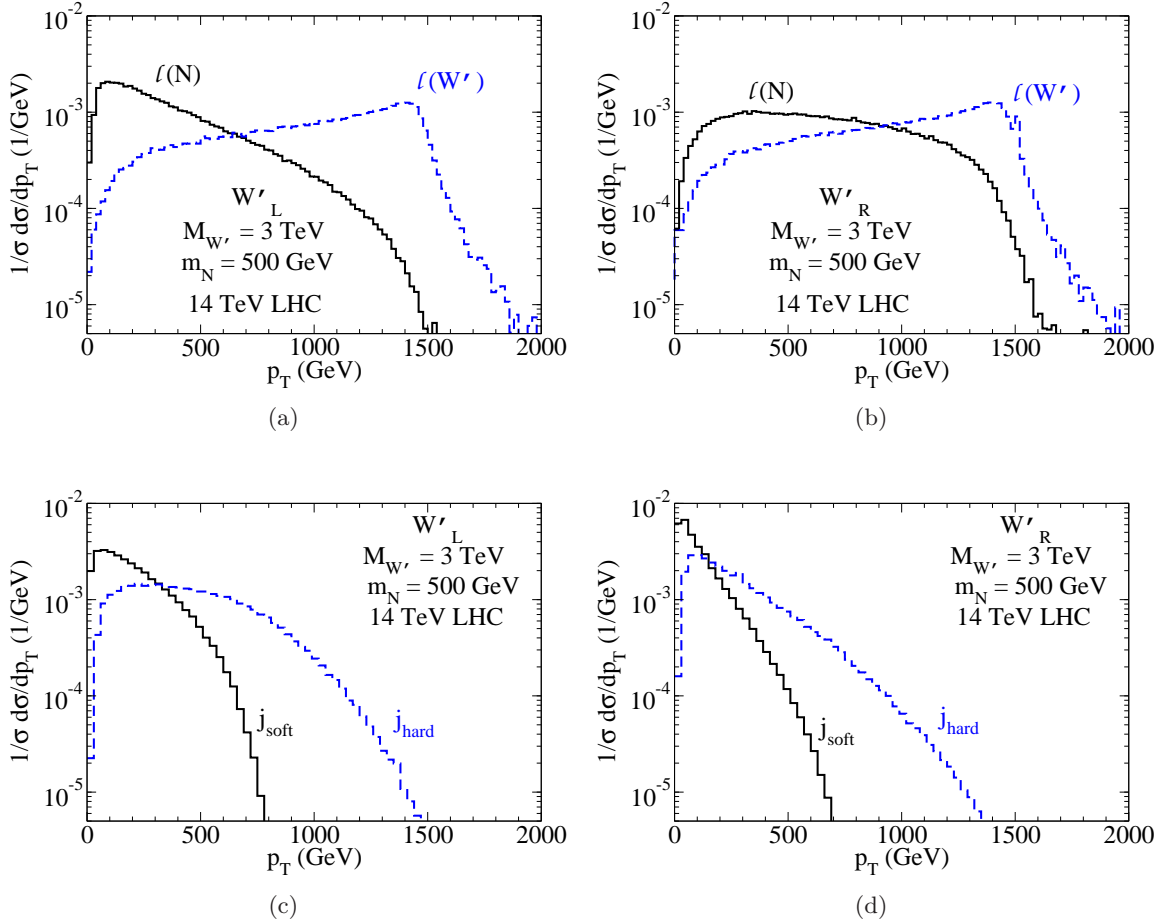

\centering
\subfigure[]{
	\label{ptlLH.FIG}
	\includegraphics[width=0.45\textwidth]{ptlLH500_3.eps}
	}
\subfigure[]{
	\includegraphics[width=0.45\textwidth]{ptlRH500_3.eps}
	\label{ptlRH.FIG}
}\vspace{.15in}\\
\subfigure[]{
       \includegraphics[width=0.45\textwidth]{ptjLH500_3.eps}
       \label{ptjLH.FIG}
}
\subfigure[]{
       \includegraphics[width=0.45\textwidth]{ptjRH500_3.eps}
}
\caption{ Transverse momentum distributions for (a,b) the lepton identified as originating from the $\wpri$ (dashed) and neutrino (solid), 
and (c,d) the hardest (dashed) and softest (solid) jets in $pp\to W'\to \ell^+ \ell^+ j j$ production.  
The $\wpri_L$ case is represented in (a,c) and the $\wpri_R$ case in (b,d). The energy smearing has been applied. 
}
\label{ptdist.FIG}
\end{figure}

Independent of reconstructing $N$, the charged lepton associated with the $W'$ decay can be identified 
by analyzing the transverse momentum, $p_{T}$, distributions of our final-state objects.
In Fig.~\ref{ptdist.FIG}, the $p_{T}$ distributions of the charged leptons (a,b) and jets (c,d) for the $\wpri_L$ (a,c) and $\wpri_R$ (b,d) gauge states.
As expected, the lepton identified as originating from the $W'$ has a Jacobian peak around $M_{W'}/2$ for both the $\wpri_L$ and $\wpri_R$ cases.   
To understand the other distributions, we consider spin correlations.

\begin{figure}[tb]
\begin{center}
\subfigure[]{
\includegraphics[width=0.35\textwidth,clip=true]{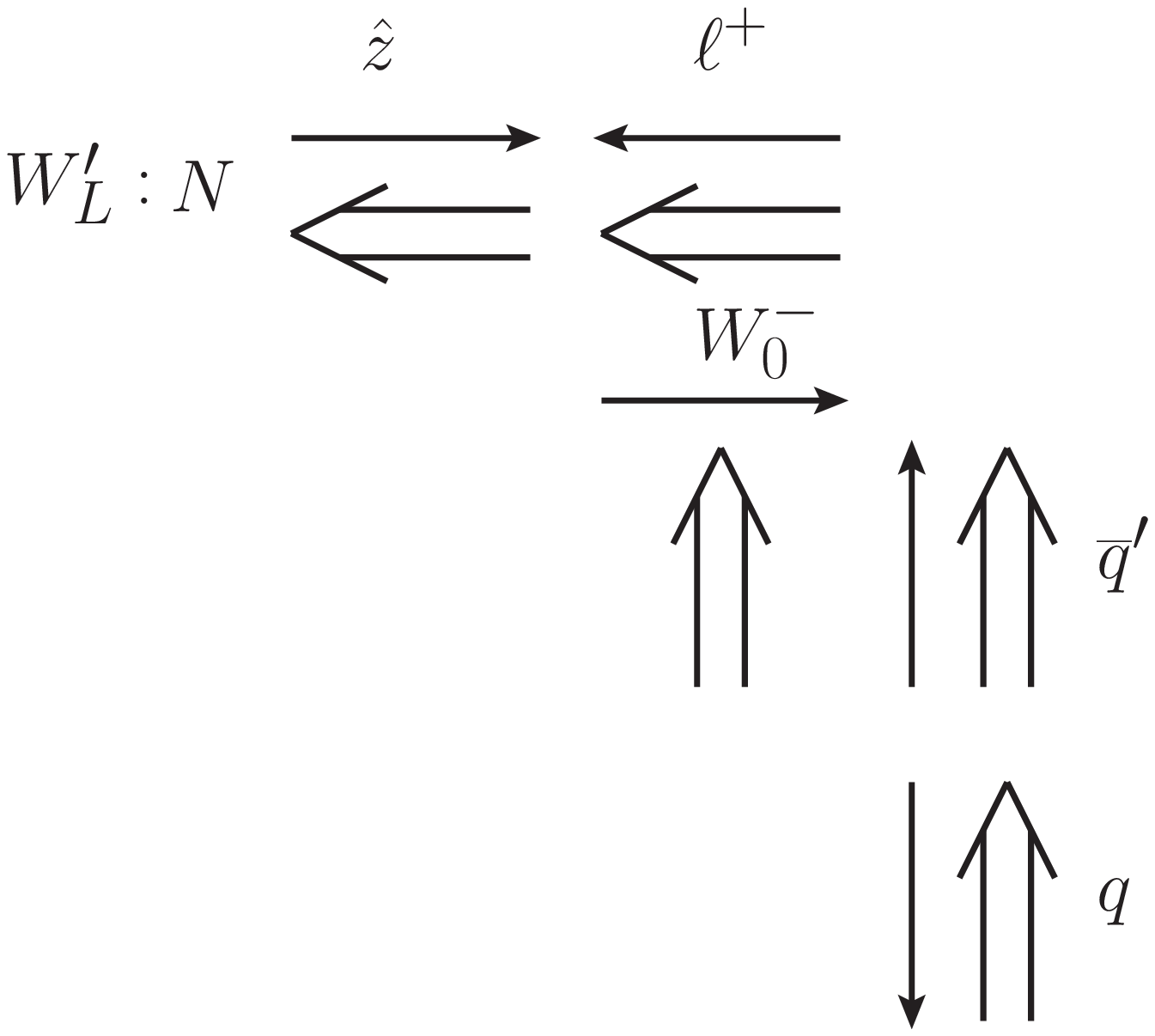}
\label{LspinLH.FIG}
        }
\subfigure[]{
\includegraphics[width=0.35\textwidth,clip=true]{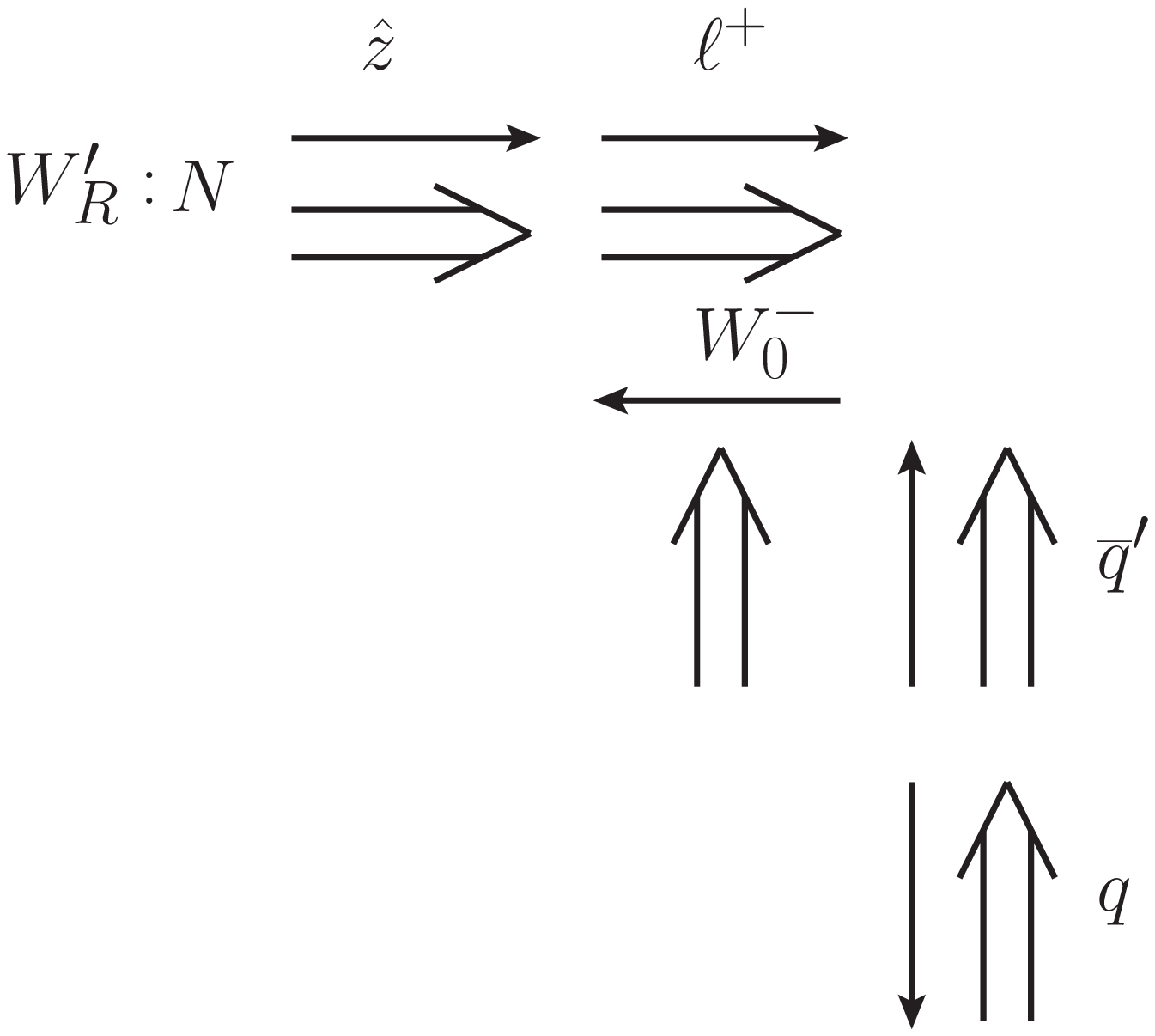}
\label{LspinRH.FIG}
        }\\
\subfigure[]{
\includegraphics[width=0.35\textwidth,clip=true]{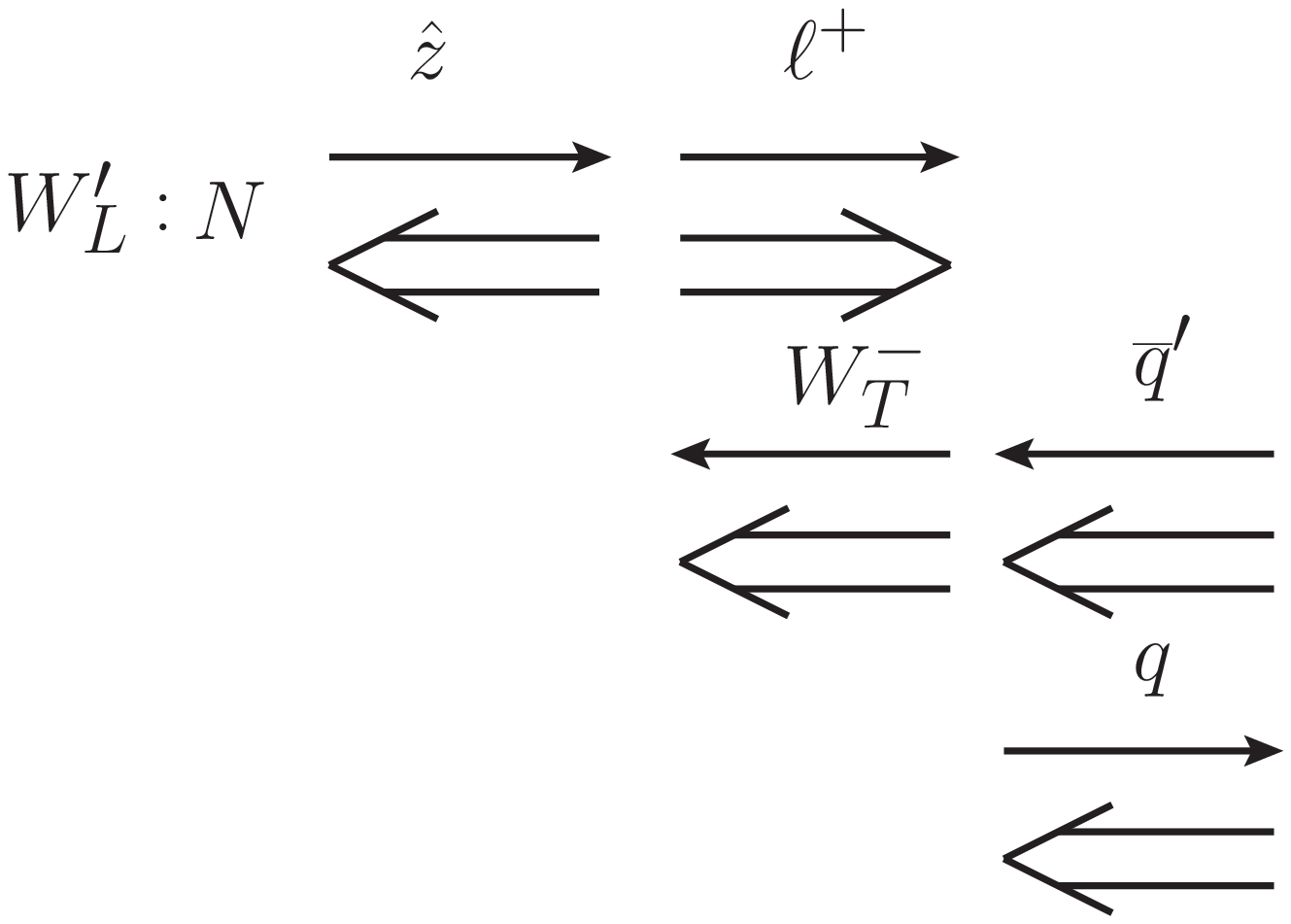}
\label{TspinLH.FIG}
        }
\subfigure[]{
\includegraphics[width=0.35\textwidth,clip=true]{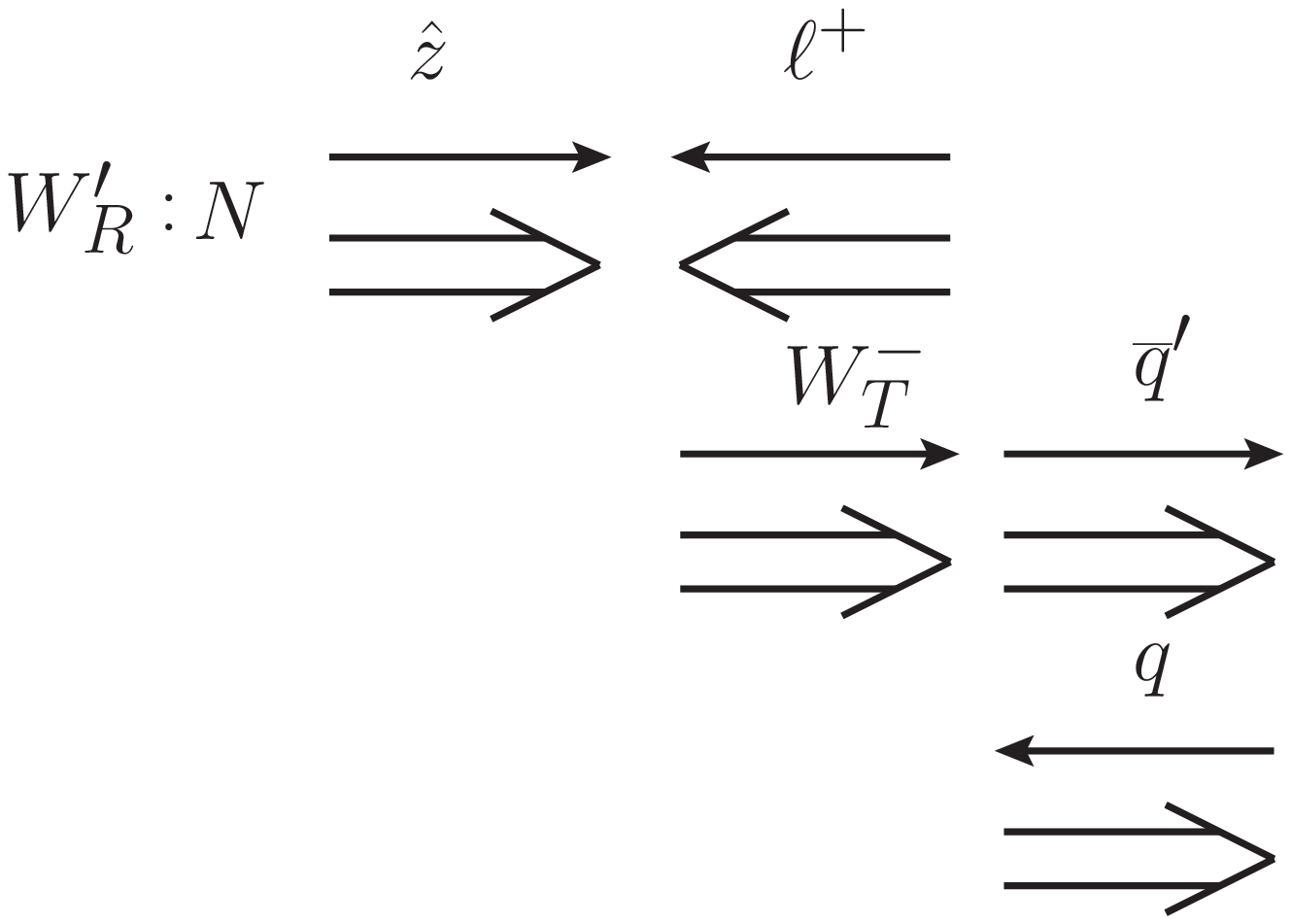}
\label{TspinRH.FIG}
        }
\caption{Helicity and spin correlations in the chains $N_{L,R} \to \ell^+ W^{-} \to \ell^+ q \overline{q}'$ 
from $\wpri_L$ decay in  (a),~(c); and from $\wpri_R$ decay in (b),~(d).
Figures (a) and (b) are for longitudinally polarized SM $W$'s, and Figs.~(c) and (d) are for 
transversely polarized SM $W$'s.  The decay goes from left to right as labeled by the particle names. 
The momenta (single arrow lines) and spins (double arrow lines) are in the parent rest-frame 
in the direction of the heavy neutrino's motion ($\hat{z}$) in the $\wpri$ rest-frame.
}
\label{Lspin.FIG}
\end{center}
\end{figure}

Figure \ref{Lspin.FIG} shows the spin correlations of the process in Eq.~(\ref{sig.EQ}) with the single arrowed lines representing momentum direction and double arrowed lines spin. 
The direction $\hat{z}$ is defined as the direction of motion of the neutrino in the $W'$ rest-frame.  
Each column indicates the spin and momentum of the particles in their parents' rest-frame with the first column in the neutrino rest-frame.  
Note that for the $W'_R$ ($W'_L$) the heavy neutrino is in a mostly right-(left-) handed helicity state.  
Hence, for the $W'_R$ ($W'_L$) the neutrino spin points with (against) the $\hat{z}$ direction.  
The decays of the neutrino through longitudinal $W$ are shown in Fig.~\ref{LspinLH.FIG} and \ref{LspinRH.FIG} for $W'_L$ and $W'_R$, respectively,
 and the decays through a transversely polarized $W$ are shown in Fig.~\ref{TspinLH.FIG} for $\wpri_L$ and Fig.~\ref{TspinRH.FIG} for $\wpri_R$.

  As shown in Fig.~\ref{NW.FIG}, $500$~GeV neutrino preferentially decays into longitudinally polarized $W$'s.  
  We therefore focus on that case.
For the $\wpri_R$, the lepton from the heavy neutrino decay moves preferentially along the $\hat{z}$ direction.  
Hence, the boost into the partonic c.m.~frame will be along the charged lepton's momentum.  
In the $\wpri_L$ case, the charged lepton moves in negative $\hat{z}$ direction and the boost into the partonic 
c.m.~frame is against the lepton's momentum.  
Therefore, the lepton from the heavy neutrino decay is harder in the $\wpri_R$ case than in the $\wpri_L$ case. 
The contribution from decay into transversely polarized $W$'s is in the opposite direction.  
However, as noted previously, this contribution is smaller than the decays into longitudinally polarized $W$'s. 
Similar arguments can be made to explain that the two jets are softer in the $\wpri_R$ case than in the $\wpri_L$ case.

\begin{figure}[tb]
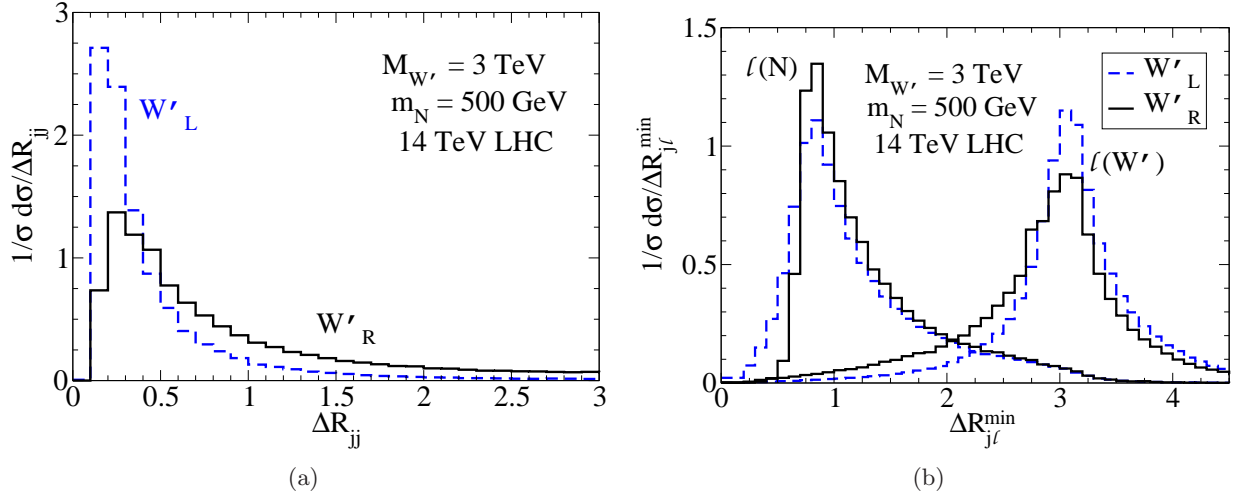

\centering
\subfigure[]{
        \includegraphics[width=0.48\textwidth]{drjj500_3.eps}
        \label{drjj.FIG}
}
\subfigure[]{
        \includegraphics[width=0.48\textwidth]{drjlmin500_3.eps}
        \label{drjlmin.FIG}
}
\caption{(a) $\Delta R_{jj}$ distribution and 
(b) $\Delta R^{\rm min}_{\ell j}$ distributions for both the lepton identified as originating from $N$ and $\wpri$.  
The solid lines are for the $\wpri_R$ case and dashed lines $\wpri_L$. Energy smearing has been applied.}
\label{deltar.FIG}
\end{figure}

As previously stated, identifying well-separated objects in our event is paramount to measuring our observables.
For 14 TeV LHC collisions, Fig.~\ref{deltar.FIG} shows (a) the separation between the two jets, $\Delta R_{jj}$, 
and (b) the minimum separation between the leptons identifed as originating from the heavy neutrino and $\wpri$ and the two jets defined by 
\begin{eqnarray}
\Delta R^{\rm min}_{\ell_{i} j}=\min_{k=1,2} \Delta R_{\ell_i j_k},
\label{Rmin.EQ}
\end{eqnarray}
where $i=\wpri$ for the lepton coming from the $\wpri$ and $i=N$ for the lepton coming from the neutrino decay.
 The solid lines are for $\wpri_R$ and the dashed lines for $\wpri_L$.  The $\Delta R_{jj}$ distributions peak at low values for both the left- and right-handed cases. 
 This is due to the $W$ from the heavy neutrino decay being highly boosted and its decay products therefore collimated.  
  Also, as can be seen from Fig.~\ref{ptdist.FIG}, in the $\wpri_R$ case the lepton from the neutrino decay is harder and hence the SM $W$ softer than in the $\wpri_L$ case.  
  Since the SM $W$ is less boosted in the right-handed case, the jets are less collimated and the $\Delta R_{jj}$ distribution has a longer tail for $\wpri_R$ than for $\wpri_L$.  
   Also, since the neutrino is highly boosted, its decay products are expected to land opposite in the transverse plane from the lepton from $\wpri$ decay. 
   Hence, $\Delta R^{\rm min}_{\ell_{\wpri} j}$ peaks near $\pi$ for both the the left-handed and right-handed case.  
   Finally, $\Delta R^{\rm min}_{\ell_N j}$ is peaked near $2m_N/E_N\approx 0.7$ for both the $\wpri_L$ and $\wpri_R$ cases.   
 The $\Delta R$ distributions at the 8 TeV LHC are peaked at similar values, but are more narrow than the 14 TeV distributions. 
  Based on these arguments, we define the isolation cuts given by Eq.~(\ref{cuts2.EQ}).

The isolation cuts more severely affect the $\wpri_L$ cross section since the $\Delta R_{jj}$ distribution is strongly peaked at low values for $\wpri_L$.
As the mass of the $\wpri$ increases, the SM $W$ from the heavy neutrino decay will become more boosted.  
Hence, the two jets will become more collimated and the effects of the isolation cuts will be even more significant. 
Since we will only be interested in the angular distributions of the lepton, it is possible to relax the $\Delta R_{jj}$ cut and look for one or two jets with two like sign leptons.  
Also, the separation between the lepton and jets from the heavy neutrino decay depend on the ratio of $m_N/M_{\wpri}$.  
As $m_N/M_{\wpri}$ increases (decreases) the lepton and jets become more (less) well separated.

\subsection{Background Reduction and Statistical Significance}
\label{BkgdRedux.SEC}

\begin{figure}[tb]
\centering
\includegraphics[width=0.55\textwidth]{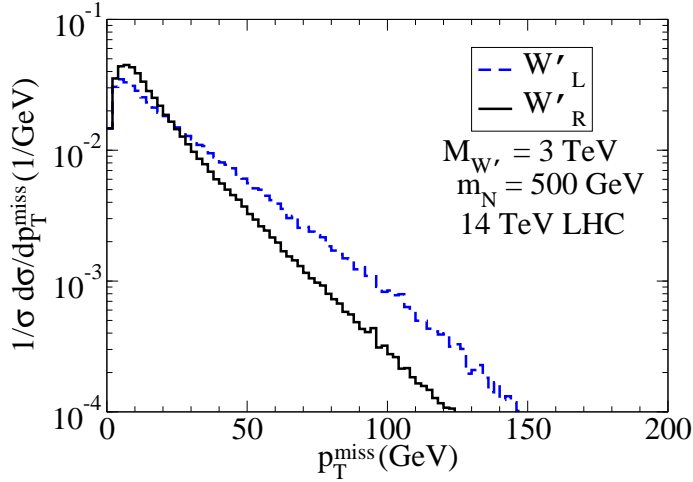}
\caption{Missing energy distribution for $pp\rightarrow {W'}_{L,R}^+\rightarrow \mu^+\mu^+ q\overline{q}'$ at the LHC.  
Energy smearing has been applied.}
\label{etmiss.FIG}
\end{figure}

The SM background for our $\ell^{+}\ell^{+}jj$ signature has been thoroughly studied for the 14 TeV LHC by Ref.~\cite{Atre:2009rg}.
The largest background to our process was found to be from $t\bar{t}$ events with the cascade decays,
\begin{equation}
 t\rightarrow W^+b\rightarrow\ell^+\nu_m b,\quad\bar{t}\rightarrow W^-\bar{b}\rightarrow W^-\bar{c}\nu_m \ell^+,
\end{equation}
and was also found to be greatly suppressed by the lepton isolation cuts in Eq.~(\ref{cuts2.EQ}).  
The background can be further suppressed by noting that leptonic $t\bar{t}$ events contain a final state light neutrino and 
therefore a considerable amount of missing transverse energy, $\etmiss$.
This is in direct comparison with our signal where all the $\etmiss$ is due to detector resolution effects.  
The $\etmiss$ for our like-sign leptons + dijet events is shown in Fig.~\ref{etmiss.FIG} for both the right- (solid) and left-handed (dashed) $W'$ cases.  
Furthermore, the two jets in our process originate from a SM $W$ whereas the jets in the top background do not.
Hence $\etmiss$ and dijet invariant mass, $m_{jj}$, cuts are also applied:
\begin{eqnarray}
\etmiss<30~\rm{GeV},~~~60~{\rm GeV}<&m_{jj}&<100~{\rm GeV}.
\label{cuts3.EQ}
\end{eqnarray}
The effect of these cuts on the signal rate are seen in the fourth line of Table~\ref{WpNxsect.TAB}. 

Having obtained a measurement of $m_{N}$ from Eq.~(\ref{mNPermu.EQ}) and $M_{W'}$ from the $W'$'s Jacobian peak, if desired, 
invariant mass cuts on $m_{\ell_{N}jj}$ and $\hat{s}$ can be imposed to further isolate the signal:
\begin{eqnarray}
|m_{\ell_{N}jj}-m_N|\leq 0.1~ m_N~~{\rm and}~~|\hat{s}-M_{W'}|\leq 0.1~M_{\wpri}.
\label{cuts4.EQ}
\end{eqnarray}
The effects of these cuts are shown in the fifth line of Table~\ref{WpNxsect.TAB}.

 As $\sqrt{s}$ increases from 8 TeV to 14 TeV, the percentage of events passing the selection cuts also increases. See the final line of Table~\ref{WpNxsect.TAB}. 
 In particular, we note that relatively fewer events are failing the cuts imposed on the reconstructed masses (Eq.~(\ref{cuts4.EQ})). 
 To understand this effect, consider that increasing the c.m.~energy also enlarges the phase space. Consequently, our internal propagators are more likely to be on-shell.

 \begin{figure}[tb]
\centering
\subfigure[]{
	\includegraphics[width=0.45\textwidth]{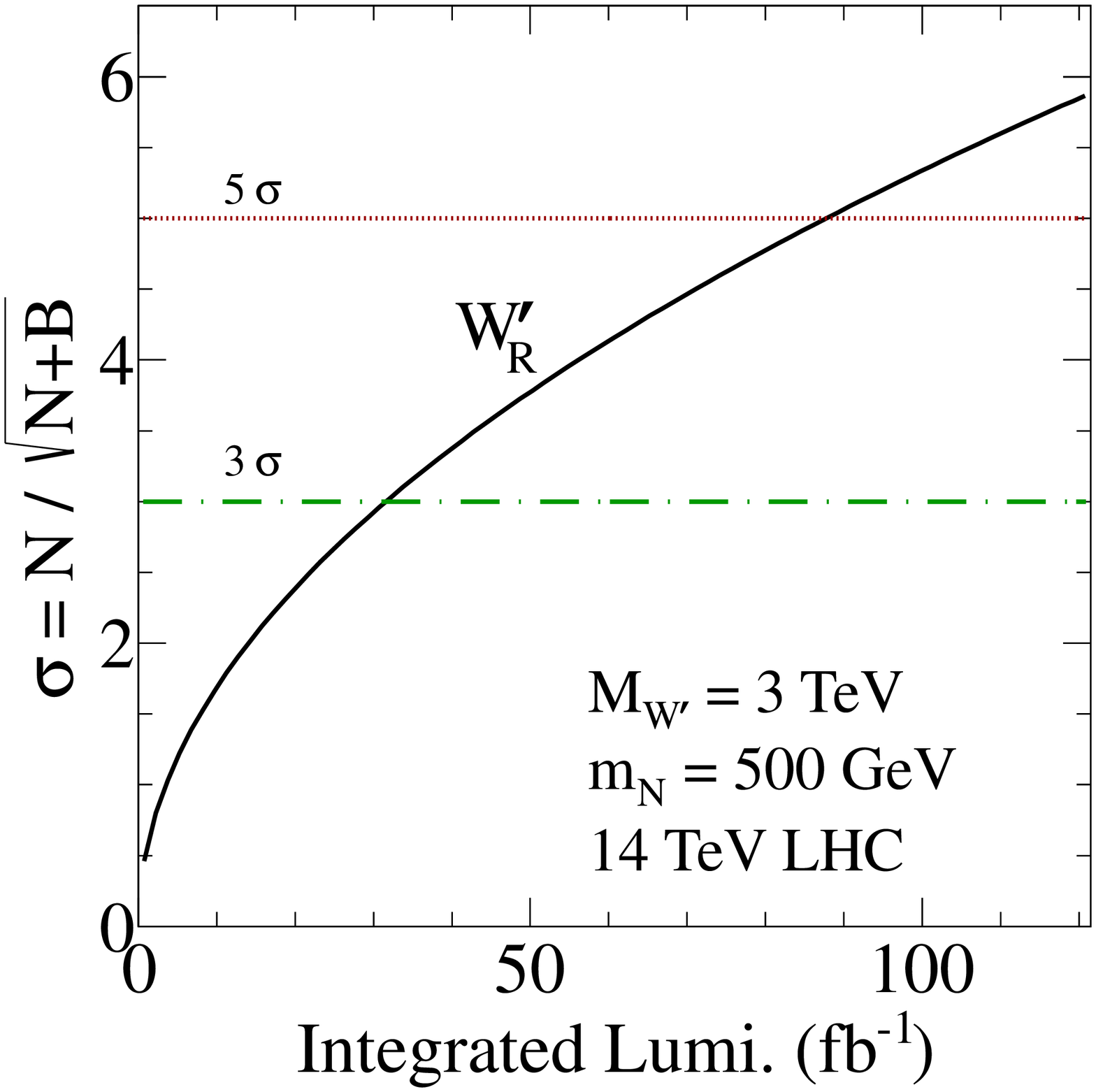}}
\subfigure[]{
		\includegraphics[width=0.45\textwidth]{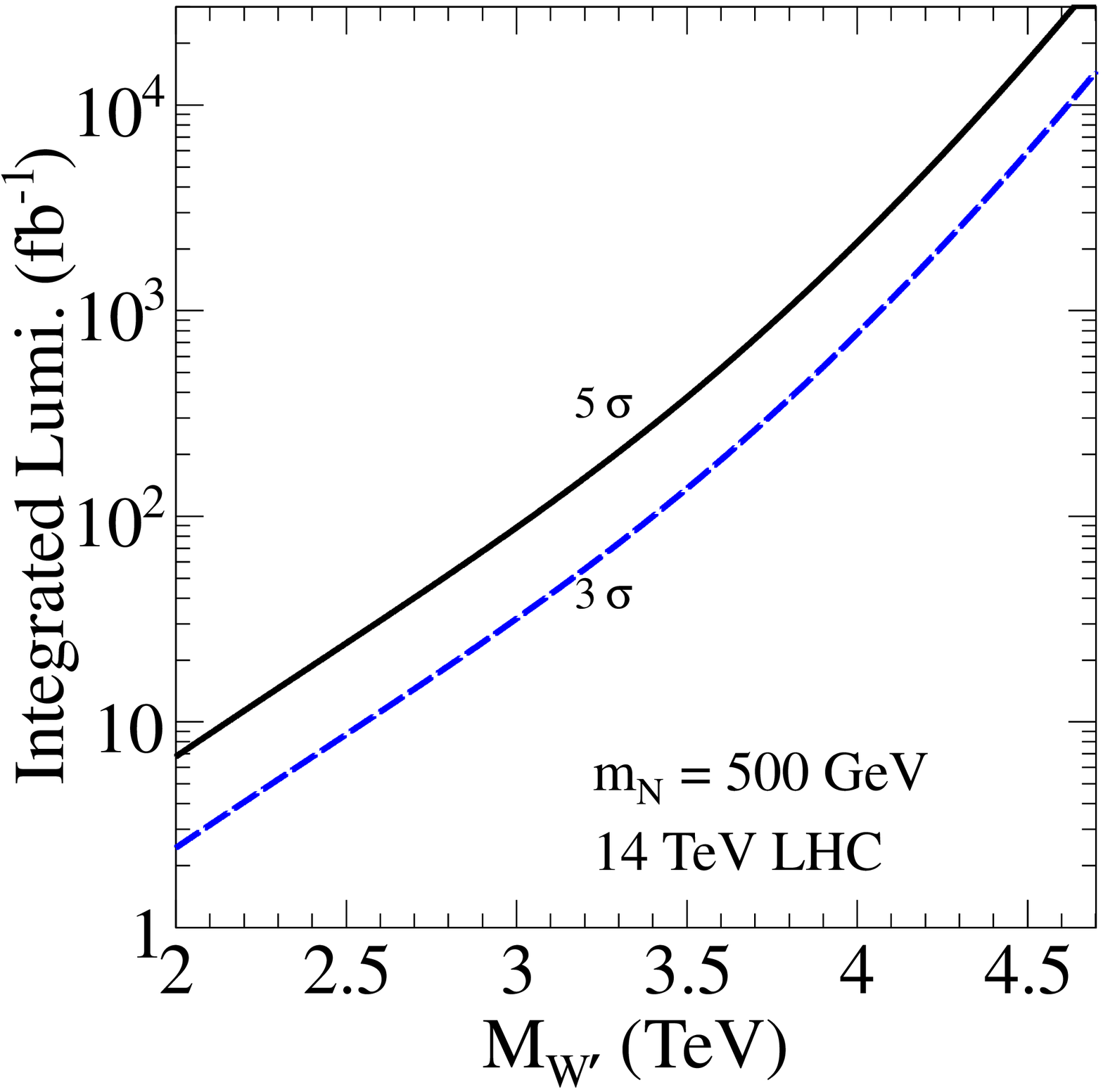}}
\caption{Integrated luminosity needed at 14 TeV LHC for (a) achievable statistical significance for $W'_R$ with $M_{W'}=3$ TeV and $m_{N}=500$ GeV, and (b) reachable $W'_R$ mass at $3\sigma$ and $5\sigma$ sensitivity. }
\label{significanceVsLumi.FIG}
\end{figure}

The contribution from the irreducible background for our $\ell^{\pm}\ell^{\pm}jj$ signal,  
\begin{eqnarray}
pp\rightarrow W^{\pm}W^{\pm}W^{\mp},&&pp\rightarrow W^{\pm}W^{\pm}jj,\quad\quad pp\rightarrow t\overline{t}
\end{eqnarray}
events and
\begin{eqnarray}
pp\rightarrow jjZZ,&&pp\rightarrow jjZW,
\end{eqnarray}
wherein leptons from the $Z$ boson escape from a detector, 
are estimated~\cite{Atre:2009rg} to be at most $\sigma=0.08$~fb using a comparable list of selection cuts.
However, this previous analysis does not impose any restriction on the invariant mass of the system as done in Eq.~(\ref{cuts4.EQ}), 
and therefore, realistically, the background will be much less than $0.08$~fb.
In either case, our $W'_{R}$ signal is clearly above background. 
Using $\sigma=0.08$~fb as an estimation for our background, 
we calculate the significance and reachability of our $W'_{R}$ signal at the 14 TeV LHC as shown in 
Fig. ~\ref{significanceVsLumi.FIG}. With 100 fb$^{-1}$ integrated luminosity, 
a $W'_R$ signal via the lepton-number violating process can be observed at a $5\sigma$ level up to a mass of 3 TeV.
As evident, the required integrated luminosity for a discovery at the LHC grows rapidly with increasing $M_{W'_R}$. 
This is expected if we again consider that the $W$ boson becomes increasingly boosted as $M_{W'_R}$ grows.
A more boosted $W$ leads to more collimated jets, which have more difficulty passing the isolation cuts (Eq.~(\ref{cuts2.EQ}))
than their less collimated counterparts.

\section{$W'$ Chiral Couplings From Angular Correlations at the LHC}
\label{WpriAC.SEC}
Once a new gauge boson $W'$ is observed at the LHC, it is of fundamental importance to determined the nature of its coupling to the SM fermions. 
Here, we identify various kinematical quantities that depend on the chiral couplings of the fermions to a $\wpri$.  
Each quantity will have a different dependence on the $\wpri$ chiral couplings and so will provide independent measurements of the chiral couplings.

\begin{figure}[tb]
\begin{center}
\subfigure[]{
      \includegraphics[width=0.29\textwidth,clip=true]{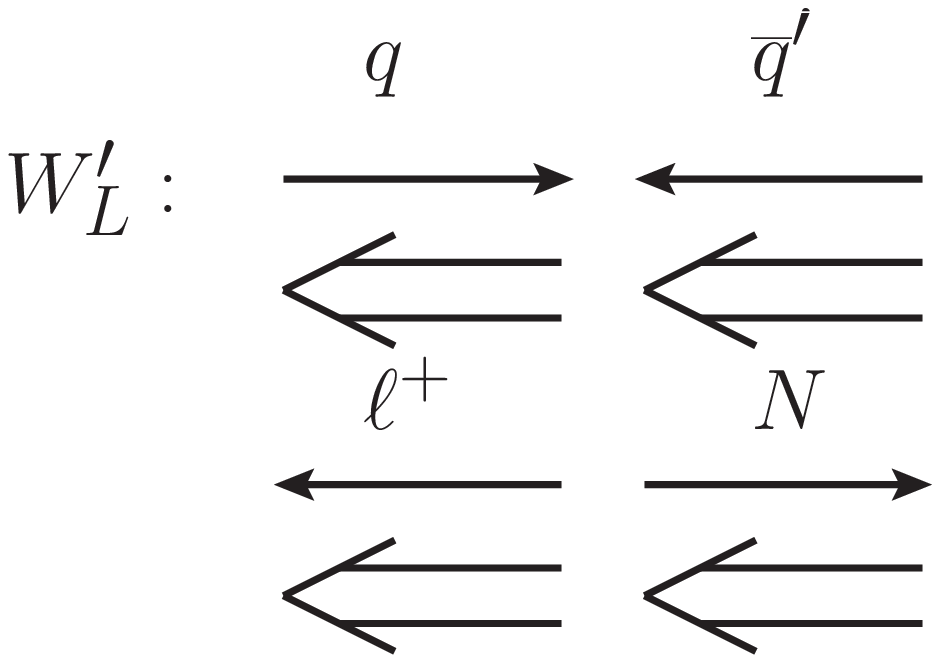}
}
\subfigure[]{
      \includegraphics[width=0.29\textwidth,clip=true]{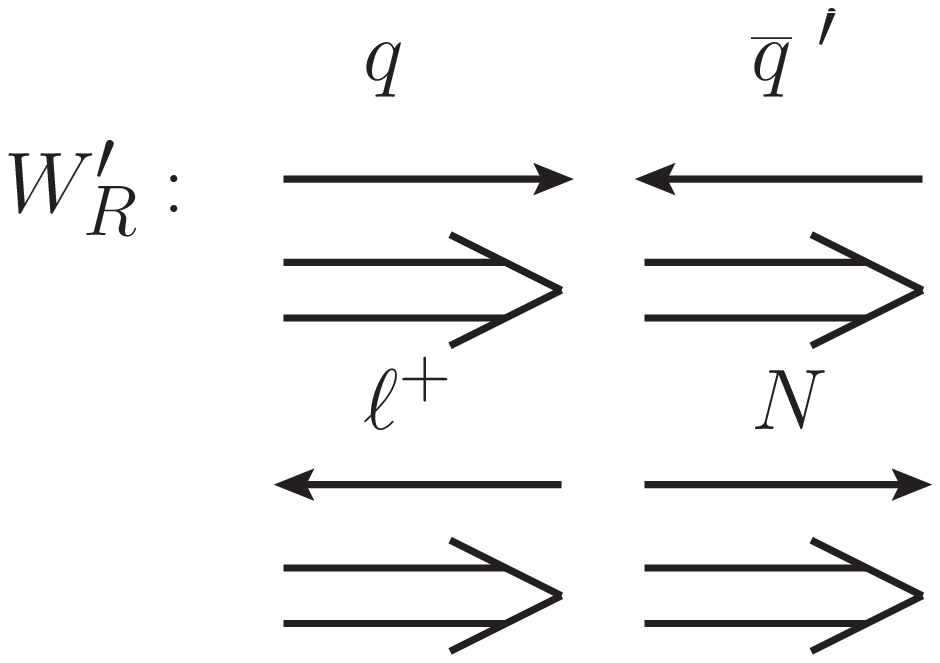}
}
\end{center}
\caption{Spin correlations for $q\bar{q}'\rightarrow W'\rightarrow N\ell^+$ for (a) left-handed and (b) right-handed couplings.  Single arrow lines represent momentum directions and double arrow lines represent spin directions.}
\label{Nspin.FIG}
\end{figure}

\subsection{$W'$ Chiral Couplings To Leptons}
\label{Wptspin.SEC}
Figure \ref{Nspin.FIG} shows the spin correlations for the process $q\bar{q}'\rightarrow W'\rightarrow N\ell^+$ in the partonic c.m.~frame for both 
the (a) left-handed and (b) right-handed cases.  
Double arrowed lines represent spin and single arrowed lines momentum.  As it is well-known, although the 
preferred charged lepton momentum direction leads to a clear distribution of parity violation, it cannot reveal more detailed nature of the chiral coupling. 
On the other hand, the nature of the $W'$ leptonic chiral couplings is encoded in polarization of the heavy neutrino, i.e., 
in the $\wpri_R$ ($\wpri_L$) case the heavy neutrino is preferentially right-handed (left-handed).  
Hence, if the polarization of the neutrino can be determined, the left-handed and right-handed cases can be distinguished. 
Spin observables such as $\left<\hat{s}_N\cdot \hat{a}\right>$, where $s_N$ is the spin of the heavy neutrino and $\hat{a}$ is an arbitrary spin quantization axis, 
are sensitive to the polarization of the heavy neutrino.  
Defining the angle $\theta^*$ between the $\hat a$ and the direction of motion of the charged lepton originating from the heavy neutrino decay, $\hat{p}_{\ell_2},$ 
the angular distribution of the partial width of the neutrino decaying into a charged lepton and two jets is~\cite{Tait:2000sh}
\bea
\displaystyle \frac{1}{\Gamma}\frac{d\Gamma }{d\cos\theta^*}(N\rightarrow \ell^{\pm} jj)=\frac{1}{2}
\left(1+2~A^{\ell^\pm}~\cos\theta^*\right),
\label{Ndec.EQ}
\eea
where $A^{\ell^+}=-A^{\ell^-}\equiv A$ due to the CP invariance.
The coefficient $A$ is related to $\left<\hat{s}_N\cdot \hat{a}\right>$ and is the forward-backward asymmetry of the charged lepton with respect to the direction $\hat a$.  
We will refer to $A$ as the analyzing power.  
The angular distribution of either of the two jets from the neutrino decay will also have a similar linear form and may be used to perform this analysis, 
although uncertainties in jet measurements may cause more complications.  

A highly boosted neutrino from a heavy $W'$ decay will be produced mostly in a helicity state; 
hence, it is natural to choose $\hat a = \hat{p}_N$, 
the direction of motion of the neutrino in the partonic c.m.~frame, and measure $\hat{p}_{\ell_2}$ in the neutrino rest-frame.   
At the partonic level, the angular distribution of the lepton from neutrino decay in the reconstructible neutrino rest-frame is (See App.~\ref{appendME.APP})
\begin{eqnarray}
\frac{d\hat{\sigma}(u\bar{d}\rightarrow\ell^+_1\ell^+_2 W^-)}{d\cos\theta_{\ell_{2}}}
=\frac{\hat{\sigma}_{Tot.}}{2}\left[1+\left(\frac{\hat{\sigma}(W_{0})-\hat{\sigma}(W_{T})}{\hat{\sigma}(W_{0})+\hat{\sigma}(W_{T})}\right)\left(\frac{2-\mu_{N}^{2}}{2+\mu_{N}^{2}}\right)
\left(\frac
{g_{R}^{\ell\:2}\vert Y_{\ell_1 N}\vert^{2}-g_{L}^{\ell\:2}\vert V_{\ell_1 N}\vert^{2}}
{g_{R}^{\ell\:2}\vert Y_{\ell_1 N}\vert^{2}+g_{L}^{\ell\:2}\vert V_{\ell_1 N}\vert^{2}}\right)\cos\theta_{\ell_{2}}\right].
\label{Ang.EQ}
\end{eqnarray}
Here $\hat\sigma(W_0)$ and $\hat\sigma(W_T)$ are the partonic level $u\overline{d}\rightarrow W^{'+}\rightarrow \ell^+_1\ell^+_2 W^{-}_{\lambda}$ cross sections 
with $N$ decaying into longitudinally $(\lambda=0)$ and transversely $(\lambda=T)$ polarized $W$'s, respectively. They are given by
\begin{eqnarray}
\hat{\sigma}(W_{0})&\equiv&\hat{\sigma}(u\bar{d}\rightarrow \ell^+_1 N\rightarrow \ell^+_1\ell^+_2 W^-_0)\\
&=&\frac{1}{9}\frac{1}{2^{10}}\frac{g^2}{\pi^2}\frac{ \vert V^{CKM'}_{ud}\vert^{2} \vert V_{\ell_{2}N}\vert^2}{(1+\delta_{\ell_{1}\ell_{2}})}
\left(g^{q~2}_{R} + g^{q~2}_{L}\right)
\left(g^{\ell\:2}_{R}\vert Y_{\ell_{1}N}\vert^{2} + g^{\ell\:2}_{L}\vert V_{\ell_{1}N}\vert^{2} \right)\left(\frac{m_{N}}{\Gamma_{N}}\right)\nonumber\\
&\times&
\frac{\hat{s}}{\left[(\hat{s}-M_{W'}^{2})^{2}+(\Gamma_{W'}M_{W'})^{2}\right]}
(1-y_{W}^{2})^{2}(1-\mu_{N}^{2})^{2}(2+\mu_{N}^{2})
\left(\frac{1}{2y_{W}^{2}}\right)\\
\hat{\sigma}(W_{T})&\equiv&\hat{\sigma}(u\bar{d}\rightarrow \ell^+_1 N\rightarrow \ell^+_1\ell^+_2 W^-_T)\\
&=& \hat{\sigma}(W_{0})\times 2y_{W}^{2}.
\end{eqnarray}
where $\mu_N=m_N/\sqrt{\hat s}$, $y_W=M_W/m_N$, and $\hat{\sigma}_{Tot.}=\left(\hat\sigma(W_0)+\hat\sigma(W_T)\right)\times{\rm BR}(W\rightarrow q\bar{q}')$ 
is the total partonic cross section.
As $W'$ comes on-shell, $\mu_N\rightarrow x_N$.
In this reference frame, $\theta^{*}$ from Eq.~(\ref{Ndec.EQ}) satisfies
\begin{equation}
 \cos\theta^{*} = \cos\theta_{\ell_2} \equiv \hat{p}_{\ell_2} \cdot \hat{p}_{N},
 \label{thetaEll2.Eq}
\end{equation}
where, again, $\hat{p}_{\ell_2}$ is measured in the neutrino rest-frame and $\hat{p}_{N}$ is measured in the partonic c.m. frame.

For an on-shell $\wpri$, the analyzing power at the partonic and hadronic level are the same.  
In such a case, after comparing Eqs.~(\ref{Ndec.EQ}) and (\ref{Ang.EQ}), we find that the analyzing power is
\begin{eqnarray}
A&=&
\frac{1}{2}
\left(\frac{\hat\sigma(W_0)-\hat\sigma(W_T)}{\hat\sigma(W_0)+\hat\sigma(W_T)}\right)
\left(\frac{2-x^2_N}{2+x^2_N}\right)
\left(\frac
{g_{R}^{\ell\:2}\vert Y_{\ell_1 N}\vert^{2}-g_{L}^{\ell\:2}\vert V_{\ell_1 N}\vert^{2}}
{g_{R}^{\ell\:2}\vert Y_{\ell_1 N}\vert^{2}+g_{L}^{\ell\:2}\vert V_{\ell_1 N}\vert^{2}}\right)\nonumber\\
&=&
\frac{1}{2}
\left(\frac{1-2y^2_W}{1+2y^2_W}\right)
\left(\frac{2-x^2_N}{2+x^2_N}\right)
\left(\frac
{g_{R}^{\ell\:2}\vert Y_{\ell_1 N}\vert^{2}-g_{L}^{\ell\:2}\vert V_{\ell_1 N}\vert^{2}}
{g_{R}^{\ell\:2}\vert Y_{\ell_1 N}\vert^{2}+g_{L}^{\ell\:2}\vert V_{\ell_1 N}\vert^{2}}\right).
\label{Apart.EQ}
\end{eqnarray}

The different signs for the analyzing power between the neutrino decays to the two different $W$ polarizations and between the $\wpri_{L,R}$ cases can be understood via the spin correlation in Fig.~\ref{Lspin.FIG}.  
For the $\wpri_R$ case, a heavy neutrino decaying to a longitudinal (transverse) $W$ will have 
the charged lepton preferentially moving with (against) $\hat{p}_{N}$.
For the $\wpri_L$ case the helicity of the neutrino, and therefore the direction of the charged lepton, is reversed. 
Hence the analyzing power is proportional to $(\hat\sigma(W_0)-\hat\sigma(W_T))(g_{R}^{\ell\:2}\vert Y_{\ell_1 N}\vert^{2}-g_{L}^{\ell\:2}\vert V_{\ell_1 N}\vert^{2})$.  

In the analysis of Fig.~\ref{Lspin.FIG}, the left- and right-chiral neutrinos at the $\wpri\rightarrow N\ell^+$ vertex 
are approximated as the left-handed and right-handed helicity states in the partonic c.m.~frame. 
As the neutrino becomes more massive relative to the $\wpri$, the approximation of the chiral basis by the helicity basis begins to break down, 
i.e., the left- (right-) helicity state makes a larger contribution to the right- (left-) chiral state.  
In Eq.~(\ref{Ang.EQ}), this is reflected by the $\cos\theta_{\ell_2}$ ($\cos\theta_\ell$ for simplicity) coefficient  
\begin{equation}
\frac{2-x^2_N}{2+x^2_N} = \frac{2M_{W'}^{2}-m_{N}^{2}}{2M_{W'}^{2}+m_{N}^{2}}.
\end{equation}
As $x_N$ increases, the distribution flattens due to the right-handed (left-handed) neutrino helicity state, 
thereby making a larger contribution to the $\wpri_L$ ($\wpri_R$) distributions.

\begin{figure}
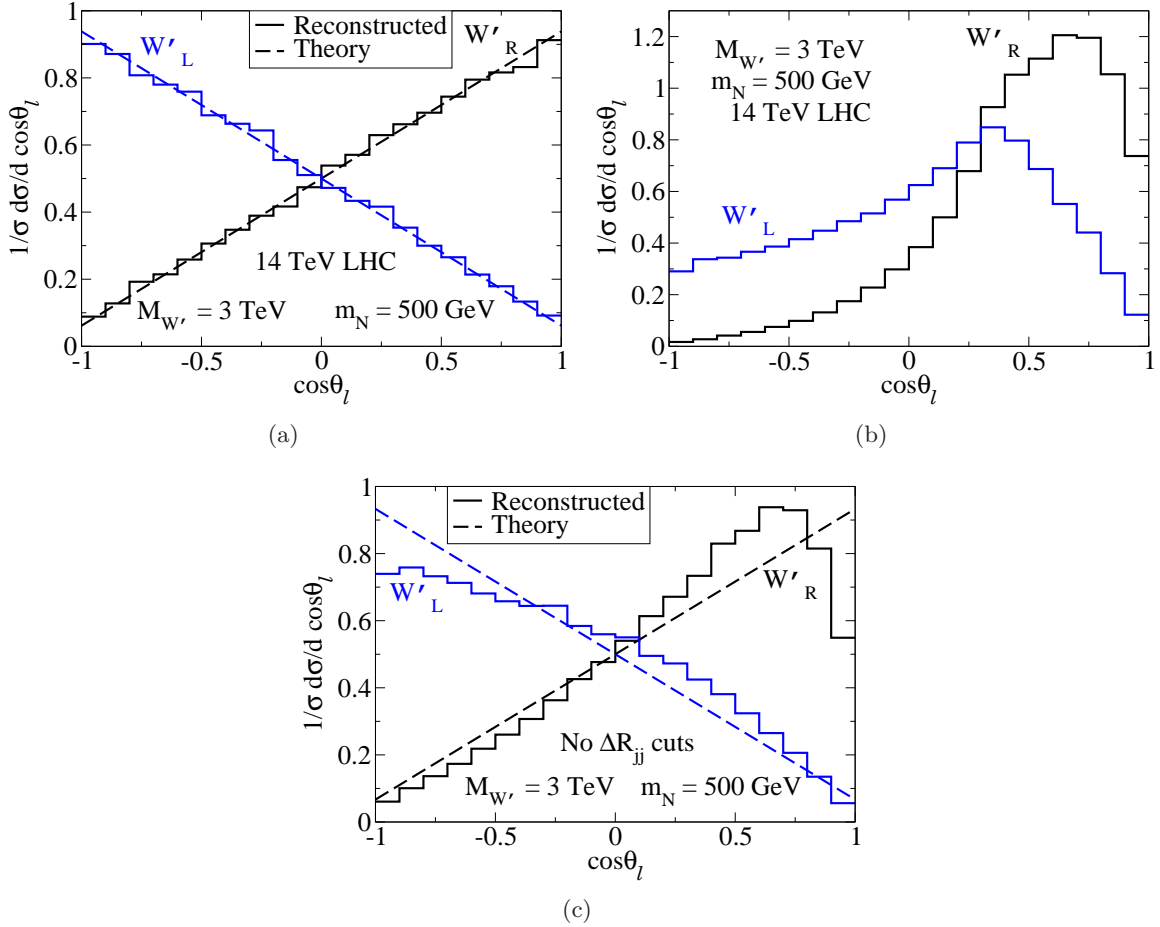

\begin{center}
\subfigure[]{
      \includegraphics[width=0.45\textwidth,clip]{CthlNCNS500_3.eps}
\label{wpriacNS.FIG}}
\subfigure[]{
      \includegraphics[width=0.45\textwidth,clip]{CthlAll500_3.eps}
\label{wpriacSM.FIG}}\\
\subfigure[]{
      \includegraphics[width=0.45\textwidth,clip]{CthlAll500_3NoJIso.eps}
\label{wpriacNoJIso.FIG}}
\caption{
The angular distribution of the charged lepton originating from neutrino decay 
in the heavy neutrino rest-frame with respect to the neutrino moving direction in the partonic c.m.~frame at the LHC with $M_{W'}$, $m_N$ set by Eq.~(\ref{benchParam.EQ}).
Distribution (a) without smearing or cuts, (b) with energy smearing and cuts 
in Eqs.~(\ref{cuts1.EQ}), (\ref{cuts2.EQ}), (\ref{cuts3.EQ}), and (\ref{cuts4.EQ})
, and (c) with all cuts applied to (b) except the $\Delta R_{jj}$ cuts in Eq.~(\ref{cuts2.EQ}). 
The solid lines are for the Monte Carlo simulation results and in (a) and (c) the dashed lines are for the analytical result in Eq.~(\ref{Ang.EQ}).
}
\label{wpriac.FIG}
\end{center}
\end{figure}

Figure \ref{wpriac.FIG} shows the hadronic level angular distribution of the lepton in the neutrino's rest-frame for both $\wpri_L$ and $\wpri_R$ at the LHC.  
The case without smearing or cuts is shown in Fig.~\ref{wpriacNS.FIG}, and contains both the analytical results (dashed line) and Monte Carlo simulation (solid line) histograms.  
As can be clearly seen, the analytical and numerical results are in good agreement.
 Figure \ref{wpriacSM.FIG} shows the leptonic angular distribution after energy smearing and cuts in Eqs.~(\ref{cuts1.EQ}), (\ref{cuts2.EQ}), (\ref{cuts3.EQ}), and (\ref{cuts4.EQ}).  
 Notice that there is a small depletion of events for $\cos\theta_\ell \approx 1$ and a large depletion when $\cos\theta_\ell<0$.  
 First, when $\cos\theta_\ell\approx 1$ the charged lepton is moving with and the jets against the direction of motion of the neutrino in the partonic c.m.~frame. 
 Hence, with boost back to the partonic c.m.~frame, the jets are softest at this point and the jet $p_T$ cuts in Eq.~(\ref{cuts1.EQ}) lead to a depletion of event in this region. 
When $\cos\theta_\ell < 0$, the lepton is moving against and the SM $W$ is moving with the neutrino's direction of motion.  
Hence, with the boost back to the partonic c.m.~frame, the $W$ is boosted and its decay products highly collimated.  
Consequently, the $\Delta R_{jj}$ cuts in Eq.~(\ref{cuts2.EQ}) lead to a large depletion of events.  
Figure \ref{wpriacNoJIso.FIG} shows lepton angular distribution with the same cuts as Fig.~\ref{wpriacSM.FIG} except the $\Delta R_{jj}$ cuts.  
For comparison, both the Monte Carlo simulation with cuts (solid) and analytical results without cuts (dashed) are shown.  
It is clear that the discriminating power of the lepton angular distribution would increases and the Monte Carlo distribution approaches the analytical results if the jet isolation cuts are relaxed.

The analyzing power in Eq.~(\ref{Apart.EQ}) can additionally be related to the forward backward asymmetry
\bea
{\cal A}\, = \, \frac{\sigma(\cos\theta_\ell\geq 0)-\sigma(\cos\theta_\ell < 0)}{\sigma(\cos\theta_\ell\geq 0)+\sigma(\cos\theta_\ell < 0)}.
\eea
Without cuts or smearing, ${\cal A}\,=\,A$; and for the values of $m_N$, $M_{W'}$ stipulated in Eq.~(\ref{benchParam.EQ}), 
\begin{equation}
 A=\left\{
\begin{matrix}
+0.43, & W'=W'_{R}\\ 
-0.43, & W'=W'_{L}
\end{matrix}\right. .
\end{equation}
The simulated values for the forward backward asymmetry with consecutive cuts are shown in Table~\ref{WpNAFB.TAB}. 
Again, simulations are in good agreement with the theoretical prediction for the forward backward asymmetry for no smearing or cuts. 
  As the cuts become more severe, the simulated and theoretical values deviate more, however the $\wpri_L$ and $\wpri_R$ cases can still be distinguished clearly.  
  Furthermore, as shown in the final row, if the $\Delta R_{jj}$ cuts in Eq.~(\ref{cuts2.EQ}) are relaxed, 
  the discriminating power of the asymmetry is greatly increased, and the theory and simulation are in much better agreement.

\begin{table}
\begin{center}
\begin{tabular}{|l|c|c|c|c|} \hline \hline
                         \multirow{2}{*}{ ~~~~~~~~~~~~~~~~~~~~~~~~~~~~~~~~~~~${\cal A}$ }&\multicolumn{2}{c|} {8 TeV}& \multicolumn{2}{c|} {14 TeV} \\\cline{2-5}
                                   &$\wpri_L$ &$\wpri_R$ & $\wpri_L$ & $\wpri_R$   \\ \hline \hline
~~~~~~~~~~Reco.~without Cuts or Smearing  & $-0.42$  &$0.42$ & $-0.43$ & $0.43$ \\ \hline
+~Smearing~+~Fiducial~+~Kinematics (Eq.~(\ref{cuts1.EQ}))~~& $-0.46$  &$0.33$ & $-0.47$ &$0.34$\\ \hline
~~~~~~~~~~~~~~~~~+~Isolation~(Eq.~(\ref{cuts2.EQ})) & $-0.11$& $0.59$  &$0.083$&$0.72$ \\ \hline
~~~~~~~~~~+$\not\!\!E_{T}$~+~$m_{jj} $~Requirements~(Eq.~(\ref{cuts3.EQ})) & $-0.078$ & $0.62$ & $0.11$ & $0.75$ \\\hline
~~~~~~~~~~~~~~~~~+~Mass~Reco.~(Eq.~(\ref{cuts4.EQ})) & $0.16$ & $0.77$& $0.18$&  $0.77$  \\ \hline
~~~~~~~~~~~~~~~~~$-\Delta R_{jj}$            &$-0.34$  & $0.49$& $-0.34$ & $0.49$ \\\hline \hline 
\end{tabular} 
\caption{Forward-backward asymmetry for $pp\rightarrow {W'}_{L,R}^+\rightarrow \mu^+\mu^+ q\overline{q}'$ with consecutive cuts at 8 and 14 TeV LHC.  The last row has the same cuts applied as the previous row with the removal of the $\Delta R_{jj}$ cuts in Eq.~(\ref{cuts2.EQ}).}
\label{WpNAFB.TAB}
\end{center}
\end{table}

\subsection{$\wpri$ Chiral Couplings to Initial-State Quarks}

Thus far, we have only presented the results to test the chiral coupling of $W'$ to the final state leptons. It is equally important to examine its couplings to the initial state quarks. 
Define an azimuthal angle 
\begin{equation}
\displaystyle \cos\Phi =
\frac{\hat{p}_N\times{\vec p}_{\ell_{2}}}{|\hat{p}_N\times{\vec p}_{\ell_{2}}|}
\cdot
\frac{\hat{p}_N\times{\vec p}_q}{|\hat{p}_N\times{\vec p}_q|}, 
\label{AziDef.EQ}
\end{equation}
as the angle between the $qq'\rightarrow N\ell^{+}_{1}$ production plane and $N\rightarrow W^-\ell^{+}_{2}$ decay plane in the neutrino rest-frame,  
where ${\vec p}_{\ell_{2}}$ is the three momentum of $\ell_{2}$, the charged lepton identified as originating from the neutrino;
$\hat{p}_N$ is the direction of motion of the neutrino in the partonic c.m.~frame; and ${\vec p}_q$ is the initial-state quark momentum.  
The definition of $\Phi$ is invariant under boosts along $\hat{p}_N$, 
hence the quark and charged lepton momenta can be evaluated either in the partonic c.m.~or the neutrino rest-frame.
The angular distribution between the two planes is thus calculated to be
\bea
\displaystyle\frac{d\hat\sigma}{d\Phi}=\frac{\sigma_{Tot.}}{2\pi}\left[
1+\frac{3\pi^2}{16} \ \frac{\mu_N}{2+\mu^2_N}
\left(\frac{\hat\sigma(W_0)-\hat\sigma(W_T)}{\hat\sigma(W_0)+\hat\sigma(W_T)}\right)
\left(\frac
{g_{R}^{q\:2}-g_{L}^{q\:2}}
{g_{R}^{q\:2}+g_{L}^{q\:2}}\right)
\cos\Phi\right].
\label{Azim.EQ}
\eea
The distribution for $\wpri_L$ is $180^\circ$ out of phase with the $\wpri_R$ distribution and the slope  
only depends on the $W'$ chiral coupling to the initial-state quarks.  
Hence, the phase of this distribution determines the chirality of the initial-state quarks couplings to the $\wpri$ $independently$ of the leptonic chiral couplings to the $\wpri$.

\begin{figure}[tb]
\begin{center}
\subfigure[]{
      \includegraphics[width=0.35\textwidth,clip=true]{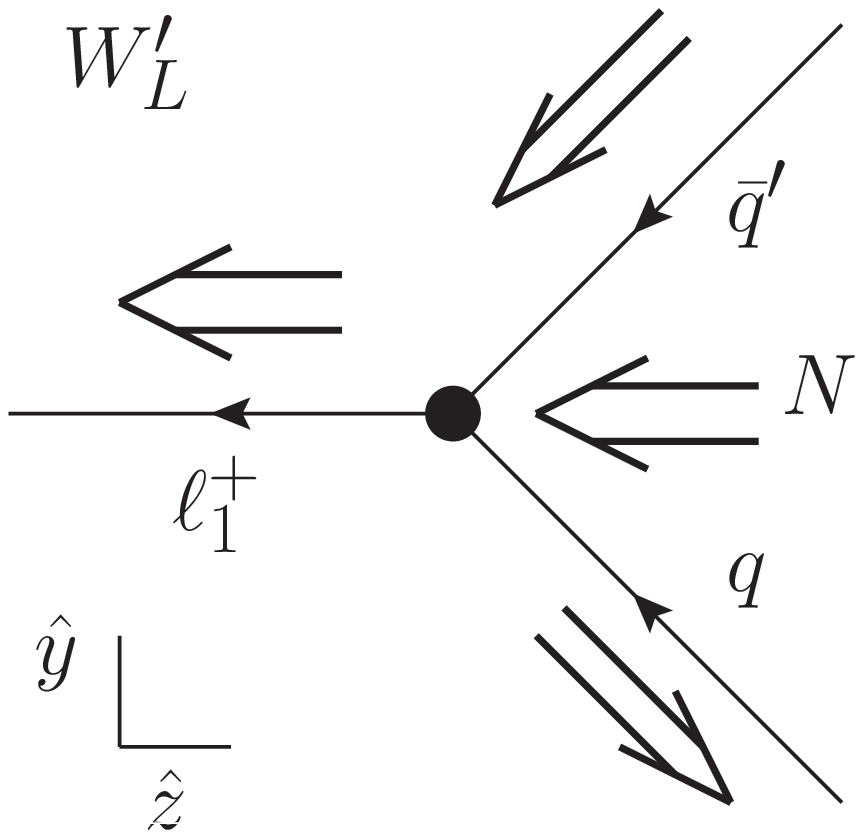}
\label{prodR.FIG}
}
\subfigure[]{
      \includegraphics[width=0.35\textwidth,clip=true]{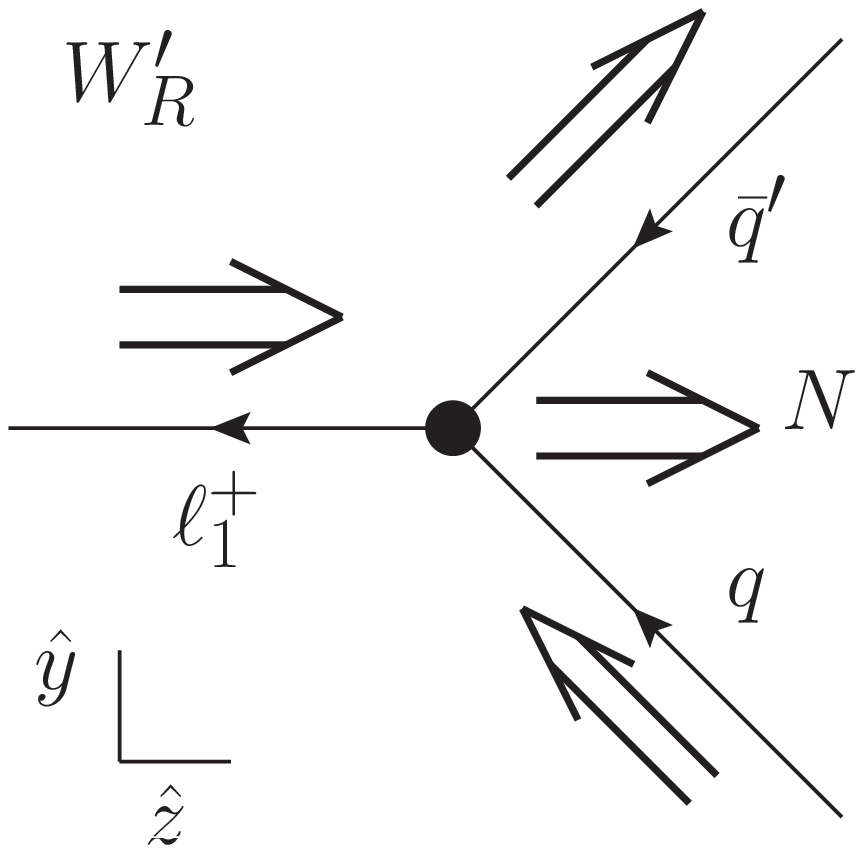}
\label{prodL.FIG}
}
\end{center}
\caption{Spin correlations for neutrino production in the neutrino rest-frame.  Single arrowed lines represent momentum and double arrowed lines represent spin in the helicity basis.  The $\hat{z}$-axis is defined to be the neutrino's direction of motion in the partonic c.m.~frame and the $\hat{y}$-axis is defined such that $y$-component of the initial-state quark momentum is always positive.}
\label{Kin.FIG}
\end{figure}

To understand the distribution in Eq.~(\ref{Azim.EQ}), we consider the spin correlations between the initial and final states.  
As noted previously, the angle $\Phi$ is invariant under the boosts along $\hat{p}_{N}$.
So for simplicity, we consider the spin correlations in the heavy neutrino rest-frame.  
Figure \ref{Kin.FIG} shows the spin correlations of the neutrino production in the neutrino's rest-frame for both the (a) $\wpri_L$ and (b) $\wpri_R$ cases.  
Like before, single arrowed lines represent momentum directions and double arrowed lines spin in the helicity basis.   
Also, we define the production plane to be oriented in the $\hat{y}-\hat{z}$ plane 
such that the $\hat{y}$-component of the quark momentum always points along the positive $\hat y$-axis and that $\hat{z} = \hat{p}_{N}$. 
With this axis convention, $\Phi=-\phi_{\ell_{2}}$, where $\phi_{\ell_{2}}$ is the azimuthal angle of $\ell_{2}$ as measured from the positive $\hat{y}-$axis.

\begin{figure}[tb]
\centering
\subfigure[]{
      \includegraphics[width=0.48\textwidth,clip=true]{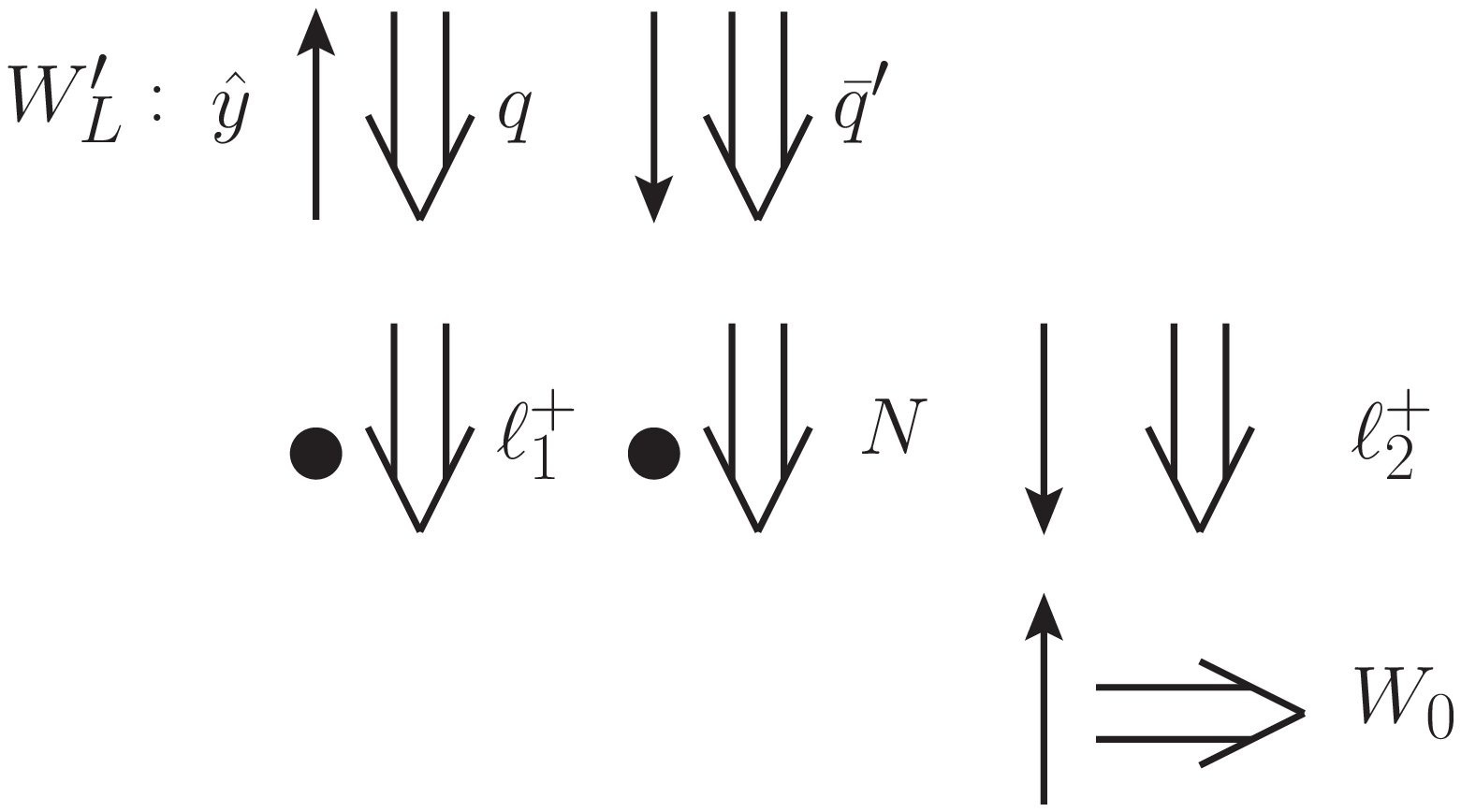}
}
\subfigure[]{
      \includegraphics[width=0.48\textwidth,clip=true]{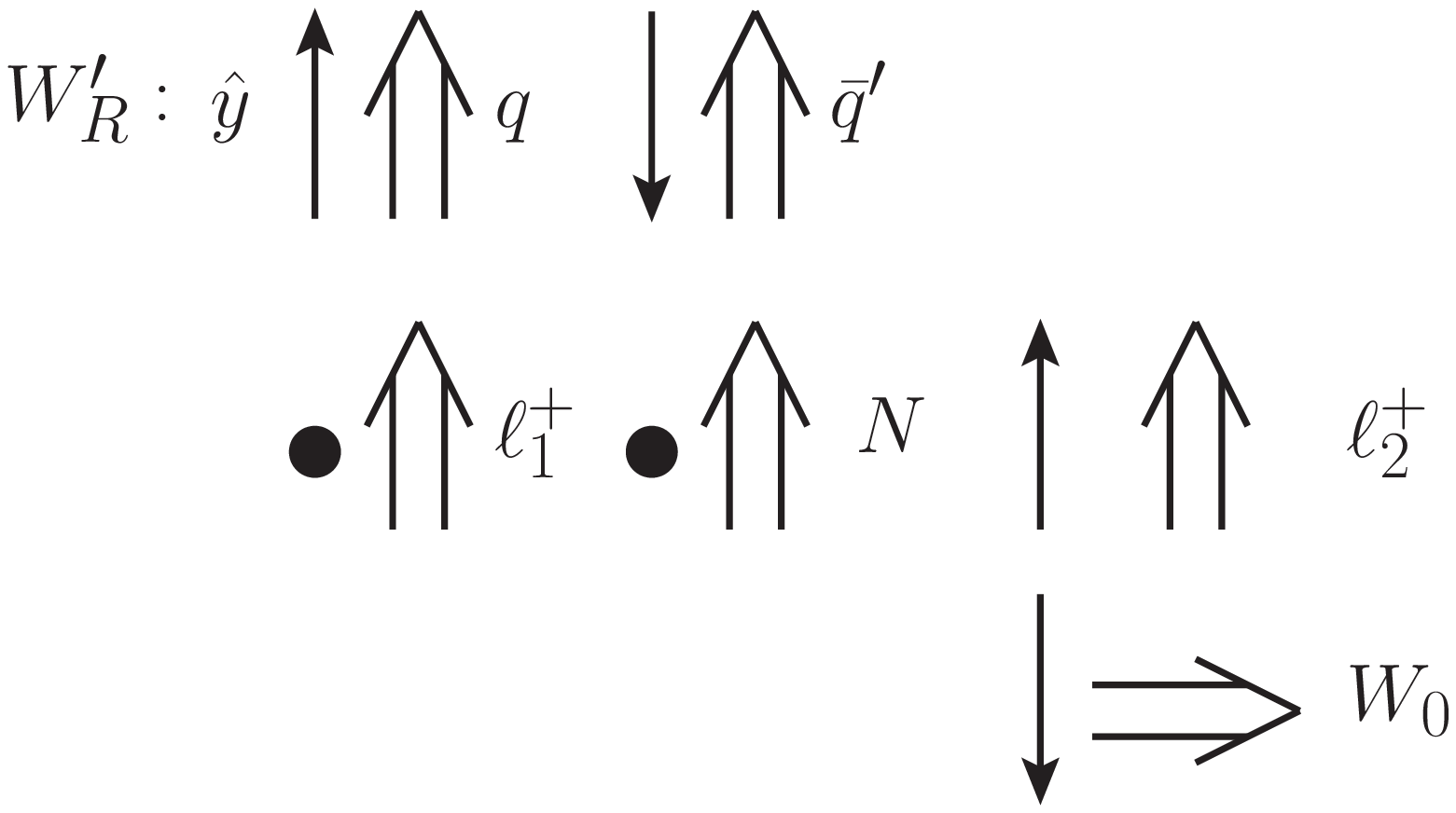}
}\\\vspace{0.3in}
\subfigure[]{
      \includegraphics[width=0.48\textwidth,clip=true]{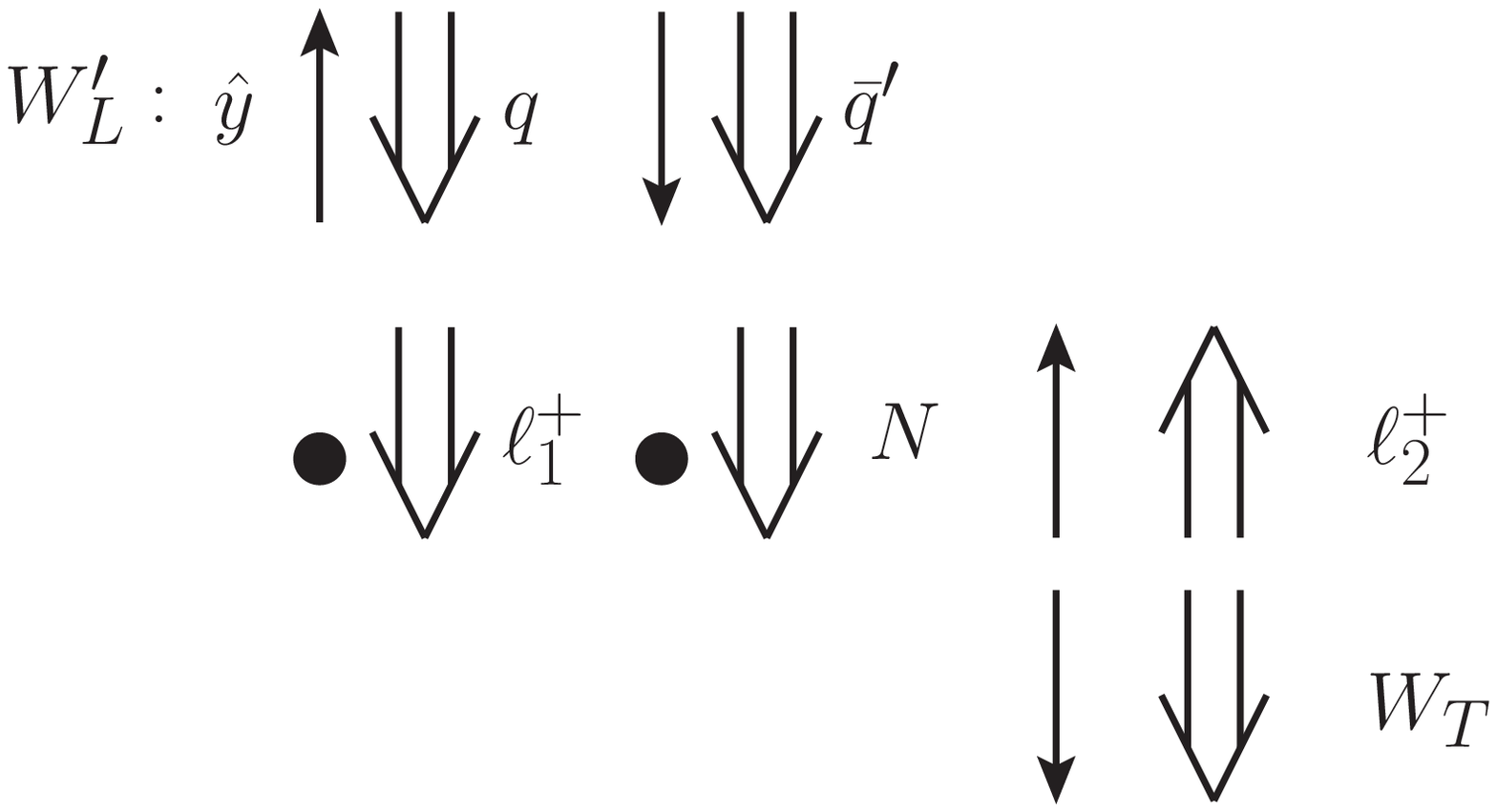}
}
\subfigure[]{
      \includegraphics[width=0.48\textwidth,clip=true]{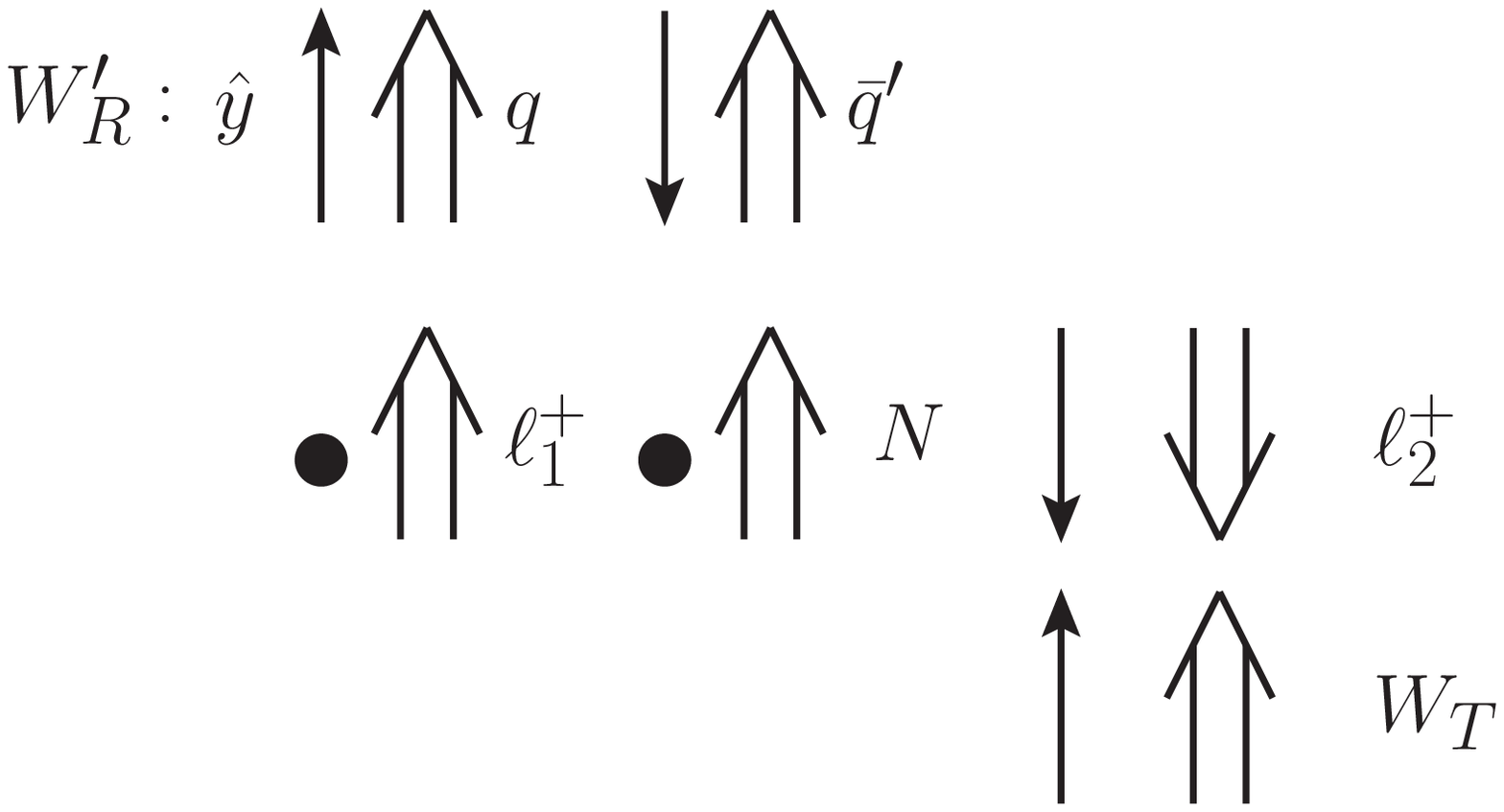}
}
\caption{Spin correlations in the neutrino rest-frame as described in Fig.~\ref{Kin.FIG}.  Double arrowed lines represent spin with $\hat{y}$ being the quantization axis and single arrowed lines are the $\hat{y}$ component of the particles.}
\label{PhiSpin.FIG}
\end{figure}

Figure \ref{PhiSpin.FIG} shows the spin correlations for the heavy neutrino production and decay with the spin quantization axis chosen to be the $\hat{y}$ direction as defined above.   
The $\wpri_L$ case is shown in Figs.~\ref{PhiSpin.FIG}(a,c) and the $\wpri_R$ case in (b,d). 
The solid dots next to the $N$ and $\ell_1$ indicate that they have no momentum in the $\hat{y}$-direction.  
In the $\wpri_R$ case, the initial-state quark must be right-handed and the initial-state antiquark left-handed.  
Hence, the total spin of the initial-state points in the positive $\hat{y}$-direction, causing the spin of the neutrino to also point in the positive $\hat{y}-$direction.  
When the neutrino decays to a longitudinal or transverse $W$, the lepton from the neutrino decay has spin along or against the $\hat{y}$-axis, respectively.  
For the $\wpri_R$ case, figures \ref{PhiSpin.FIG}(b) and (d) show the decay into longitudinal and transverse $W$'s, respectively.  
Therefore, for the decay into $W_0$ ($W_T$)  case, the lepton prefers to move in the same (opposite) direction as the initial-state quark and $\Phi$ peaks at $0$ ($\pm\pi$).  
In the $\wpri_L$ case, the direction of motion of $\ell_2$ 
relative to the direction of motion of the initial-state quark is reversed and the peaks in the $\Phi$ distribution are shifted by $\pi$.  
This explains the $180^\circ$ phase difference in the angular distribution, Eq.~(\ref{Azim.EQ}), between the $\wpri_L$ and $\wpri_R$ cases, and between the neutrino decay to $W_0$ and $W_T$.  
Also, notice that this argument only relies on the $\wpri-q-q'$ coupling and {\it not} the $\wpri-N-\ell$ chiral couplings.  
Hence, measuring the distribution of the angle between the $qq'\rightarrow N\ell_{1}$ production and the $N\rightarrow {\ell_2}^+ W^-$ decay planes can determine the chiral couplings of a $\wpri$ to light quarks independently from the chiral couplings of the $\wpri$ to leptons.

\begin{figure}[tb]
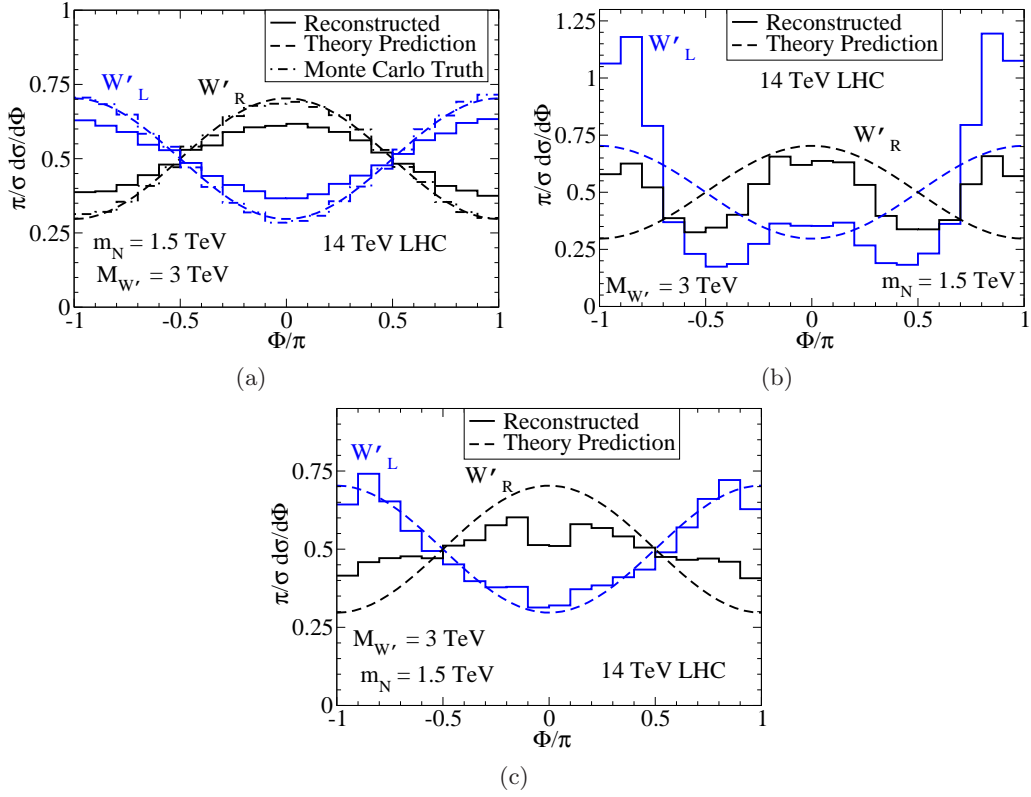

\centering
\subfigure[]{
       \includegraphics[width=0.4\textwidth,clip]{PhiNCNS1500_3.eps}
\label{PhiNCNS.FIG}}
\subfigure[]{
       \includegraphics[width=0.4\textwidth,clip]{PhiAll1500_3.eps}
\label{philrecl.FIG}}
\subfigure[]{
       \includegraphics[width=0.4\textwidth,clip]{PhiAllNoJIso1500_3.eps}
\label{philrecNoJIso.FIG}}
\caption{$\Phi$ distributions at the 14 TeV LHC with $M_\wpri=3$~TeV and $m_N=1.5$~TeV for fully reconstructed events (solid), 
the analytical result in Eq.~(\ref{Azim.EQ}) (dashed), and Monte Carlo truth (dash-dot).  
Figure (a) is without energy smearing or cuts, (b) with energy smearing and cuts in Eqs.~(\ref{cuts1.EQ}), (\ref{cuts2.EQ}), (\ref{cuts3.EQ}), and (\ref{cuts4.EQ}),
 and (c) with the same cuts as (b) without the $\Delta R_{jj}$ cut in Eq.~(\ref{cuts2.EQ}).
}
\label{Phi.FIG}
\end{figure}

Most of the angular definition and analysis depend on the initial state quark momentum direction.  
Since the LHC is a symmetric $pp$ machine, this is not known {\it a priori}.  However, at the LHC $u$ and $d$ quarks are valence and antiquarks are sea.  
Hence, the initial-state quark generally has a larger momentum fraction than the initial-state antiquark; 
and the initial-state quark direction can be identified as the direction of motion of the fully reconstructed partonic c.m.~frame.  
Similar techniques have been used for studying forward-backward asymmetries associated with new heavy gauge bosons~\cite{Gopalakrishna:2010,Langacker:1984dc}.

Figure \ref{Phi.FIG} shows the $\Phi$ distributions at the $14$~TeV LHC with $M_\wpri=3$~TeV for both $\wpri_L$ and $\wpri_R$.  
From Eq.~(\ref{Azim.EQ}), the amplitude of the $\Phi$ distribution depends on the ratio $m_N/M_\wpri$, and therefore increase $m_N$ to $1.5$~TeV.  
The solid line is the $\Phi$ distribution with the initial state quark moving direction identified as the partonic c.m.~frame boost direction; 
the dashed lines is the theoretical distribution given in Eq.~(\ref{Azim.EQ}); and in (a) the dash-dot lines are the Monte Carlo truth,~i.e. 
using the known direction of the initial-state quark.  

Figure \ref{PhiNCNS.FIG} does not include cuts or smearing; as can be seen, the Monte Carlo truth and theoretical calculation agree very well.
The reconstructed distribution has a smaller amplitude than the theoretical distribution due to the direction of the initial-state quark being misidentified.  
Figure \ref{philrecl.FIG} shows the theoretical prediction and reconstructed distribution with smearing and the cuts in Eqs.~(\ref{cuts1.EQ},\ref{cuts2.EQ},\ref{cuts3.EQ},\ref{cuts4.EQ}) applied.  
For $\Phi=0$, the SM $W$ is maximally boosted and its decay products are maximally collimated.  
Consequently, the $\Delta R_{jj}$ cut in Eq.~(\ref{cuts2.EQ}) causes a large depletion of events in the central region.  
Figure \ref{philrecNoJIso.FIG} shows the reconstructed distribution with the same cuts as (b) minus the $\Delta R_{jj}$ cut.  
With the relaxation of this cut, the $\wpri_L$ and $\wpri_R$ cases become reasonably discernible with the $\wpri_L$ distribution nearly the same as the theoretical prediction.  
The continued depletion of events at $\Phi=0$ and $\Phi=\pm\pi$ are due to the rapidity cuts on leptons and jets, respectively.

\section{Unlike-Sign Dilepton Angular Distributions}
\label{Opp.SEC}

Intrinsically, Majorana neutrinos can decay to positively or negatively charged leptons, and therefore also contribute to the $L$-conserving process
\bea
pp\rightarrow \wpri \rightarrow \ell_1^+ \ell_2^- jj.
\eea
  These events can be reconstructed similarly to the method described in Sec.~\ref{Like.SEC}.  
  However, the SM backgrounds for this process, particularly $pp\rightarrow Z jj$, will be larger.  
  Our purpose here is not to do a full signal versus backgrounds study, but to comment on the differences between the like-sign and unlike-sign lepton cases. 
  Again, $u\bar{d}$ has a larger parton luminosity than $d\bar{u}$, so we focus only on $W'^+$ production:
\bea
pp\rightarrow {\wpri}^+\rightarrow N\ell^+_1 \rightarrow \ell_1^+ \ell_2^- jj
\eea

\subsection{$W'$ Chiral Coupling from Angular Distributions}
\label{Same.SEC}

\begin{figure}
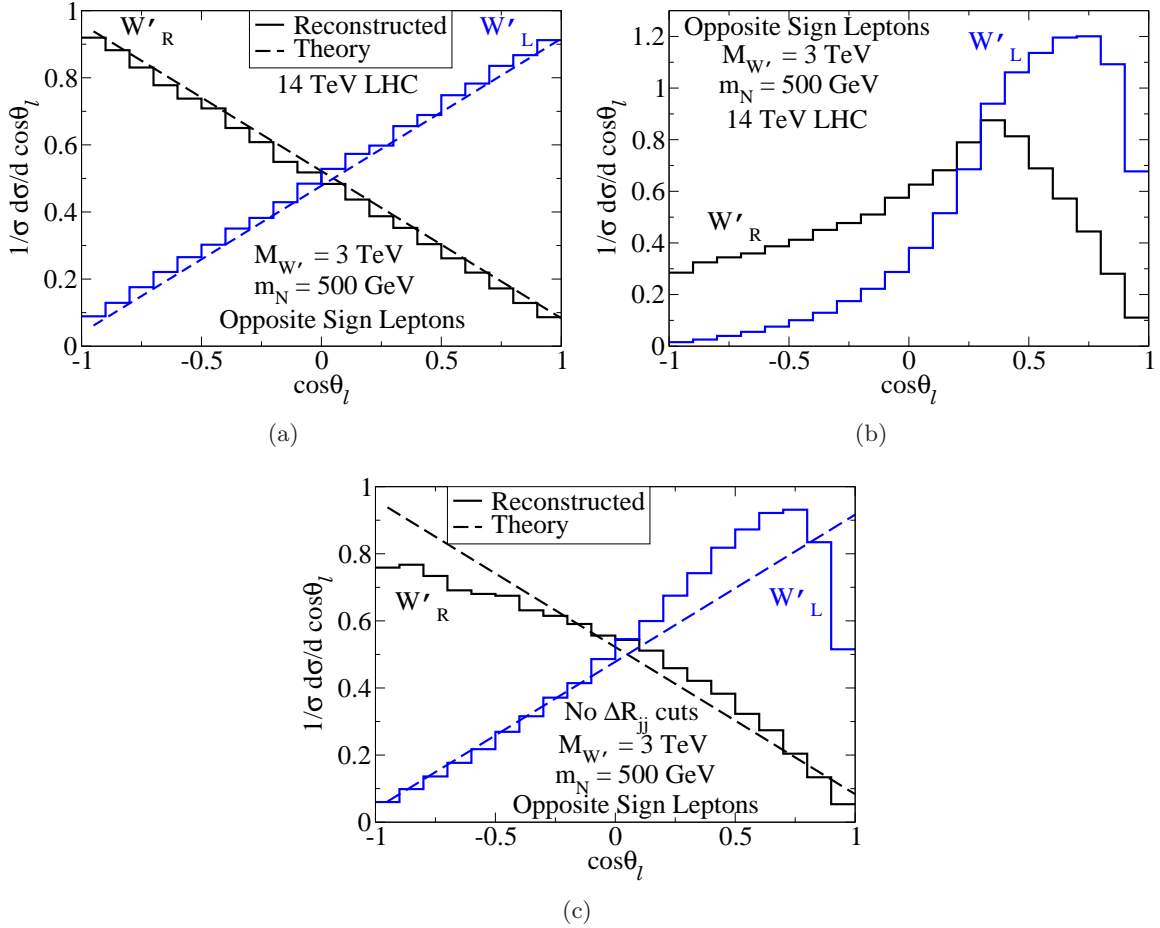

\begin{center}
\subfigure[]{
      \includegraphics[width=0.45\textwidth,clip]{CthlNCNS500_3_opp.eps}
\label{wpriacNS_opp.FIG}}
\subfigure[]{
      \includegraphics[width=0.45\textwidth,clip]{CthlAll500_3_opp.eps}
\label{wpriacSM_opp.FIG}}\\
\subfigure[]{
      \includegraphics[width=0.45\textwidth,clip]{CthlAll500_3NoJIso_opp.eps}
\label{wpriacNoJIso_opp.FIG}}
\caption{
For the opposite sign lepton case, the angular distribution of the charged lepton originating from neutrino decay  
in the heavy neutrino rest-frame with respect to the neutrino moving direction in the partonic c.m.~frame at the LHC with $M_{W'}$, $m_N$ set by Eq.~(\ref{benchParam.EQ}).
Distribution (a) without smearing or cuts, (b) with energy smearing and cuts 
in Eqs.~(\ref{cuts1.EQ}), (\ref{cuts2.EQ}), (\ref{cuts3.EQ}), and (\ref{cuts4.EQ})
, and (c) with all cuts applied to (b) except the $\Delta R_{jj}$ cuts in Eq.~(\ref{cuts2.EQ}). 
The solid lines are for the Monte Carlo simulation results and in (a) and (c) the dashed lines are for the analytical result in Eq.~(\ref{Ang.EQ}).}
\label{wpriac_opp.FIG}
\end{center}
\end{figure}

For the unlike-sign case, we mimic our entire like-sign analysis and 
reconstruct the polar angular distribution of the lepton originating from neutrino decay in the heavy neutrino rest-frame (App.~\ref{DiracAngDist.APP}).
Respectively, the polar and azimuthal distributions are similar to those in 
Eqs.~(\ref{Ang.EQ}) and (\ref{Azim.EQ}) up to a opposite sign in front of the angular dependence.
\begin{equation}
\frac{d\hat{\sigma}}{d\cos\theta_{\ell_{2}}}=
\frac{\hat{\sigma}_{Tot.}}{2}\left[1-\left(\frac{\hat{\sigma}(W_{0})-\hat{\sigma}(W_{T})}{\hat{\sigma}(W_{0})+\hat{\sigma}(W_{T})}\right)\left(\frac{2-\mu_{N}^{2}}{2+\mu_{N}^{2}}\right)
\left(\frac
{g_{R}^{\ell\:2}\vert Y_{\ell_{1}N}\vert^{2}-g_{L}^{\ell\:2}\vert V_{\ell_{1}N}\vert^{2}}
{g_{R}^{\ell\:2}\vert Y_{\ell_{1}N}\vert^{2}+g_{L}^{\ell\:2}\vert V_{\ell_{1}N}\vert^{2}}\right)\cos\theta_{\ell_{2}}\right],
\end{equation}
\begin{equation}
\frac{d\hat{\sigma}}{d\Phi}
=\frac{\hat{\sigma}_{Tot.}}{2\pi}\left[1-\frac{3\pi^{2}}{16}\frac{\mu_{N}}{2+\mu_{N}^{2}}\left(\frac{\hat{\sigma}(W_{0})-\hat{\sigma}(W_{T})}{\hat{\sigma}(W_{0})+\hat{\sigma}(W_{T})}\right)\left(\frac{g_{R}^{q\:2}-g_{L}^{q\:2}}{g_{R}^{q\:2}+g_{L}^{q\:2}}\right)\cos\Phi\right].
\label{AzimUnlikeSig.EQ}
\end{equation}
Figure~\ref{wpriac_opp.FIG} shows the $\Phi$ distributions for the unlike-sign process and follows the identical procedure as for the like-sign case. 
The solid line is the $\Phi$ distribution with the initial-state quark propagation direction identified as the partonic c.m.~frame boost direction; 
the dashed lines are the theoretical distributions given by Eq.~(\ref{AzimUnlikeSig.EQ}); 
and in (a) the dashed-dotted lines are the Monte Carlo truth, i.e., using the known direction of the initial-state quark. 
Figure~\ref{wpriacNS_opp.FIG} does not include cuts or smearing. 
Figure~\ref{wpriacSM_opp.FIG} shows the theoretical prediction and reconstructed distribution with smearing and cuts 
in Eqs.~(\ref{cuts1.EQ}),~(\ref{cuts2.EQ}),~(\ref{cuts3.EQ}), and~(\ref{cuts4.EQ}) applied.
Figure~\ref{wpriacNoJIso_opp.FIG} shows the reconstructed distribution with the same cuts as~\ref{wpriacSM_opp.FIG} minus the $\Delta R_{jj}$ isolation cut.

To understand why the sign of the slope for the $L$-conserving distributions differ from the $L$-violating distributions, we turn to spin correlations.
For $W'^+$, the spin correlations for $u\bar{d}\rightarrow W'^+\rightarrow N\ell^+$ 
are shown in Fig.~\ref{Nspin.FIG} without yet specifying $N$'s decay.
However, we only need to analyze the angular correlation in the neutrino decay.  
 The spin correlations are simply obtained by replacing the right-handed antilepton in Fig.~\ref{Lspin.FIG} with a left-handed lepton.  
 Since the direction of the spin of the lepton is completely determined by the neutrino spin, which is unchanged between the two cases, 
 the effect of the helicity flip is to reverse the direction of the final state lepton momentum relative to the $\hat z$ direction.  
 Therefore, the slopes of the lepton angular distribution are opposite for the like-sign and unlike-sign lepton cases. 
 These same arguments can be made to show that the phases of the 
 $\Phi$ distribution in Eqs.~(\ref{Azim.EQ}) and~(\ref{AzimUnlikeSig.EQ}) differ by $180^\circ$.   
  
The analysis of the two cases also reveals that, unlike the angular distributions, 
the total cross section is independent of having like-sign or unlike-sign leptons in the final state.
This may be understood by recognizing that the difference between the two final states is tantamount to a charge conjugation.
Having integrated out the angular dependence, the total cross section is invariant under parity inversion.
Consequently, by CP-invariance, the total rate is invariant under charge conjugation.
This behavior is evident in Eq.~(\ref{NWidth.EQ}) and Fig.~\ref{NW.FIG}, 
which show that $N$ decays to $\ell^{+}W^{-}$ and $\ell^{-}W^{+}$ equally.

\section{Summary}
\label{Conc.SEC}

The nature of the neutrino mass remains one of most profound puzzles in particle physics. 
The possibility of its being Majorana-like is an extremely interesting aspect since it may have far-reaching consequences in particle physics, nuclear physics and cosmology. 

Given the outstanding performance of the LHC, we are motivated to study the observability for a heavy Majorana neutrino $N$ along with a new charged gauge boson $W'$ at the LHC. 
We first parameterized their couplings in a model-independent approach in Sec.~\ref{ThFrame.SEC} and presented the current constraints on the mass and coupling parameters. 

We studied the production and decay of $W'$ and $N$ at the LHC, and optimized the observability of the like-sign dilepton signal over the SM backgrounds. 
We emphasized the complementarity of these two particles by exploiting the characteristic kinematical distributions resulting from spin-correlations to unambiguously determine their properties. Our phenomenological results can be summarized as follows.

\begin{itemize}
\item[1.] The heavy neutrino is likely to have a large R.H.~component and thus the $W'_R$ would likely yield a larger signal rate 
than that for $W'_L$, governed by the mixing parameters as discussed in Sec.~\ref{ThFrame.SEC}.
Under these assumptions, we found that at the 14 TeV LHC a 5$\sigma$ signal,  via the clean channels
$\ell^\pm \ell^\pm  jj$, 
may be reached for $M_{W'_R}=3$ TeV  ($4$ TeV) with 90 fb$^{-1}$ (1 ab$^{-1}$) integrated luminosity, as seen in Fig.~\ref{significanceVsLumi.FIG}.
\item[2.] 
The chiral coupling of $W'$ to the leptons can be inferred by the polar angle distribution of the leptons in the reconstructed neutrino frame, as seen in Fig.~\ref{wpriac.FIG}, 
owing to the spin correlation from the intermediate state $N$. 
\item[3.]
The chiral coupling of $W'$ to the initial state quarks can be inferred by the azimuthal angular distribution of the neutrino production and decay planes, 
as seen in Fig.~\ref{Phi.FIG}.
\item[4.]  
The kinematical distributions for the like-sign and unlike-sign cases have been found to be quite sensitive to spin correlations and are complementary. 
In particular, the angular distributions differ by a minus sign and provide qualitative differences for a Majorana and a Dirac $N$.
Thus in addition to observing final states that violate lepton-number, comparison of the two scenarios provides a means to differentiate the Majorana nature of $N$. 
\end{itemize}

Overall, if the LHC serves as a discovery machine for a new gauge boson $W'$, then  its properties and much rich physics will await to be explored. 
Perhaps a Majorana nature of a heavy neutrino may be first established associated with $W'$ physics. 

\vspace*{0.3cm}

\noindent {\it Acknowledgements:}
We would like to thank Brian Yencho for useful discussions.
We would also like to thank the Aspen Center for Physics 
and the Kavli Institute for the Physics and Mathematics of the Universe, where part of the work was completed, for their hospitality.
This work was supported in part by the U.S.~Department of Energy 
under grant No.~DE-FG02-95ER40896 and in part by PITT PACC.
IL is supported by the U.S.~Department of Energy under grant No.~DE-AC02-98CH10886.
RR acknowledges support from the University of Wisconsin, the University of Pittsburgh, and the NSF under Grant No. OISE-1210244.
Z.Si is supported in part by NSFC under grant No.11275114 
and in part by NSF of Shandong province under grant JQ200902.

\appendix

\section{Neutrino Mixing Formalism and $W'$ Couplings}
\label{appendNeu.APP}

\subsection*{A1. Neutrino flavor mixing}
We assume that there are three left-handed (L.H.) neutrinos (denoted by $\nu_{aL}, a=1,2,3$) 
with three corresponding light mass eigenstates (denoted by $m$),
and $n$ right-handed (R.H.) neutrinos (denoted by $N_{a'R}, \  a'=1,\dots,\: n$)
with $n$ corresponding heavy mass eigenstates (denoted by $m'$).
The mixing between chiral states and mass eigenstates may then be parameterized~\cite{Atre:2009rg} by
\begin{equation}
\left(\begin{array}{c}
\nu_{L}\\
N_{L}^{c}
\end{array}\right)=\left(\begin{array}{cc}
U_{3\times3} & V_{3\times n}\\
X_{n\times3} & Y_{n\times n}
\end{array}\right)\left(\begin{array}{c}
\nu_{m}\\
N_{m'}^{c}
\end{array}\right), 
\label{appNuMix.EQ}
\end{equation}
where $\psi^{c}=\mathcal{C}\overline{\psi}^{T}$ denotes the charge conjugate of the spinor field $\psi$, 
with $\mathcal{C}$ labeling the charge conjugation operator,
and the chiral states satisfy $\psi_{L}^{c}\equiv(\psi^{c})_{L}=(\psi_{R})^{c}.$ 
Expanding the L.H. and R.H. chiral states, we obtain:
\begin{eqnarray}
\overline{\nu_{aL}}=\sum_{m=1}^{3}\overline{\nu_{m}}U_{ma}^{*}+\sum_{m'=4}^{n+3}\overline{N_{m'}^{c}}V_{m'a}^{*},
&\qquad&
\overline{N_{a'L}^{c}}=\sum_{m=1}^{3}\overline{\nu_{m}}X_{ma'}^{*}+\sum_{m'=4}^{n+3}\overline{N_{m'}^{c}}Y_{m'a'}^{*} \\
\overline{\nu_{aR}^{c}}=\sum_{m=1}^{3}\overline{\nu_{m}^{c}}U_{ma}+\sum_{m'=4}^{n+3}\overline{N_{m'}}V_{m'a},
&\qquad&
\overline{N_{a'R}}=\sum_{m=1}^{3}\overline{\nu_{m}^{c}}X_{ma'}+\sum_{m'=4}^{n+3}\overline{N_{m'}}Y_{m'a'}. 
\end{eqnarray}
Under this formalism,
one expects diagonal mixing of order $1$, 
\begin{equation}
UU^{\dagger}\:\text{and}\: YY^{\dagger}\sim\mathcal{O}(1); 
\end{equation}
and suppressed off-diagonal mixing,
\begin{equation}
VV^{\dagger}\:\text{and}\: XX^{\dagger}\sim\mathcal{O}(m_{m}/m_{m'}). 
\end{equation}


\subsection*{A2. Model-Independent $W'$ Charged Current Couplings}

The goal of this paper is to explore the feasibility of quantifying the properties of a new charged gauge boson, $W'$, at the LHC. 
For this purpose, we relax the $W'$ interactions to include both left-handed and right-handed leptons,
\beq 
L_{aL}
=\left(
\begin{array}{c} \nu_a\\
l_a
\end{array}
\right)_L ,\quad   
R_{b R}
=\left(
\begin{array}{c} N_{b}\\
l_{b}
\end{array}
\right)_R ,
\eeq
with $a,b=1,\:2,\:3$. The L.H.~neutrinos and charged leptons that are members of
SU$(2)_{L}$ doublets in the Standard Model (SM) are denoted by $\nu_{aL}$ and $l_{a}$.
The R.H.~neutrinos, which are SM singlets, and R.H.~charged leptons are denoted by $N_{bR}$ and $l_{b}$.
To grasp the feature of Left-Right symmetric models for a $W'$, we pair $N_{bR}$ and $l_{b}$ into the an SU$(2)_R$ doublet. 
Though there may be more ``sterile'' neutrinos, i.e., $b>3$, we consider only $b=3$ and one new mass eigenstate in our phenomenological presentation.
The mass mixing matrix in Eq.~(\ref{appNuMix.EQ}), in the present case, becomes a $6\times6$ matrix with several repeating entries.

With this assignment, the resulting charged current interactions are 
\begin{eqnarray}
\mathcal{L}=\left( -\frac{g_{L}^{\ell}}{\sqrt{2}}W_{\mu L}^{'+} \
\sum_{a=1}^{3} \overline{\nu_{aL}} \gamma^{\mu}P_{L}l_{a}^{-} -
\frac{g_{R}^{\ell}}{\sqrt{2}}W_{\mu R}^{'+} \ 
\sum_{b=1}^{3} \overline{N_{b R}} \gamma^{\mu}P_{R}l_{b}^{-} \right)
+h.c.
\end{eqnarray}
We have explicitly included the couplings of left- and right-charged currents with new gauge interactions via $W'_{L,R}$. 

The gauge state leptons, $l_{a}$ and $l_{b}$, may be rotated into the mass eigenstates, which are defined to be the flavors eigenstates $\ell=e,\mu,\tau$. 
This amounts to the rotation 
\begin{equation}
 l_{a}^{-}=\sum_{\ell=e}^{\tau}O_{a\ell}\ell^{-}.
\end{equation}
With the SM-like simplest Higgs mechanism, this transformation is trivial and we will make it implicit without loss of generality. 
By simultaneously expanding into the neutrinos' mass basis and into the charged leptons' flavor basis, we obtain 
\begin{eqnarray}
\mathcal{L}=&-&\sum_{\ell=e}^{\tau}\frac{g_{L}^{\ell}}{\sqrt{2}}W_{\mu}^{'+}\left[\sum_{m=1}^{3}\overline{\nu_{m}} U_{m\ell}^{*}+\sum_{m'=4}^{n+3}\overline{N_{m'}^{c}} V_{m'\ell}^{*}\right]\gamma^{\mu}P_{L}\ell^{-} +{\it h.c.}
\nonumber\\
&-&\sum_{\ell=e}^{\tau}\frac{g_{R}^{\ell}}{\sqrt{2}}W_{\mu}^{'+}\left[\sum_{m=1}^{3}\overline{\nu_{m}^{c}} X_{m\ell}+\sum_{m'=4}^{n+3}\overline{N_{m'}} Y_{m'\ell}\right]\gamma^{\mu}P_{R}\ell^{-} +{\it h.c.},
\label{appModIndLag.EQ}
\end{eqnarray}
where
\begin{equation}
U_{m\ell}^{*} \equiv \sum_{a=1}^{3} U_{ma}^{*} O_{a\ell}, \quad V_{m'\ell}^{*} \equiv \sum_{a=1}^{3} V_{m'a}^{*} O_{a\ell}, \quad
X_{m\ell}     \equiv \sum_{b=1}^{3} X_{mb}^{*} O_{b\ell}, \quad Y_{m'\ell}^{*} \equiv \sum_{b=1}^{3} Y_{m'b} O_{b\ell}.
\label{appRotDefs.EQ}
\end{equation}

These are the general couplings for the $W'$ charged currents that we follow in this study.
Leptonic couplings to the SM $W^{\pm}$ boson can be recovered from
Eq.~(\ref{appModIndLag.EQ}) by identifying $W^{'\pm}\to W^{\pm}$ and by setting 
\begin{equation}
g^{\ell}_{L} = g\ {\rm and}\  g^{\ell}_{R} = 0,
\end{equation}
where $g$ is the SU$(2)_{L}$ coupling constant in the SM.
Similarly, we arrive at the SU$(2)_{R}$ charged current coupling by identifying $W' \to W^{\pm}_{R}$ and by setting 
\begin{equation}
g^{\ell}_{L} = 0\ {\rm and}\  g^{\ell}_{R} \neq 0.
\end{equation}

In the quark sector, we do not plan to go through a fully-fledged construction for the charged current couplings.
Instead, we take the simplest approach and just parameterize the model-independent $W'$ Lagrangian by
\begin{equation}
 \mathcal{L}=\frac{-1}{\sqrt{2}}\sum_{i,j=1}^{3}W_{\mu}^{'+}\overline{u_{i}}V_{ij}^{CKM'}\gamma^{\mu}\left[g_{L}^{q}P_{L}+g_{R}^{q}P_{R}\right]d_{j}+{\it h.c.},
\end{equation}
where $V^{CKM'}$ is an unknown flavor mixing matrix.
\section{Derivation of Partonic Level Angular Distributions}
\label{appendME.APP}
\begin{figure}
\begin{center}
\includegraphics[clip,width=0.7\textwidth]{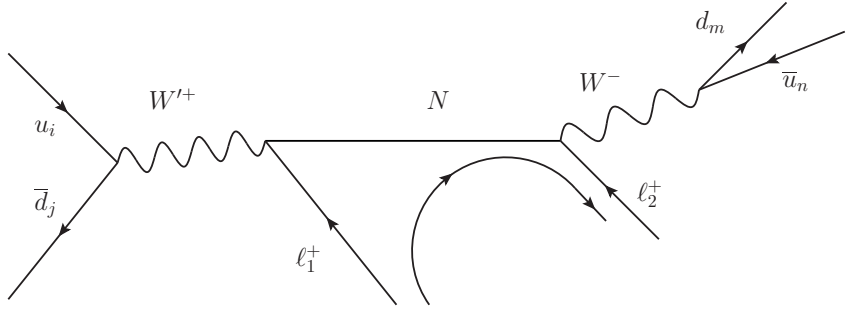}
\caption{The partonic-level process for heavy $W^{'+}$ production and decay into like-sign leptons and quarks in hadronic collisions. 
The longer, black arrow not touching the Feynman diagram denotes fermion flow (FF).}
\label{qqWPllW_Arrow.FIG}
\end{center}
\end{figure}

We strive clarify a few subtleties that arise when calculating observables involving Majorana fermions. 
To do so, we present a detailed derivation of the matrix element for the lepton-number $(L)$ violating process:
\begin{equation}
u_{i}(p_{A})+\overline{d}_{j}(p_{B})\rightarrow W'^{+}\rightarrow\ell_{1}^{+}(p_{1})+\ell_{2}^{+}(p_{2})+q_{m}(p_{3})+\overline{q}_{n}(p_{4}),
\end{equation}
with an intermediate Majorana neutrino of mass $m_{N}$, and 
governed by the Lagrangian given in Section~\ref{ThFrame.SEC}. 
  As discussed in Section \ref{Like.SEC}, and shown in Fig.~\ref{qq2Wp2llqq.FIG}, there are two interfering Feynman diagrams
associated with our $2\ell^{+}2j$ final state.  The interference term may be neglect safely when 
calculating the amplitude squared, $|\mathcal{M}|^2$, since the heavy neutrino's width is very narrow and thus 
the interference is expected to be small. When constructing 
and evaluating $|\mathcal{M}|^2$, we focus on only a single diagram (Fig.~\ref{qqWPllW_Arrow.FIG}) but stress that 
the two diagrams can be treated identically. Additionally, the narrowness of the SM $W$ boson's width allows us to 
further apply the Narrow Width Approximation (NWA). The NWA stipulates that, due to its small width 
compared to its mass, the $W$ boson will dominantly be produced on-shell, and further implies
\begin{equation}
\hat{\sigma}(u_{i}\overline{d}_{j}\rightarrow\ell_{1}^{+}\ell_{2}^{+}q\overline{q}')\approx\hat{\sigma}(u_{i}\overline{d}_{j}\rightarrow\ell_{1}^{+}\ell_{2}^{+}W^{-})\times BR(W\rightarrow q\overline{q}'),
\end{equation}
where $BR(X\rightarrow Y)$ is the branching fraction of $X$ going into $Y$. Since $BR(W\rightarrow q\overline{q}')$ 
is well-known, our work is reduced to determining the analytical expression for 
\begin{equation}
 \hat{\sigma}(u_{i}\overline{d}_{j}\rightarrow\ell_{1}^{+}\ell_{2}^{+}W^{-}).
\end{equation}

\subsection*{B1. Determination of the Spin-Summed, Polarization-Dependent, Squared Matrix Element}
The usefulness of Feynman rules stems from the ability to assign specific multiplicative 
factors to each component of a Feynman diagram. However, Dirac field Feynman
rules are dependent on Wick's Theorem, which is a statement on field contractions.
For Dirac fields, only combinations of the form $\overline{\psi}\psi$ can contract,
where as for Majorana fields, $\psi\psi$ and $\overline{\psi}\overline{\psi}$ are 
allowed to contract. In short, Feynman rules for Dirac fermions do not account for
all possible Majorana interactions. 

We therefore adopt the Feynman rules developed in Ref.~\cite{Denner:1992} for a two-fold reason. 
The first is that the rules for diagram segments not involving Majorana fermions do not change. 
The second is that for parts that do involve Majorana fermions, 
the new Feynman rules reduce to (a) treating the Majorana fermion like a Dirac fermion 
and modifying the vertex factor for an ordinary Dirac fermion 
with an appropriately placed factor of $-1$, and/or (b) making a single 
$u\leftrightarrow v$ spinor substitution. The placement of the additional 
minus sign and possible spinor substitution is based on the direction of fermion flow (FF)
relative to the traditionally chosen fermion {\em{number}} flow (FNF). 
When the fermion flow and fermion number flow are equal, the newer rules simplify to the usual rules.
Computationally, these rules provide a desirable technique that can be automated in a straight forward manner. 

In the present case, we identify the relevant FF as being identical to the lepton number-changing current. 
The FF current starts at $\ell_{1}$, the charged lepton produced in the $W'$ boson decay, 
and points anti-parallel to $\ell_{1}$'s momentum; the current then continues parallel to the Majorana
neutrino's momentum; and finally terminates at $\ell_{2}$,  the charged lepton produced in the $N$ decay,
and points parallel to $\ell_{2}$'s momentum.
See the curved black arrow in Fig.~\ref{qqWPllW_Arrow.FIG}. With this orientation, the FF is 
parallel to the FNF at the $W'\ell_{1}N$ vertex, and anti-parallel to it at the $N\ell_{2}W$ vertex. 
This change in relative current orientation causes two modifications, the first of which is to the 
spinor of the outgoing lepton originating from the $N\ell_{2}W$ vertex:
\begin{equation}
 \overline{v}_{\ell_2}(p_2)\rightarrow\overline{u}_{\ell_2}(p_2),
\end{equation}
and accounts explicitly for the change in lepton number. The second modification is to the 
$N\ell_{2}W$ vertex itself and occurs in the following way:
\begin{equation}
C^{\rho}_{N\ell_{2}}=\frac{-ig}{\sqrt{2}}V_{\ell_{2}N}\gamma^{\rho}P_{L}\rightarrow C'^{\rho}_{N\ell_{2}}=(-1)^2\frac{ig}{\sqrt{2}}V_{\ell_{2}N}\gamma^{\rho}P_{R},
\label{nChiralFlip.EQ}
\end{equation}
where $g$ is the SM SU$(2)_{L}$ coupling constant, 
$P_{R,L}\equiv\frac{1}{2}(1\pm\gamma_5)$, 
and, as defined in Ref.~\cite{Denner:1992}, the primed-vertex convention indicates
\begin{equation}
\Gamma'\equiv\mathcal{C}\Gamma^{T}\mathcal{C}^{-1}=\eta\Gamma,
\end{equation}
where $\mathcal{C}$ is the charge conjugation operator and for which
\begin{equation}
\eta=\begin{cases}
1, & \Gamma\in\{1,i\gamma^{5},\gamma^{\mu}\gamma^{5}\}\\
-1, & \Gamma\in\{\gamma^{\mu},\sigma^{\mu\nu}\}
\end{cases}.
\label{majEta.EQ}
\end{equation}

As a result, we find that the matrix element describing the $u_{i}\overline{d}_{j}\rightarrow\ell_{1}^{+}\ell_{2}^{+}W_{\lambda}^{-}$
scattering process, for an outgoing SM $W^{-}$ boson with polarization
$\lambda$, and in the Feynman Gauge, is
\begin{eqnarray}
i\mathcal{M}_{\lambda}
&=& \varepsilon_{\lambda\rho}^{*}(p_{W})\cdot\frac{\left[\overline{v}_{Bj}A_{ji}^{\mu}u_{Ai}\right]\cdot\left[\overline{u}_{2}C'^{\rho}_{N\ell_{2}}(\not\! p_{N}+m_{N})B_{\mu\:\ell_{1}N}v_{1}\right]}{\left(\hat{s}-M_{W'}^{2}+i\Gamma_{W'}M_{W'}\right)\left(p_{N}^{2}-m_{N}^{2}+i\Gamma_{N}m_{N}\right)},
\label{udWllME.EQ}
\end{eqnarray}
where the vertex terms are given by
\begin{eqnarray}
A_{ji}^{\mu}&=&\frac{1}{\sqrt{2}}V^{CKM'}_{ji}\gamma^{\mu}~\left[g_{R}^{q}P_{R}+g_{L}^{q}P_{L}\right]\\
B_{\ell_{1} N}^{\nu}&=&\frac{1}{\sqrt{2}}\gamma^{\nu}\left[g_{R}^{\ell}P_{R}Y_{\ell_{1}N}+g_{L}^{\ell}P_{L}V_{\ell_{1}N}^{*}\right]
\end{eqnarray}
To be explicit: $\varepsilon_{\lambda\rho}^{*}(p_W)$ denotes the 
outgoing polarization vector of the on-shell $W$ boson with momentum $p_W$, mass $M_W$, and polarization 
$\lambda$; $\overline{v}_{Bj}$ represents the the spinor $\overline{v}$ of an initial-state 
antiquark of flavor $j$ and momentum $p_B$; similarly, $u_{Ai}$ represents the spinor $u$ of 
an initial-state quark of flavor $i$ and momentum $p_A$; $\overline{u}_{2}$ denotes the spinor 
of our final-state $\it{antilepton}$ with flavor $\ell_{2}$ and momentum $p_2$; and
likewise, $v_{1}$ denotes the spinor of our final-state antilepton with flavor $\ell_{1}$ 
and momentum $p_1$. The $W'$ mass, width, and momentum-squared are respectively given by $M_{W'}$, 
$\Gamma_{W'}$, and the Mandelstam variable 
\begin{equation}
\hat{s}=(p_A+p_B)^2=(p_1+p_2+p_W)^2. 
\end{equation}
The heavy 
neutrino's mass, width, and momentum are similarly given by $m_{N}$, $\Gamma_N$, and 
\begin{equation}
p_N=p_A+p_B-p_1=p_W+p_2.
\end{equation}

After squaring and summing over external spins, diagrams, 
 and colors ($N_{C}$), but not external boson 
polarizations ($\lambda$), the polarization-dependant squared amplitude is
\begin{eqnarray}
\sum\vert\mathcal{M}_{\lambda}\vert^{2} =
\frac{4N^2_{C}~g^{2}~\vert V_{ji}^{CKM'}\vert^{2}~\vert V_{\ell_{2}N}\vert^{2}~\text{Tr}\left[\not\! p_{A}\gamma^{\sigma}\not\!p_{B}\gamma^{\mu}\left(g_{R}^{q~2}P_{R}+g_{L}^{q~2}P_{L}\right)\right]}{2^{3}(1+\delta_{\ell_{1}\ell_{2}})\left[\left(\hat{s}-M_{W'}^{2}\right)^{2}+\left(\Gamma_{W'}M_{W'}\right)^{2}\right]\left[\left(p_{N}^{2}-m_{N}^{2}\right)^{2}+\left(\Gamma_{N}m_{N}\right)^{2}\right]}\nonumber\\
\times \text{Tr}\left[\not\!p_{1}\gamma_{\sigma}\left(\not\! p_{N}+m_{N}\right)\not\!\varepsilon_{\lambda}\not\! p_{2}\not\!\varepsilon_{\lambda}^{*}P_{R}\left(\not\! p_{N}+m_{N}\right)\gamma_{\mu}\left(g_{R}^{\ell~2}P_{R}\vert Y_{\ell_{1}N}\vert^{2}+g_{L}^{\ell~2}P_{L}\vert V_{\ell_{1}N}\vert^{2}\right)\right]\\
=\frac{2^{3}N^2_{C}~g^{2}~\vert V_{ji}^{CKM'}\vert^{2}~\vert V_{\ell_{2}N}\vert^{2}}{(1+\delta_{\ell_{1}\ell_{2}})\left[\left(\hat{s}-M_{W'}^{2}\right)^{2}+\left(\Gamma_{W'}M_{W'}\right)^{2}\right]\left[\left(p_{N}^{2}-m_{N}^{2}\right)^{2}+\left(\Gamma_{N}m_{N}\right)^{2}\right]}\nonumber\\
\times \left[
 \vert Y_{\ell_{1}N}\vert^{2}\left(g_{R}^{q}g_{R}^{\ell}\right)^{2}\mathcal{A}_{\lambda}
+\vert Y_{\ell_{1}N}\vert^{2}\left(g_{L}^{q}g_{R}^{\ell}\right)^{2}\mathcal{B}_{\lambda}
+\vert V_{\ell_{1}N}\vert^{2}\left(g_{L}^{q}g_{L}^{\ell}\right)^{2}\mathcal{C}_{\lambda}
+\vert V_{\ell_{1}N}\vert^{2}\left(g_{R}^{q}g_{L}^{\ell}\right)^{2}\mathcal{D}_{\lambda}\right],
\end{eqnarray}

where
\begin{eqnarray}
\mathcal{A}_{\lambda}&=&2(p_{A}\cdot p_{1})(p_{B}\cdot p_{N})\left[(p_{N}\cdot p_{2})+2(p_{N}\cdot\varepsilon_{\lambda})(\varepsilon_{\lambda}\cdot p_{2})\right]\nonumber\\
&-& m_{N}^{2}(p_{A}\cdot p_{1})\left[(p_{B}\cdot p_{2})+2(p_{B}\cdot\varepsilon_{\lambda})(\varepsilon_{\lambda}\cdot p_{2})\right],\\
\mathcal{B}_{\lambda}&=&2(p_{B}\cdot p_{1})(p_{A}\cdot p_{N})\left[(p_{N}\cdot p_{2})+2(p_{N}\cdot\varepsilon_{\lambda})(\varepsilon_{\lambda}\cdot p_{2})\right]\nonumber\\
&-& m_{N}^{2}(p_{B}\cdot p_{1})\left[(p_{A}\cdot p_{2})+2(p_{A}\cdot\varepsilon_{\lambda})(\varepsilon_{\lambda}\cdot p_{2})\right],\\
\mathcal{C}_{\lambda}&=&m_{N}^{2}(p_{A}\cdot p_{1})\left[(p_{B}\cdot p_{2})+2(p_{B}\cdot\varepsilon_{\lambda})(\varepsilon_{\lambda}\cdot p_{2})\right],\\
\mathcal{D}_{\lambda}&=&m_{N}^{2}(p_{B}\cdot p_{1})\left[(p_{A}\cdot p_{2})+2(p_{A}\cdot\varepsilon_{\lambda})(\varepsilon_{\lambda}\cdot p_{2})\right],
\end{eqnarray}
and $\varepsilon_{\lambda}$ is taken to be real. 

The Majorana neutrino's width, $\Gamma_{N}$, is expected to be very small. 
Therefore, to simplify analytic integration, we again apply the Narrow Width Approximation such that
\begin{equation}
\frac{1}{(p_{N}^{2}-m_{N}^{2})^{2}+(\Gamma_{N}m_{N})^{2}}
\approx \frac{\pi}{\Gamma_{N}m_{N}}~\delta\left(p_{N}^{2}-m_{N}^{2}\right).
\end{equation}
We are motivated to make this additional approximation to highlight and emphasize the 
analyzing power of the angular distributions. Our reported numerical 
results do not reflect this extra stipulation (see Eq.~(\ref{diffXSecNoNWA.EQ})).
Consequentially, the squared and summed amplitude becomes
\begin{equation}
\sum\vert\mathcal{M}_{\lambda}\vert^{2}\approx
\frac{2^{3}\pi N_{C}\: g^{2}~\vert V_{ji}^{CKM'}\vert^{2}~\vert V_{\ell_{2}N}\vert^{2}\:\delta\left(p_{N}^{2}-m_{N}^{2}\right)}{(1+\delta_{\ell_{1}\ell_{2}})\left(\Gamma_{N}m_{N}\right)\left[\left(\hat{s}-M_{W'}^{2}\right)^{2}+\left(\Gamma_{W'}M_{W'}\right)^{2}\right]}
\label{sqME.EQ}
\end{equation}
\begin{equation}
\times \left[
 \vert Y_{\ell_{1}N}\vert^{2}\left(g_{R}^{q}g_{R}^{\ell}\right)^{2}\mathcal{A}_{\lambda}
+\vert Y_{\ell_{1}N}\vert^{2}\left(g_{L}^{q}g_{R}^{\ell}\right)^{2}\mathcal{B}_{\lambda}
+\vert V_{\ell_{1}N}\vert^{2}\left(g_{L}^{q}g_{L}^{\ell}\right)^{2}\mathcal{C}_{\lambda}
+\vert V_{\ell_{1}N}\vert^{2}\left(g_{R}^{q}g_{L}^{\ell}\right)^{2}\mathcal{D}_{\lambda}\right]\nonumber.
\end{equation}

\subsection*{B2. Phase Space Volume Element}
\label{PhaseSpace.APP}

We calculate the partonic-level cross section using the usual formula,
\begin{equation}
d\hat{\sigma}=\frac{1}{2\hat{s}}\frac{1}{4N_{C}^{2}}\sum\vert\mathcal{M}\vert^{2}\cdot dPS_{n}.
\label{sigmaHat.EQ}
\end{equation}
Here, the factor of $4N_{C}^{2}$ comes from averaging over initial-state
colors and spins. The factor $dPS_{n}$ represents the
\emph{n}-body phase space volume element, 
\begin{equation}
 dPS_{n}(P;p_{1}\dots p_{n}) = 
\prod_{k=1}^{n}\frac{d^{3}p_k}{(2\pi)^{3}2E_k} ~(2\pi)^{4}\delta^{4}\left(P-p_1-\dots-p_n\right),
\end{equation}
which can be decomposed using the recursion formula
\begin{equation}
dPS_{n}(P;p_{1},\dots,p_{n})=
dPS_{n-1}(P;p_{1},\dots,p_{n-1,n})\times 
dPS_{2}(p_{n-1,n};p_{n-1},p_{n})\times
\frac{d\: p_{n-1,n}^{2}}{2\pi},
\end{equation}
where $P=\sum_{m=1}^{n}p_{m}$ and $p_{i,j}=p_{i}+p_{j}$. In the present case, 
$dPS_{3}$ is expressible as
\begin{equation}
dPS_{3}(p_{A}+p_{B};p_{1},p_{2},p_{W})=
dPS_{2}(p_{A}+p_{B};p_{1},p_{N})\times 
dPS_{2}(p_{N};p_{2},p_{W})\times
\frac{d\: p_{N}^{2}}{2\pi}.
\label{phaseSpace1.EQ}
\end{equation}

Since each $dPS_{k}$ is individually Lorentz invariant,
the two phase space elements in Eq.~(\ref{phaseSpace1.EQ}) can be evaluated in different
reference frames. When $dPS_{2}(p_{1},p_{N})$ is evaluated
in the partonic c.m.~frame and $dPS_{2}(p_{2},p_{W})$
in the neutrino rest-frame, the full volume element is found to be
\begin{equation}
dPS_{3}(p_{A}+p_{B};p_{1},p_{2},p_{W})=d\Omega_{N}
\frac{(1-\tilde{\mu}_{N}^{2})}{2(4\pi)^{2}}\times d\Omega_{\ell_{2}}\frac{(1-\rho_{W}^{2})}{2(4\pi)^{2}}
\times\frac{d\: p_{N}^{2}}{2\pi},
\label{phaseSpace2.EQ}
\end{equation}
with 
\begin{equation}
 \mu_{N}^2=\frac{m_N^2}{\hat{s}},\quad\quad
 \tilde{\mu}_{N}^{2} = \frac{p_N^2}{\hat{s}},\quad\quad
 \rho_{W}^2=\frac{M_{W}}{p_{N}^{2}},
\end{equation}
and, in the on-shell limit,
\begin{equation}
 \mu_{N},\tilde{\mu}_{N}\rightarrow x_N=\frac{m_N}{M_{W'}},\quad\quad
  \rho_{W}\rightarrow y_{W}=\frac{M_{W}}{m_{N}}.
\end{equation}
The solid angle element $d\Omega_{N}$ is defined as the angle made by 
$N$ with respect to the direction of propagation of the initial-state 
quark in the c.m.~frame; $d\Omega_{\ell_{2}}$ is defined as the angle 
made by $\ell_{2}^{+}$ with respect to the heavy neutrino spin 
axis in the neutrino's rest-frame. 

\subsection*{B3. Partonic-Level Angular Distributions}
\label{AngDist.APP}

The angular distribution of the charged lepton from the neutrino decay
is most efficiently determined by evaluating $\sum\vert\mathcal{M}\vert^{2}$ 
in the neutrino rest-frame. 
Like individual $dPS_{k}$ volume
elements, $\vert\mathcal{M}\vert^{2}$ is separately Lorentz invariant
and thus can be evaluated in its own reference frame. 

In order to evaluate Eq.~(\ref{sqME.EQ}) in the neutrino rest-frame, we 
must first rotate and boost the four-momenta of the initial-state 
quarks from the c.m.~frame. Without the loss of generality, we assume 
that the initial-state (anti)quark is originally traveling in the 
positive (negative) $\hat{z}-$axis and that the $\ell_{1}^{+}N$ pair propagate in 
$\hat{y}-\hat{z}$ plane. This allows us to rotate the 
entire $2\rightarrow2$ system such that the neutrino's momentum 
is aligned with the $\hat{z}-$axis, and then boost into the neutrino rest-frame. 
Since we are applying the NWA and immediately integrating over $dp_{N}^{2}$, we will take $N$ to be on-shell.
After boosting, our four-momenta are:
\begin{eqnarray}
p_{A}&=&\frac{\hat{s}}{4m_{N}}\left((1-\cos\theta_{N})+\mu_{N}^{2}(1+\cos\theta_{N}),\:0,-2\mu_{N}\sin\theta_{N},\: \mu_{N}^{2}(1+\cos\theta_{N})-(1-\cos\theta_{N})\right),\nonumber\\
p_{B}&=&\frac{\hat{s}}{4m_{N}}\left((1+\cos\theta_{N})+\mu_{N}^{2}(1-\cos\theta_{N}),\:0,~~2\mu_{N}\sin\theta_{N}	,\: \mu_{N}^{2}(1-\cos\theta_{N})-(1+\cos\theta_{N})\right),\nonumber
\end{eqnarray}
\begin{equation}
 p_{N}=(m_{N},0,0,0),\quad\text{and}\quad p_{1}=\frac{\hat{s}}{2m_{N}}(1-\mu_{N}^{2})(1,0,0,-1),
\end{equation}
where $\theta_{N}$ represents the polar angle between $\vec{p}_{N}$
and $\vec{p}_{A}$ in the c.m.~frame. In the neutrino rest-frame, 
the $N\rightarrow\ell_{2}^{+}W^{-}$ decay products' 
momenta are 
\begin{eqnarray}
p_{2}&=&\vert\vec{p}_{2}\vert\left(1,\sin\theta_{\ell_{2}}\cos\phi_{\ell_{2}},\sin\theta_{\ell_{2}}\sin\phi_{\ell_{2}},\cos\theta_{\ell_{2}}\right),\quad\vert\vec{p}_{2}\vert=\vert\vec{p}_{W}\vert=\frac{m_N}{2}(1-y_{W}^{2}),\nonumber\\
p_{W}&=&\vert\vec{p}_{2}\vert\left(\frac{E_{W}}{\vert\vec{p}_{2}\vert},-\sin\theta_{\ell_{2}}\cos\phi_{\ell_{2}},-\sin\theta_{\ell_{2}}\sin\phi_{\ell_{2}},-\cos\theta_{\ell_{2}}\right),\; E_{W}=\frac{m_N}{2}(1+y_{W}^{2}),
\end{eqnarray}
where $\theta_{\ell_{2}}$ and $\phi_{\ell_{2}}$ are defined with respect to the neutrino spin axis in the c.m.~frame.
Explicitly, $\hat{z}=\hat{p}_N$, where $\hat{p}_N =\vec{{p}}_{N} / \vert \vec{p}_{N}\vert$ is measured in the c.m.~frame, and $\phi_{\ell_2}$ w.r.t.~to the $+\hat{y}$ axis.
This is consistent with Eq.~(\ref{phaseSpace2.EQ}). 
The polarization vectors for the SM $W$ boson are subsequently: 
\begin{eqnarray}
\varepsilon_{0}^{\mu}(p_{W})&=&\frac{E_{W}}{m_{W}}\left(\frac{\vert\vec{p}_{2}\vert}{E_{W}},-\sin\theta_{\ell_{2}}\cos\phi_{\ell_{2}},-\sin\theta_{\ell_{2}}\sin\phi_{\ell_{2}},-\cos\theta_{\ell_{2}}\right),\nonumber\\ 
\varepsilon_{T1}^{\mu}(p_{W})&=&\left(0,-\cos\theta_{\ell_{2}}\cos\phi_{\ell_{2}},-\cos\theta_{\ell_{2}}\sin\phi_{\ell_{2}},\sin\theta_{\ell_{2}}\right),\nonumber\\
\varepsilon_{T2}^{\mu}(p_{W})&=&\left(0,\sin\phi_{\ell_{2}},-\cos\phi_{\ell_{2}},0\right).
\end{eqnarray}
Here the labels $0$, $T1,$ and $T2$ denote the longitudinal
and transverse polarizations of the outgoing vector boson. After combining Eqs.~
(\ref{sqME.EQ}), (\ref{sigmaHat.EQ}), (\ref{phaseSpace2.EQ}), and integrating 
over $dp_{N}^{2}$, as well as $d\Omega_{N},$ for the $L$-violating 
process $u_{i}\overline{d}_{j} \rightarrow \ell_{1}^{+}N \rightarrow \ell_{1}^{+}\ell_{2}^{+}W^{-}$ 
with a longitudinally polarized $W^{-}$ boson the angular distribution is
\begin{eqnarray}
\frac{d\hat{\sigma}_{0}}{d\Omega_{\ell_{2}}}&=&\frac{\hat{\sigma}(W_{0})}{2^{4}\pi}\times\{4\left[1+\left(\frac{2-\mu_{N}^{2}}{2+\mu_{N}^{2}}\right)\left(\frac{g_{R}^{\ell\:2}\vert Y_{\ell_{1}N}\vert^{2}-g_{L}^{\ell\:2}\vert V_{\ell_{1}N}\vert^{2}}{g_{R}^{\ell\:2}\vert Y_{\ell_{1}N}\vert^{2}+g_{L}^{\ell\:2}\vert V_{\ell_{1}N}\vert^{2}}\right)\cos\theta_{\ell_{2}}\right]\nonumber\\
&-&\frac{3\pi~\mu_{N}}{\left(2+\mu_{N}^{2}\right)}\left(\frac{g_{R}^{q\:2}-g_{L}^{q\:2}}{g_{R}^{q\:2}+g_{L}^{q\:2}}\right)\sin\theta_{\ell_{2}}\cos\phi_{\ell_{2}}\}.
\end{eqnarray}
Accordingly, for transversely polarized $W$ bosons the angular distributions are
\begin{eqnarray}
\frac{d\hat{\sigma}_{T1}}{d\Omega_{\ell_{2}}}=\frac{d\hat{\sigma}_{T2}}{d\Omega_{\ell_{2}}}&=&\frac{\hat{\sigma}(W_{T})}{2^{5}\pi}\times
\{
4\left[1-\left(\frac{2-\mu_{N}^{2}}{2+\mu_{N}^{2}}\right)\left(\frac
{g_{R}^{\ell\:2}\vert Y_{\ell_{1}N}\vert^{2}-g_{L}^{\ell\:2}\vert V_{\ell_{1}N}\vert^{2}}
{g_{R}^{\ell\:2}\vert Y_{\ell_{1}N}\vert^{2}+g_{L}^{\ell\:2}\vert V_{\ell_{1}N}\vert^{2}}\right)\cos\theta_{\ell_{2}}\right]\nonumber\\
&+&\frac{3\pi~\mu_{N}}{\left(2+\mu_{N}^{2}\right)}
\left(\frac{g_{R}^{q\:2}-g_{L}^{q\:2}}{g_{R}^{q\:2}+g_{L}^{q\:2}}\right)
\sin\theta_{\ell_{2}}\cos\phi_{\ell_{2}}\}.
\end{eqnarray}
In the preceding lines, we have used the following quantities
\begin{eqnarray}
\hat{\sigma}(W_{0})&\equiv&\hat{\sigma}(u\bar{d}\rightarrow \ell^+_1 N\rightarrow \ell^+_1\ell^+_2 W^{-}_{0})\nonumber\\
&=&
\frac{g^{2}~ \vert V^{CKM'}_{ji}\vert^{2}~ \vert V_{\ell_{2}N} \vert^{2}}{3N_{C}~2^{10}~\pi^{2}~(1+\delta_{\ell_{1}\ell_{2}})}
\left(g^{q~2}_{R} + g^{q~2}_{L}\right)
\left(g^{\ell\:2}_{R}\vert Y_{\ell_{1}N}\vert^{2} + g^{\ell\:2}_{L}\vert V_{\ell_{1}N}\vert^{2} \right)\nonumber\\
&\times&
\frac{m_{N}}{\Gamma_{N}}
\frac{\hat{s}}{\left[(\hat{s}-M_{W'}^{2})^{2}+(\Gamma_{W'}M_{W'})^{2}\right]}
(1-\mu_{N}^{2})^{2}(1-y_{W}^{2})^{2}(2+\mu_{N}^{2})
\left(\frac{1}{2y_{W}^{2}}\right)\\
\hat{\sigma}(W_{T})&\equiv&\hat{\sigma}(u\bar{d}\rightarrow \ell^+_1 N\rightarrow \ell^+_1\ell^+_2 W^{-}_{T})\nonumber\\
&=& \hat{\sigma}(W_{0})\times 2y_{W}^{2} 
\label{sigTTil.EQ}
\end{eqnarray}
Integrating over the azimuthal angle, the polar distributions are calculated to be
\begin{equation}
 \frac{d\hat{\sigma}_{0}}{d\cos\theta_{\ell_{2}}}=\frac{\hat{\sigma}(W_{0})}{2}\left[1+\left(\frac{2-\mu_{N}^{2}}{2+\mu_{N}^{2}}\right)\left(\frac
{g_{R}^{\ell\:2}\vert Y_{\ell_{1}N}\vert^{2}-g_{L}^{\ell\:2}\vert V_{\ell_{1}N}\vert^{2}}
{g_{R}^{\ell\:2}\vert Y_{\ell_{1}N}\vert^{2}+g_{L}^{\ell\:2}\vert V_{\ell_{1}N}\vert^{2}}\right)\cos\theta_{\ell_{2}}\right]
\end{equation}
and
\begin{equation}
 \frac{d\hat{\sigma}_{T}}{d\cos\theta_{\ell_{2}}}\equiv\frac{d(\hat{\sigma}_{T1}+\hat{\sigma}_{T1})}{d\cos\theta_{\ell_{2}}}=\frac{\hat{\sigma}(W_{T})}{2}\left[1-\left(\frac{2-\mu_{N}^{2}}{2+\mu_{N}^{2}}\right)\left(\frac
{g_{R}^{\ell\:2}\vert Y_{\ell_{1}N}\vert^{2}-g_{L}^{\ell\:2}\vert V_{\ell_{1}N}\vert^{2}}
{g_{R}^{\ell\:2}\vert Y_{\ell_{1}N}\vert^{2}+g_{L}^{\ell\:2}\vert V_{\ell_{1}N}\vert^{2}}\right)\cos\theta_{\ell_{2}}\right]
\end{equation}
After combining the two, we find that the polarization-summed polar distribution for the full 
$u_{i}\overline{d}_{j}\rightarrow \ell_{1}^{+}\ell_{2}^{+}q\overline{q}'$ 
process is
\begin{eqnarray}
\frac{d\hat{\sigma}_{Tot.}}{d\cos\theta_{\ell_{2}}}
&\equiv&\frac{d(\hat{\sigma}_{0}+\hat{\sigma}_{T})}{d\cos\theta_{\ell_{2}}}\nonumber\\
&=&\frac{\hat{\sigma}_{Tot.}}{2}\left[1+\frac{\hat{\sigma}(W_{0})-\hat{\sigma}(W_{T})}{\hat{\sigma}(W_{0})+\hat{\sigma}(W_{T})}\left(\frac{2-\mu_{N}^{2}}{2+\mu_{N}^{2}}\right)
\left(\frac
{g_{R}^{\ell\:2}\vert Y_{\ell_{1}N}\vert^{2}-g_{L}^{\ell\:2}\vert V_{\ell_{1}N}\vert^{2}}
{g_{R}^{\ell\:2}\vert Y_{\ell_{1}N}\vert^{2}+g_{L}^{\ell\:2}\vert V_{\ell_{1}N}\vert^{2}}\right)\cos\theta_{\ell_{2}}\right],
\label{angDistAppend.EQ}
\end{eqnarray}
where
\begin{equation}
\frac{\hat{\sigma}(W_{0})-\hat{\sigma}(W_{T})}{\hat{\sigma}(W_{0})+\hat{\sigma}(W_{T})}=
\frac{\hat{\sigma}(W_{0})-2y_{W}^{2}\hat{\sigma}(W_{0})}{\hat{\sigma}(W_{0})+2y_{W}^{2}\hat{\sigma}(W_{0})}
=\frac{1-2y_{W}^{2}}{1+2y_{W}^{2}},
\end{equation}
and the total partonic-level cross section is
\begin{eqnarray}
\hat{\sigma}_{Tot.}&\equiv&\hat{\sigma}(u_{i}\overline{d}_{j}\rightarrow \ell_{1}^{+}\ell_{2}^{+}q\overline{q}') \\
&=&(\hat{\sigma}(W_{0})+\hat{\sigma}(W_{T}))\times BR(W\rightarrow q\overline{q}')\\
&=&\hat{\sigma}(W_{0})(1+2y_{W}^{2})\times BR(W\rightarrow q\overline{q}')\\
&=&\frac{g^{2}~ \vert V^{CKM'}_{ji}\vert^{2}~ \vert V_{\ell_{2}N} \vert^{2}}{3N_{C}~2^{10}~\pi^{2}~(1+\delta_{\ell_{1}\ell_{2}})}
\left(g^{q~2}_{R} + g^{q~2}_{L}\right)
\left(g^{\ell\:2}_{R}\vert Y_{\ell_{1}N}\vert^{2} + g^{\ell~2}_{L}\vert V_{\ell_{1}N}\vert^{2}\right)
\left(\frac{m_{N}}{\Gamma_{N}}\right)\nonumber\\
&\times&
\frac{\hat{s}~(1-y_{W}^{2})^{2}(1-y_{W}^{2})^{2}(2+\mu_{N}^{2})}{\left[(\hat{s}-M_{W'}^{2})^{2}+(\Gamma_{W'}M_{W'})^{2}\right]}
\left(\frac{1+2y_{W}^{2}}{2y_{W}^{2}}\right)\times BR(W\rightarrow q\overline{q}').
\label{sig0Til.EQ}
\end{eqnarray}

Having instead chosen to integrate first over the polar angle before the azimuthal angle, 
the polarization-dependent azimuthal distributions for the 
$u_{i}\overline{d}_{j} \rightarrow \ell_{1}^{+}N \rightarrow \ell_{1}^{+}\ell_{2}^{+}W^{-}$ 
process are
\begin{equation}
\frac{d\hat{\sigma}_{0}}{d\phi_{\ell_{2}}}=\frac{\hat{\sigma}(W_{0})}{2\pi}\left[1-\frac{3\pi^{2}}{16}\frac{\mu_{N}}{\left(2+\mu_{N}^{2}\right)}\left(\frac{g_{R}^{q\:2}-g_{L}^{q\:2}}{g_{R}^{q\:2}+g_{L}^{q\:2}}\right)\cos\phi_{\ell_{2}}\right],
\end{equation}
and
\begin{equation}
\frac{d\hat{\sigma}_{T}}{d\phi_{\ell_{2}}} 
\equiv\frac{(d\hat{\sigma}_{T_{1}}+\hat{\sigma}_{T_{2}})}{d\phi_{\ell_{2}}}
=\frac{\hat{\sigma}(W_{T})}{2\pi}\left[1+\frac{3\pi^{2}}{16}\frac{\mu_{N}}{\left(2+\mu_{N}^{2}\right)}\left(\frac{g_{R}^{q\:2}-g_{L}^{q\:2}}{g_{R}^{q\:2}+g_{L}^{q\:2}}\right)\cos\phi_{\ell_{2}}\right].
\end{equation}
Similarly, after combining the azimuthal distributions, the total 
polarization-summed azimuthal distribution for the full 
$u_{i}\overline{d}_{j}\rightarrow \ell_{1}^{+}\ell_{2}^{+}q\overline{q}'$ 
process is
\begin{equation}
\frac{d\hat{\sigma}_{Tot.}}{d\phi_{\ell_{2}}}=\frac{\hat{\sigma}_{Tot.}}{2\pi}\left[1-\frac{3\pi^{2}}{16}\frac{\mu_{N}}{\left(2+\mu_{N}^{2}\right)}\left(\frac{\hat{\sigma}(W_{0})-\hat{\sigma}(W_{T})}{\hat{\sigma}(W_{0})+\hat{\sigma}(W_{T})}\right)\left(\frac{g_{R}^{q\:2}-g_{L}^{q\:2}}{g_{R}^{q\:2}+g_{L}^{q\:2}}\right)\cos\phi_{\ell_{2}}\right].
\end{equation}
Under the definition of the azimuthal angle, $\Phi,$ in Eq.~(\ref{AziDef.EQ}),
we have $\Phi=-\phi_{\ell_{2}},$ and consequentially recover Eq.~(\ref{Azim.EQ}):
\begin{equation}
\frac{d\hat{\sigma}_{Tot.}}{d\Phi}=\frac{\hat{\sigma}_{Tot.}}{2\pi}\left[1-\frac{3\pi^{2}}{16}\frac{\mu_{N}}{\left(2+\mu_{N}^{2}\right)}\left(\frac{\hat{\sigma}(W_{0})-\hat{\sigma}(W_{T})}{\hat{\sigma}(W_{0})+\hat{\sigma}(W_{T})}\right)\left(\frac{g_{R}^{q\:2}-g_{L}^{q\:2}}{g_{R}^{q\:2}+g_{L}^{q\:2}}\right)\cos\Phi\right].
\label{phiDistAppend.EQ}
\end{equation}

Lastly, were the NWA never applied to $N$, the differential cross section 
for the $u_{i}\overline{d}_{j}\rightarrow \ell_{1}^{+}\ell_{2}^{+}W^{-}$
process is
\begin{eqnarray}
\frac{d\hat{\sigma}}{dp_{N}^{2}} &=&
\frac{g^{2}~ \vert V^{CKM'}_{ji}\vert^{2}~ \vert V_{\ell_{2}N} \vert^{2}}{3N_{C}~2^{11}~\pi^{3}~M_{W}^{2}~(1+\delta_{\ell_{1}\ell_{2}})}
\left(g^{q~2}_{R} + g^{q~2}_{L}\right)
\left(p_{N}^{2}g_{R}^{\ell\:2}\vert Y_{\ell_{1}N}\vert^{2} + m_{N}^{2}\vert V_{\ell_{1}N}\vert^{2} g^{\ell~2}_{L}\right)
\nonumber\\
&\times&
\frac{\hat{s}~(1-\tilde{\mu}_{N}^{2})^{2}(2+\tilde{\mu}_{N}^{2})}{\left[(\hat{s}-M_{W'}^{2})^{2}+(\Gamma_{W'}M_{W'})^{2}\right]}
\frac{p_{N}^{2}~(1-\rho_{W}^{2})^{2}(1+2\rho_{W}^{2})}{\left[(p_{N}^{2}-m_{N}^{2})^{2}+(\Gamma_{N}m_{N})^{2}\right]},
\label{diffXSecNoNWA.EQ}
\end{eqnarray}
where $\tilde{\mu}_{N}^{2}\equiv p_{N}^2/\hat{s}$ and $\rho_{W}^{2}\equiv M_{W}^{2}/p_{N}^{2}$.

\subsection*{B4. Partonic-Level Angular Distributions: L-Conserving Case}
\label{DiracAngDist.APP}

For comparison, we consider the case where the heavy neutrino decays through the
following $L$-conserving process:
\begin{equation}
 u_{i}\overline{d}_{j}\rightarrow W' \rightarrow \ell_{1}^{+}N\rightarrow\ell_{1}^{+}\ell_{2}^{-}W^{+}.
\end{equation}
Following the identical arguments specified in the preceding appendix, the subsequent polarization-dependent angular distributions are
\begin{eqnarray}
 \frac{d\hat{\sigma}_{0}}{d\Omega_{\ell_{2}}}&=&\frac{\hat{\sigma}(W_{0})}{2^{4}\pi}\times\{4\left[1-\left(\frac{2-\mu_{N}^{2}}{2+\mu_{N}^{2}}\right)
\left(\frac
{g_{R}^{\ell\:2}\vert Y_{\ell_{1}N}\vert^{2}-g_{L}^{\ell\:2}\vert V_{\ell_{1}N}\vert^{2}}
{g_{R}^{\ell\:2}\vert Y_{\ell_{1}N}\vert^{2}+g_{L}^{\ell\:2}\vert V_{\ell_{1}N}\vert^{2}}\right)\cos\theta_{\ell_{2}}\right]\nonumber\\
&+&\frac{3\pi \mu_{N}}{\left(2+\mu_{N}^{2}\right)}\left(\frac{g_{R}^{q\:2}-g_{L}^{q\:2}}{g_{R}^{q\:2}+g_{L}^{q\:2}}\right)\sin\theta_{\ell_{2}}\cos\phi_{\ell_{2}}\},
\end{eqnarray}
and
\begin{eqnarray}
 \frac{d\hat{\sigma}_{T1}}{d\Omega_{\ell_{2}}}=\frac{d\hat{\sigma}_{T2}}{d\Omega_{\ell_{2}}}&=&\frac{\hat{\sigma}(W_{T})}{2^{4}\pi}\times\{4\left[1+\left(\frac{2-\mu_{N}^{2}}{2+\mu_{N}^{2}}\right)
\left(\frac
{g_{R}^{\ell\:2}\vert Y_{\ell_{1}N}\vert^{2}-g_{L}^{\ell\:2}\vert V_{\ell_{1}N}\vert^{2}}
{g_{R}^{\ell\:2}\vert Y_{\ell_{1}N}\vert^{2}+g_{L}^{\ell\:2}\vert V_{\ell_{1}N}\vert^{2}}\right)\cos\theta_{\ell_{2}}\right]\nonumber\\
&-&\frac{3\pi \mu_{N}}{\left(2+\mu_{N}^{2}\right)}\left(\frac{g_{R}^{q\:2}-g_{L}^{q\:2}}{g_{R}^{q\:2}+g_{L}^{q\:2}}\right)\sin\theta_{\ell_{2}}\cos\phi_{\ell_{2}}\}.
\end{eqnarray}
The polarization-summed distributions for the polar and azimuthal cases are therefore
\begin{equation}
\frac{d\hat{\sigma}_{Tot.}}{d\cos\theta_{\ell_{2}}}=
\frac{\hat{\sigma}_{Tot.}}{2}\left[1-\left(\frac{\hat{\sigma}(W_{0})-\hat{\sigma}(W_{T})}{\hat{\sigma}(W_{0})+\hat{\sigma}(W_{T})}\right)\left(\frac{2-\mu_{N}^{2}}{2+\mu_{N}^{2}}\right)
\left(\frac
{g_{R}^{\ell\:2}\vert Y_{\ell_{1}N}\vert^{2}-g_{L}^{\ell\:2}\vert V_{\ell_{1}N}\vert^{2}}
{g_{R}^{\ell\:2}\vert Y_{\ell_{1}N}\vert^{2}+g_{L}^{\ell\:2}\vert V_{\ell_{1}N}\vert^{2}}\right)\cos\theta_{\ell_{2}}\right],
\end{equation}
and
\begin{equation}
\frac{d\hat{\sigma}_{Tot.}}{d\Phi}
=\frac{\hat{\sigma}_{Tot.}}{2\pi}\left[1+\frac{3\pi^{2}}{16}\frac{\mu_{N}}{\left(2+\mu_{N}^{2}\right)}\left(\frac{\hat{\sigma}(W_{0})-\hat{\sigma}(W_{T})}{\hat{\sigma}(W_{0})+\hat{\sigma}(W_{T})}\right)\left(\frac{g_{R}^{q\:2}-g_{L}^{q\:2}}{g_{R}^{q\:2}+g_{L}^{q\:2}}\right)\cos\Phi\right],
\end{equation}
respectively, where $\sigma_{Tot.}$ is still given by Eq.(\ref{sig0Til.EQ}).
Comparison to Eqs.~(\ref{angDistAppend.EQ}) and~(\ref{phiDistAppend.EQ}) demonstrates that the slopes of the angular distributions differ in sign for the $L$-violating and $L$-conserving cases. 
Consequentially, adding the $L-$conserving and $L-$violating distributions together results in the quantitative feature
\begin{equation}
\hat{\sigma}_{Tot.}
= \frac{d\hat{\sigma}_{Tot.}^{L}}{d\cos\theta_{\ell_{2}}} + \frac{d\hat{\sigma}_{Tot.}^{\not L}}{d\cos\theta_{\ell_{2}}}
= \pi\left[
\frac{d\hat{\sigma}_{Tot.}^{L}}{d\Phi} + \frac{d\hat{\sigma}_{Tot.}^{\not L}}{d\Phi}\right],
\end{equation}
where $L$ ($\not\!\!L$) denotes the lepton number-conserving (violating) angular distributions.


\pagebreak


\end{document}